\documentclass[times, twocolumn, twocolappendix]{aastex631}
\usepackage{amsmath, amssymb, bm}
\usepackage{subfigure}
\usepackage{hyperref}
\usepackage{CJK}
\usepackage[T1]{fontenc}

\newcommand{\obja}{2M0412}
\newcommand{\objb}{2M0434}
\newcommand{\objc}{2M0436}
\newcommand{\objd}{2M0450}
\newcommand{\obje}{CIDA 12}
\newcommand{\objf}{2M0508}

\newcommand{\uat}[2]{\href{http://vocabs.ands.org.au/repository/api/lda/aas/the-unified-astronomy-thesaurus/current/resource.html?uri=http://astrothesaurus.org/uat/#1}{#2 (#1)}}
\makeatletter

\shorttitle{Disks around mid-M Stars}
\shortauthors{Shi et al.}

\begin{document}
\begin{CJK*}{UTF8}{gbsn}
\title{Small and Large Dust Cavities in Disks around mid-M Stars in Taurus}

\author[0000-0001-9277-6495]{Yangfan Shi (施杨帆)}
\affiliation{Kavli Institute for Astronomy and Astrophysics, Peking University, Beijing 100871, China}
\affiliation{Department of Astronomy, Peking University, Beijing 100871, China}
\affiliation{European Southern Observatory, Karl-Schwarzschild-Str. 2, D-85748 Garching bei M\"{u}nchen, Germany}

\author[0000-0002-7607-719X]{Feng Long (龙凤)}
\altaffiliation{NASA Hubble Fellowship Program Sagan Fellow}
\affiliation{Lunar and Planetary Laboratory, University of Arizona, Tucson, AZ 85721, USA}

\author[0000-0002-7154-6065]{Gregory J. Herczeg (沈雷歌)}
\affiliation{Kavli Institute for Astronomy and Astrophysics, Peking University, Beijing 100871, China}
\affiliation{Department of Astronomy, Peking University, Beijing 100871, China}

\author[0000-0001-6307-4195]{Daniel Harsono}
\affiliation{Institute of Astronomy, Department of Physics, National Tsing Hua University, Hsinchu, Taiwan}


\author[0000-0002-7616-666X]{Yao Liu}
\affiliation{Purple Mountain Observatory \& Key Laboratory for Radio Astronomy, Chinese Academy of Sciences, Nanjing 210023, China}

\author[0000-0001-8764-1780]{Paola Pinilla}
\affiliation{Mullard Space Science Laboratory, University College London, Holmbury St Mary, Dorking, Surrey RH5 6NT, UK}

\author[0000-0001-5378-7749]{Enrico Ragusa}
\affiliation{Dipartimento di Matematica, Universit\`a degli Studi di Milano, Via Saldini 50, 20133, Milano, Italy}
\affiliation{Univ Lyon, Univ Lyon1, Ens de Lyon, CNRS, Centre de Recherche Astrophysique de Lyon UMR5574, F-69230 Saint-Genis-Laval, France}

\author[0000-0002-6773-459X]{Doug Johnstone}
\affiliation{NRC Herzberg Astronomy and Astrophysics, 5071 West Saanich Rd, Victoria, BC, V9E 2E7, Canada}
\affiliation{Department of Physics and Astronomy, University of Victoria, Victoria, BC, V8P 5C2, Canada}

\author[0000-0001-6906-9549]{Xue-Ning Bai}
\affiliation{Institute for Advanced Study and Department of Astronomy, Tsinghua University, 100084, Beijing, China}

\author[0000-0001-7962-1683]{Ilaria Pascucci}
\affiliation{Lunar and Planetary Laboratory, the University of Arizona, Tucson, AZ 85721, USA}

\author[0000-0003-3562-262X]{Carlo F. Manara}
\affiliation{European Southern Observatory, Karl-Schwarzschild-Str. 2, D-85748 Garching bei M\"{u}nchen, Germany}

\author[0000-0002-1078-9493]{Gijs D. Mulders}
\affil{Facultad de Ingenier\'ia y Ciencias, Universidad Adolfo Ib\'a\~nez, Av.\ Diagonal las Torres 2640, Pe\~nalol\'en, Santiago, Chile}
\affil{Millennium Institute for Astrophysics, Chile}

\author[0000-0002-2828-1153]{Lucas A. Cieza}
\affiliation{Instituto de Estudios Astrof\'isicos, Facultad de Ingenier\'ia y Ciencias, Universidad Diego Portales, Av. Ej\'ercito 441, Santiago, Chile}



\begin{abstract}

High-angular resolution imaging by ALMA has revealed the near-universality and diversity of substructures in protoplanetary disks. However, disks around M-type pre-main-sequence stars are still poorly sampled, despite the prevalence of M-dwarfs in the galaxy.
Here we present high-resolution ($\sim$50 mas, 8 au) ALMA Band 6 observations of six disks around mid-M stars in Taurus.
We detect dust continuum emission in all six disks, $^{12}$CO in five disks, and $^{13}$CO line in two disks.
The size ratios between gas and dust disks range from 1.6 to 5.1. The ratio of about 5 for \objc~and \objd~indicates efficient dust radial drift.
Four disks show rings and cavities and two disks are smooth.
The cavity sizes occupy a wide range: 
60 au for \obja, and $\sim$10 au for \objb, \objc~and~\objf. 
Detailed visibility modeling indicates that small cavities of 1.7 and 5.7 au may hide in the two smooth disks 2M0450 and CIDA 12.
We perform radiative transfer fitting of the infrared SEDs to constrain the cavity sizes, finding that micron-sized dust grains may have smaller cavities than millimeter grains.
Planet-disk interactions are the preferred explanation to produce the large 60 au cavity, while other physics could be responsible for the three $\sim$10 au cavities under current observations and theories. 
Currently, disks around mid-to-late M stars in Taurus show a higher detection frequency of cavities than earlier type stars, although a more complete sample is needed to evaluate any dependence of substructure on stellar mass.

\end{abstract}



\section{Introduction} \label{sec:intro}

The discovery to date of over 5000 exoplanets reveals that planetary systems occupy a wide parameter space in architecture (\citealt{zhudong2021}).  In planetary systems around low-mass stars (e.g., lower than 0.4 M$_\odot$), small planets occur more frequently than those around solar-mass stars
(e.g., \citealt{mulders2015, hardegree-ullman2019}), while giant planets around those low-mass stars are rare \citep[but have a non-zero occurrence rate, e.g.,][]{bryant2023}.

The existence of a few giant planets around mid-to-late (later than M3) M stars has challenged the core accretion planet formation theory (e.g., \citealt{morales2019,stefansson2023}). Planet population synthesis simulations often fail to form any giant planet around these very low-mass stars, in both the planetesimal accretion case (e.g., \citealt{miguel2020}) and the pebble accretion case (e.g., \citealt{liu2019, mulders2021, burn2021, schlecker2022}). Special circumstances, such as a reduction in the type-I migration velocities by a factor of 10, are needed to form planetary cores more massive than 10 $M_\oplus$ (\citealt{burn2021, schlecker2022}), which then leads to the runaway gas accretion that enables giant planet formation (\citealt{pollack1996}). 
Besides core accretion, direct collapse in gravitationally unstable disks may also form giant planets at early stages (e.g., \citealt{boss1997, mercer2020, longarini2023, boss2023}), when those class 0/I prostellar disks own higher disk mass than their later stage counterparts (e.g., \citealt{tychoniec2020,tobin2020}). 
Possible evidence of gravitational instability has been found in some recent ALMA observations, even though they targeted higher mass stars than mid-to-late M stars focused in this work \citep[e.g.,][]{perez2016, paneque-carreno2021,weber2023}.

In the past decade, high angular-resolution observations with the Atacama Large Millimeter/submillimeter Array (ALMA) have revealed that substructures are common in protoplanetary disks. These substructures are mostly seen as rings and gaps (e.g., \citealt{almapartner2015, andrews2018, long2018, cieza2021}), asymmetric substructures like arcs and spiral arms are also found but relatively rare (e.g., \citealt{vdMarel2013, huang2018_spiral, dong2018}). These substructures are expected to be induced by various disk physical processes and planet-disk interactions (see reviews by \citealt{andrews2020, bae2023}). 
The prevalence of rings suggests the presence of pressure bumps that halt the inward radial drift of the dust (\citealt{pinilla2012}). 

Thus far these conclusions are obtained mostly from observations of solar-type stars and Herbig stars.  For disks around M stars, the dust drift should be faster \citep{pinilla2013}, so their disks may need stronger pressure bumps to trap dust particles in order to explain their millimeter emission at ages of a few Myr (e.g., \citealt{vdPlas2016, sanchis2020}).
However, only little is known about substructures in disks around mid-to-late M stars. 
Most structured disks around M-dwarf stars show rings and gaps (\citealt{gonzalez-ruilova2020, kurtovic2021, pinilla2021, cieza2021, vdMarel2022}), where the gaps are mostly central cavities or large gaps surrounding a compact inner disk (\citealt{pinilla2022}). Only one disk around a mid M-dwarf shows a clear asymmetric ring, with properties similar to asymmetries found around T Tauri and Herbig AeBe stars (\citealt{hashimoto2021}). Detailed hydrodynamical simulations toward the disk around the mid-M-dwarf CIDA 1 suggest that a planet with minimum mass of $\sim 1.4 M_{\rm Jup}$ is needed to carve out the observed cavities present in 0.9 mm and 2.1 mm continuum images (\citealt{curone2022}), challenging the core accretion planet formation theory around very low mass stars. 


Following the pilot studies above, additional deep high-resolution observations on disks around M stars (or very low mass stars) are needed to evaluate the dust morphology and test disk physical processes in extreme situations, following similar strategies applied to solar-mass stars \citep[e.g.,][]{andrews2018,cieza2019,long2019}. 
This paper presents observations from a program designed to obtain high resolution observations ($\sim 0.05^{''}$) of 16 mid-M star disks in the Taurus star-forming region.  Of the proposed targets, we have obtained observations of six disks, including the double-ringed disk of 
2MASS J04124068+2438157, which was presented in \citet{long2023}.
Here we present all six observed disks (including 2M0412 from \citealt{long2023}) to study properties of substructures in M star disks and their possible origins.

The paper is organized as follows: in Section \ref{sec:obs}, we describe our target selection, ALMA observations, data reduction and calibration. In Section \ref{sec:results}, we present our modeling method in the visibility plane and the modeling results, as well as CO gas measurement. Section \ref{sec:discuss} discusses the global properties of the six disks and possible origins for their detected substructures. In Section \ref{sec:summary} we summarize our main findings.

\section{ALMA Program and Observations} \label{sec:obs}

\subsection{Sample Selection and Properties} \label{subsec:obs_host}
We analyze disks observed in the ALMA project 2019.1.00566.S (PI: G. Herczeg), which aimed at studying the dependence of dust substructure properties on stellar mass by targeting 16 disks around M3-M5 stars selected from Taurus. The sample selection started from \citet{luhman2017} with disk identification based on excess emission in WISE W2 and W3 bands. We excluded known binaries (e.g., \citealt{kraus2011, daemgen2015}) and sources with $A_V>3$  mag or $J-$band brightness 1 mag fainter than the median to avoid edge-on disks and embedded objects. Furthermore, targets with declination $>$26 deg are excluded for visibility purposes. Like \citet{long2019}, the selection ignored the millimeter brightness of disks.
Six out of the 16 proposed targets were observed with the C43-9/10 configuration and Band 6 receivers between 2021-08-27 and 2021-09-27. 

For these six sources, their stellar properties (listed in Table~\ref{tab:host_properties}) were re-evaluated based on optical spectra when available. For 2M0450 and 2M0508, the spectral type and extinction are obtained from \citet{luhman18}, with uncertain accretion properties.
We obtained flux-calibrated low-resolution optical spectra for 2M0412, 2M0434, and 2M0436 using UH88/SNIFS (3200--10000 \AA, \citealt{lantz04}) and for 2M0507 (CIDA12) using Palomar-Hale 5m/DBSP (3200--8700 \AA\ with a gap from 5600--6250 \AA, \citealt{oke82}).  The SNIFS data reduction is described by  \citet{guo18}, while the DBSP data reduction is described by \citet{herczeg14}.  For these four targets, the spectral type, extinction, and accretion rate are calculated from a simultaneous fit, following approaches described in \citet{herczeg08} and \citet{herczeg14} and using a plane-parallel slab model for accretion developed by \citet{valenti93}.  
The properties of 2M0412 were presented in \citet{long2023}.  

The spectral type for each star is converted to temperature using the relationship of \citet{herczeg14}.  The luminosity is then calculated from the 2MASS $J$-band magnitude \citep{cutri03} and assuming zero $J$-band veiling, $J$-band bolometric corrections from \citet{pecaut13}, and $J$-band extinction from the $A_J/A_V$ ratio developed by \citet{wang19}.  All properties are calculated from distances obtained from inverting the parallax in \cite{gaiadr3}. 
The masses are then estimated using the \citet{somers20} evolutionary tracks for pre-main sequence stars with 51\% spots coverage.  These tracks lead to higher masses than standard evolutionary tracks \citep[see discussion for 2M0412 in][]{long2023}.
 




\begin{deluxetable*}{lclcccccccc}
    \label{tab:host_properties}
    \tablecaption{Properties of Host Stars}
    \tablehead{\colhead{2MASS $|$ Short Name} & \colhead{D} & \colhead{SpT} & \colhead{$A_V$} &
    \colhead{$J$} & \colhead{L$_{\rm phot}$} & \colhead{$M_*$} & \colhead{$R_*$} & \colhead{L$_{\rm acc}$} & 
    \colhead{M$_{\rm acc}$} & \colhead{Refs} \\
     & \colhead{[pc]} & & \colhead{[mag]} & & \colhead{[L$_\odot$]} & \colhead{[M$_\odot$]} & 
     \colhead{[R$_\odot$]} & \colhead{[L$_\odot$]} & \colhead{[M$_\odot$ yr$^{-1}$]}  &
    }
    \startdata
    J04124068+2438157 $|$ 2M0412  & 148.7 & M4.3  & 0.84 & 11.151 & 0.126 & 0.30 & 1.21 & 0.00204 & 3.2e-10 & L23 \\ 
    J04343128+1722201 $|$ 2M0434  & 145.7 & M4.3  & 0.30 & 11.205 & 0.115 & 0.30 & 1.16 & 0.00168 & 2.6e-10 & --  \\ 
    J04360131+1726120 $|$ 2M0436  & 144.5 & M2.7  & 1.08 & 11.105 & 0.146 & 0.58 & 1.07 & 0.0062  & 4.5e-10 & --  \\ 
    J04504003+1619460 $|$ 2M0450  & 144.4 & M4.75 & 0.0  & 11.725 & 0.054 & 0.24 & 0.84 & --      & --       & L18 \\ 
    J05075496+2500156 $|$ CIDA 12 & 170.0 & M3.7  & 0.5  & 11.415 & 0.13  & 0.39 & 1.13 & 0.0028  & 3.2e-10 &  HH14 \\ 
    J05080709+2427123 $|$ 2M0508  & 170.7 & M3.5  & 0.0  & 11.396 & 0.11  & 0.44 & 1.03 & --      & --       & L18 
    \enddata
    \tablerefs{L23: \citet{long2023}; L18: \citet{luhman18}; HH14: \citet{herczeg14}.}
\end{deluxetable*}


\subsection{ALMA Observations} \label{subsec:obs_alma}

For our ALMA observations, the receivers were configured into four spectral windows, with two centered at 217.0 and 234.4 GHz for dust continuum with bandwidths of 1.875 GHz, and the other two centered at 220.0 and 230.5 GHz targeting $^{12}$CO, $^{13}$CO, and C$^{18}$O $J=2-1$ lines. Line channel intervals are spaced at 0.244 MHz ($\sim$ 0.3 km/s). Table \ref{tab:obs_log} shows the detailed ALMA observation log.

The raw visibilities were pipeline calibrated using the specified CASA versions (6.1.1 for \obja,~\obje~\&~\objf; 6.2.1 for the rest) for each object that can be found in the QA2 report \citep{casa2022}. 
We identified residual features of atmospheric absorption correction around channel 500 in the 234.4 GHz spectral window for all six targets and flagged corresponding channels 400-600. For dust continuum imaging, we flagged the channels  within 25 $\rm km\ s^{-1}$ of the rest frame of CO lines and then averaged the dataset with a channel width of 125 MHz, which is the recommended maximum bandwidth to avoid bandwidth smearing for ALMA Band 6. 

These observations were intended to provide snapshots of these disks\footnote{Young M-stars can happen to be active in millimeter wavelength in time scale of minutes (e.g., \citealt{macgregor2018, mairs2019}), which would be an issue for disk analysis especially for compact sources like \objd~and CIDA 12. After performing time series measurement of the fluxes for \objd~and CIDA 12, no significant flux variations are found during our observations.}.  Due to the short on-source time ($\sim$15 min), the peak signal-to-noise (S/N) for each source ranges from 5--20. We attempted one round of phase-only self-calibration on sources 2M0412, 2M0434, and 2M0450. For 2M0436 and 2M0508, the emission morphology was largely altered after self-calibration, so we did not apply the solutions. CIDA 12 has too low S/N to improve from self-calibration. The self-calibration solutions are not applied to CO line emission channels, since the improvement is negligible.

\begin{figure*}
    \centering
    \includegraphics[scale=0.42]{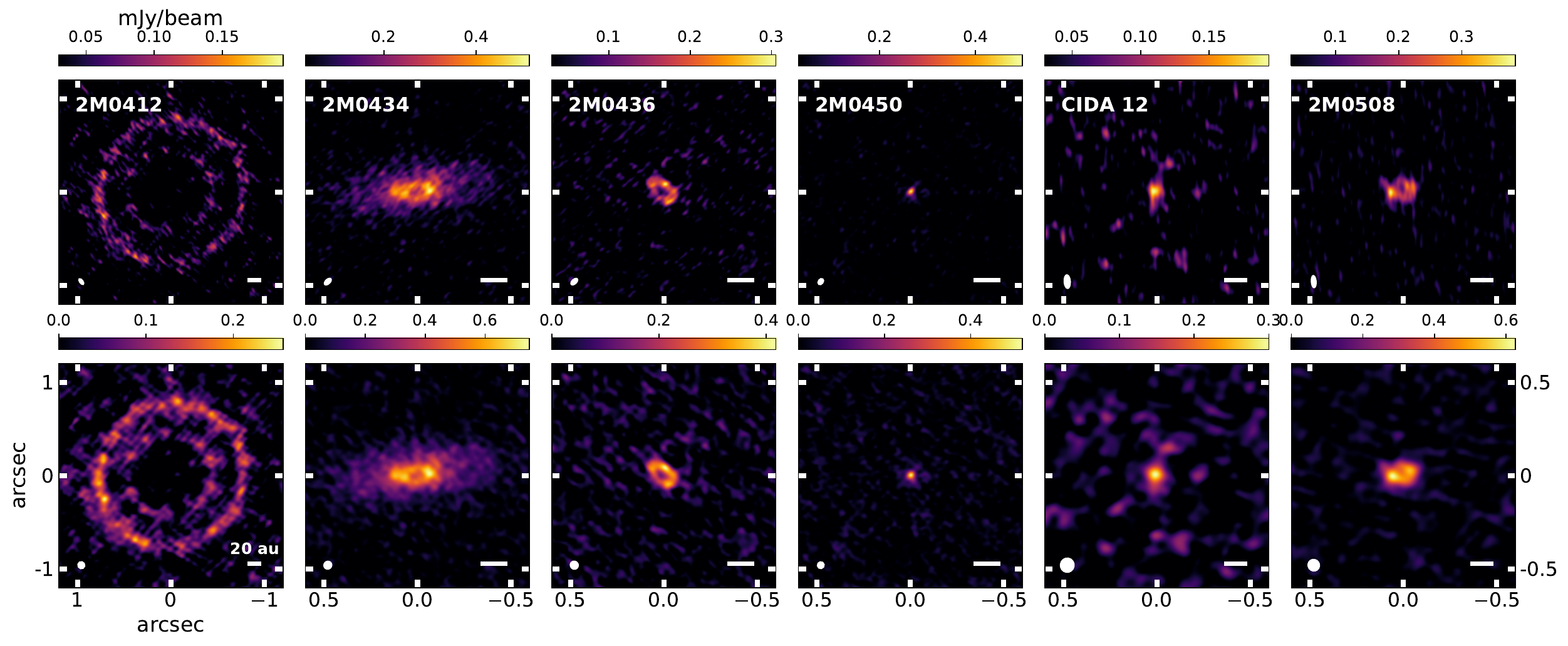}
    \caption{The top panels show ALMA continuum images of disks around mid-M stars in Taurus at 1.3mm. The box size for \obja~is $2.4''$ and the rest are $1.2''$. The white scale bars represent 20 au. The bottom panels are for the same targets with beams tapered to a more circular shape for better visualization. The beam sizes and rms noise of each image are detailed in Table \ref{tab:obs_log}.}
    \label{fig:continuum}
\end{figure*}

The continuum images were generated using the \texttt{`tclean'} task with multiscale imaging, with the Briggs \texttt{`robust'} weighting parameters being 0.5 for 2M0434, 2M0436, 2M0450 and 2M0508, 1 for 2M0412 and 2 for CIDA 12. The adopted weighting parameters are compromises between resolution and sensitivity. The uv coverage of observations produced elongated beams (aspect ratio $\sim$2). For better visualization, we performed uv-tapering to get more circular beams. Images with original beams and uv-tapered beams are shown in Figure \ref{fig:continuum}. The beam sizes are $0.03'' \sim 0.07''$ and the RMS noise is 30-40 mJy/beam, with detailed values summarized in Table \ref{tab:obs_log}.

For line images, after subtracting the continuum emission 
using \texttt{`uvcontsub'} task we applied uv-tapering to enhance the SNR of the line images. The beams are tapered to $0.2''$ for \obja\ and \objb\ and $0.1''$ for the other four targets. We detected (or marginally detected) $^{12}$CO for all six targets and $^{13}$CO for \obja\ and \objb. The channel maps in velocity and moment maps of detected $^{12}$CO and $^{13}$CO are shown in Appendix \ref{sec:app_co}.

\begin{deluxetable*}{llcccccccc}
    \label{tab:obs_log}
    \tablecaption{Summary of ALMA observations}
    \tablehead{\colhead{Source} & \colhead{Obs. Date} & \colhead{$N_{\rm ant}$} & \colhead{Baselines} & \colhead{On-source Time} & \colhead{Mean PWV} & \colhead{Peak $I_\nu$ (taper)} & \colhead{RMS Noise (taper)} & \colhead{Beam (taper)}
    \\[0.05cm] 
      & & & \colhead{[m]} & \colhead{[min]} & \colhead{[mm]} & \colhead{[mJy/beam]} & \colhead{[mJy/beam]} & \colhead{[mas$\times$mas, deg]}
     }
    \startdata
    2M0412  & 2021-08-27 & 38 & 92.1-10803.3  & 16.80 & 0.3 & 0.19 (0.26) & 0.032 (0.039) & 74$\times$35, 37 (76$\times$73, 32) \\
    2M0434  & 2021-09-27 & 45 & 70.1-14361.8  & 15.29 & 1.0 & 0.47 (0.69) & 0.039 (0.041) & 45$\times$24, -47 (43$\times$41, -41) \\
    2M0436  & 2021-09-27 & 45 & 70.1-14361.8  & 15.39 & 1.0 & 0.31 (0.42) & 0.037 (0.039) & 44$\times$24, -46 (43$\times$42, 38) \\
    2M0450  & 2021-09-28 & 47 & 70.1-14361.8  & 15.05 & 0.7 & 0.23 (0.27) & 0.029 (0.029) & 36$\times$26, -40 (36$\times$36, -35) \\
    CIDA 12 & 2021-09-13 & 37 & 178.3-12594.5 & 16.80 & 0.6 & 0.19 (0.30) & 0.031 (0.036) & 72$\times$33, 4 (76$\times$73, -9) \\
    2M0508  & 2021-09-13 & 37 & 178.3-12594.5 & 16.87 & 0.6 & 0.38 (0.63) & 0.035 (0.039) & 65$\times$25, 4 (63$\times$60, 36)
    \enddata
    \tablecomments{The numbers in parenthese in last three columns are corresponding values for uv-tapered images.}
\end{deluxetable*}

\section{Results} \label{sec:results}
From our ALMA observations, four disks in the sample show rings and cavities in their dust emission, while the other two show compact smooth emission. In this section, we characterize the dust emission through fitting the directly observable visibilities and present the measurement of dust and gas disk properties.

\subsection{Visibility Fitting of Dust Morphology}
\label{subsec:result_uvfit}
The peak fluxes of our disks range from 0.19 mJy/beam to 0.47 mJy/beam, which corresponds to peak SNRs about 6 to 12 from the non-uv-tapered images in Figure \ref{fig:continuum}. Dust emission from these disks is faint, as expected for low-mass M stars.
To quantify the dust emission morphology without biases introduced by the image reconstruction, we fit the brightness profiles in the visibility domain. The model profiles are selected based on eye identification of disk substructures. For \objd~and \obje, the function to describe their deprojected brightness is a centrally peaked axis-symmetric Gaussian profile (GP):
\begin{equation}
    \label{eq:GP}
    I_{\rm GP} (r) = 10^{f} \exp{\left(- \frac{r^2}{2\sigma^2}\right)}.
\end{equation}
where $I_{\rm GP} (r)$ is the Gaussian brightness profile as a function of disk radius $r$, with $10^f$ and $\sigma$ being the amplitude and Gaussian width.
For the other four disks showing cavities and rings, we modeled their morphologies with a radially asymmetric Gaussian ring whose inner and outer width can differ:
\begin{equation}
    \label{eq:AGR}
    I_{\rm AGR} = 
    \begin{cases}
        10^{f}\exp{\left(- \frac{(r-r_{\rm peak})^2}{2\sigma^2_1}\right)} \ (r\leq r_{\rm peak}). \\
        10^{f}\exp{\left(- \frac{(r-r_{\rm peak})^2}{2\sigma^2_2}\right)} \ (r>r_{\rm peak}).
    \end{cases}
\end{equation}
where $10^f$, $r_{\rm peak}$, and $\sigma_i$ are the amplitude, peak location and ring width for the inner and outer side.
A radially asymmetric Gaussian ring has been previously used to describe the rings in transition disks (e.g., \citealt{pinilla2017, pinilla2018, huang2020, kurtovic2021}). This model fitting is motivated by the accumulation of dust particles trapped in radial pressure bumps (\citealt{pinilla2017}). Dust particles in the outer disk are expected to take a longer time to grow and drift radially toward the pressure bump. Hence, the external width of the ring is expected to be larger than the internal width (e.g., Figure 4 in \citealt{pinilla2015}). 

Using the adopted radial profiles, we first generate corresponding face-on images.  Each model image is projected with an inclination ($i$) and a position angle (PA) and combined with a spatial offset ($\delta_{\rm RA}, \delta_{\rm Dec}$) which adds another four parameters. We generate each model image with a pixel size of 3 mas, which is far smaller than our synthesized beam ($\sim$ 40 mas). We then use \texttt{galario} \citep{tazzari2018} to transform each model image into complex visibilities sampled at the same $(u,v)$ points as those in the observations and calculate the $\chi^2 = \sum_k w_k|V_{\rm obs}(u_k, v_k) - V_{\rm mod}(u_k, v_k)|^2$, where $w_k$ is the observed visibility weight, the likelihood function is then calculated as $L \propto \exp(-\chi^2/2)$. We assumed a uniform prior probability distribution for the parameters above (for inclination the probability $\propto \sin i$ to get a uniform disk orientation).

\begin{figure*}[!th]
    \centering
    \includegraphics[scale=0.46]{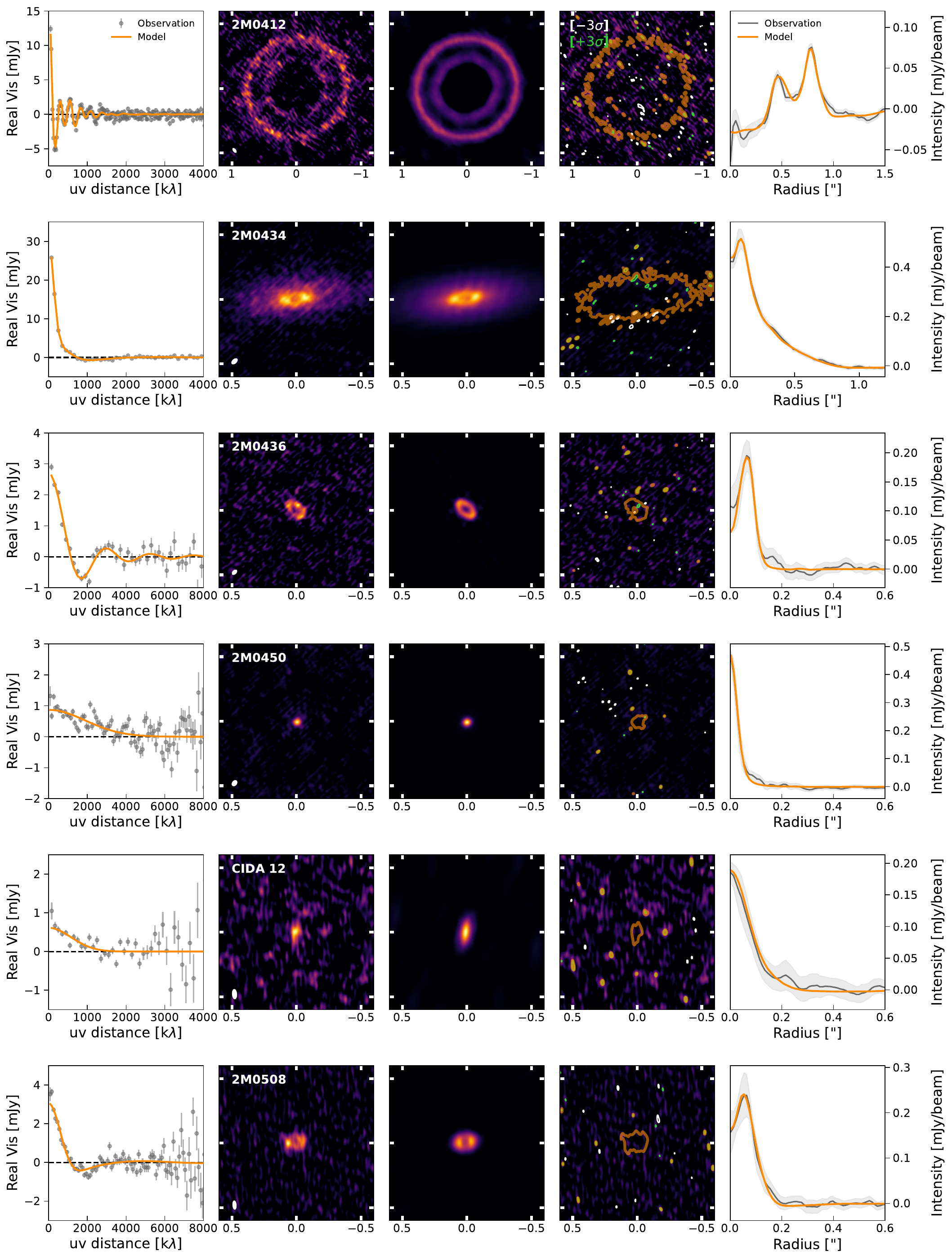}
    \caption{Best-fit visibility fitting results versus observations. \textit{Columns from left to right:} (1) the real part of deprojected and binned visibilities from observations and best-fit models, (2) observational continuum images, (3) model images convolved with the same beam as observations, (4) residual images with white contours at $-3\sigma$ and green contours at $3\sigma$, overlaid orange contours are from continuum images at $3\sigma$ level (5) azimuthally averaged radial profiles for observations and convolved model images, light gray shaded regions represent the standard deviation at each radial bin divided by the square root of the number of beams in the radial bin.}
    \label{fig:model_compare}
\end{figure*}

We sample the posterior probability distribution using a Markov Chain Monte Carlo (MCMC) method (\texttt{emcee}, \citealt{foremanMackey2013}) with 50 walkers and 10000 steps. After running MCMC, we get posterior samples after 10 times the integrated autocorrelation time to ensure stationary posterior samples. Results for each model parameter are shown in Table \ref{tab:uvfit_results}, with the reference values adopted as the median values of posterior samples and their uncertainties estimated from the 16th and 84th percentiles. The comparisons between observations and models in the visibility plane, the image plane, and deprojected azimuthally averaged radial profiles are shown in Figure \ref{fig:model_compare}. The brightness profiles of each model are shown in Figure \ref{fig:intensity}. 
 All best-fit model images show reasonable matches with observations, with no residuals above $5\sigma$. 


\subsection{Rings and cavities} 


\subsubsection{Disks with substructures in images}
\label{subsec:result_substructure}
As mentioned in Section \ref{subsec:result_uvfit}, rings are modeled as radially asymmetric Gaussian rings. Our results show that the rings generally have wider outer width than the inner, while \objc~shows the opposite. Here we summarize the peak locations and Gaussian $\sigma$ widths from the best-fit models.

\obja~has the largest rings and cavity in the sample, with the inner ring peak at 61.7 au and the outer ring peak at 113.9 au. The inner ring has an inner width of 0.9 au and outer width of 17.5 au, while the outer ring has 5.5 au inner side and 9.5 au outer side.
For the ratios between the ring outer width and inner width, best-fit model gives 19.8 for the inner ring and 1.7 for the outer ring. The outer ring is brighter than the inner ring with the outer peak intensity 2.1 times as large as the inner peak. The brightness ratio between the first ring and the exterior gap is around 6.8.

For \objb, we find a ring peak at 9.8 au, followed by radially extended emission beyond the ring. The ring has 1.5 au inner width and 8.2 au outer width. A Gaussian profile in Section \ref{subsec:result_uvfit} fits the outer broad emission well. For \objf, the best model finds a ring peak at 6.8 au, where its inner width is 0.8 au and outer width is 7.1 au. In \objc, the best-fit ring peaks at 12.2 au, with an outer width being 1.0 au and a larger inner width being 4.7 au showing different ring width behavior.

Some potential asymmetries appear in the images.
For example, the northern part of 2M0434 seems to be brighter than its southern part as shown by the residual map and a bright spot appears at the west-north of 2M0436, which are both at $3\sigma$ levels. The west part of 2M0508 is brighter though the significance is less than $3\sigma$ level.
Deeper observations are necessary to check whether above potential asymmetries exist or not. In this work, we focus on only axis-symmetric substructures.

\subsubsection{Potential cavities around \objd~and \obje}
\label{subsec:result_nuker}
Gaussian profiles (Equation \ref{eq:GP}) can only describe continuous disks with a monotonically decreasing radial profile. 
The Nuker profile, which was first introduced to describe the brightness profiles of elliptical galaxies (\citealt{lauer1995}), can also reproduce ring-like emission with a central depression. The Nuker profile is formulated as:
\begin{equation} \label{eq:nuker}
    I_\nu(r) = 10^f \left(\frac{r}{r_t}\right)^{-\gamma} \left[1+\left(\frac{r}{r_t}\right)^{\alpha}  \right]^{(\gamma-\beta)/\alpha}.
\end{equation}
where $r_t$ is the transition radius, $\alpha$ is the transition index, $\beta$ is the outer disk index, and $\gamma$ is the inner disk index. The Nuker profile behaves as $r^{-\gamma}$ at $r\ll r_t$ and $r^{-\beta}$ at $r\gg r_t$. The transition index $\alpha$ determines how smooth/sharp the transition is between these two asymptotic behaviors (see Figure 2 in \citealt{tripathi2017}). 
With $\beta > 0,\ \gamma < 0$ the Nuker profile can describe a ring-like morphology.

Appendix \ref{sec:app_nuker} presents the detailed modeling with Nuker profiles on \objd~and \obje. Interestingly, the best-fit Nuker profiles for the two smooth-appearing disks show depleted inner emission with emission peak at $\sim$ 1.7 au for \objd~and 5.7 au for \obje~(see Figure \ref{fig:nukerimage}). 

However, the Nuker profile has more parameters than the Gaussian profile, which results in a larger Bayesian Information Criterion\footnote{The Bayesian Information Criterion (BIC) is a criterion for the preference of models. It is defined as ${\rm BIC}=k\ln (n) - 2\ln (\hat{L})$, where $k$ is the number of model parameters, $n$ is the number of data points, and $\hat{L}$ is the maximized value of the likelihood function.} value. 
Hence, we still treat \objd~and \obje~as non-structured disks in this work.
Future higher angular resolution observations are needed to check their potential small cavities.

\subsection{Potential unresolved central emission in \objc} \label{subsec:result_2M0436}

\objc~shows a ring with a wider inner width, which contradicts with models of dust trapping in pressure bump. One possibility is that an external object has truncated the outer disk, 
which we expect to be substellar since no stellar companions have been detected (\citealt{kraus2011, kraus2012}). Constraining the mass of the substellar object given current knowledge would be highly uncertain and beyond the scope of this work.
Other possibilities are the presence of central compact emission blending with the ring emission and the blending of multiple unresolved rings.  Compact emission at the center is also hinted at because the model intensity of \objc~slightly underpredicts the emission of the very inner disk (right column in Figure \ref{fig:model_compare}), though this difference is not statistically significant.
An example is CIDA 1, where a model considering only one ring results in a ring with wider inner width (\citealt{kurtovic2021}), while higher-angular resolution observations reveal a compact inner disk and the inclusion of an inner disk recovers a ring with wider outer width (\citealt{pinilla2021}).

In Appendix \ref{sec:app_central_2m0436}, we show the results of modeling \objc~with a ring and a central point. The flux of the added central point emission is $0.04^{+0.03}_{-0.03}$ mJy (corner plot statistics in Figure \ref{fig:2M0436_cornerplot}). The new radial profile increases the inner emission and matches the observations better than a ring-only model, while the ring still displays a wider inner width after including a point emission.
However, the result of modeling with central point emission could be limited by current observation, deeper sensitivity and higher angular resolution are necessary for future observations.

\subsection{Millimeter Flux, Dust Disk Mass and Size}
\label{subsec:result_flux_mass_size}
The millimeter fluxes, masses and sizes of the dust disk are not explicit parameters in the models and are inferred from fitting results. Since our models are axis-symmetric, the total fluxes are obtained through $F_\nu=f_\nu(\infty)$, where $f_\nu(r)$ is the flux within radius $r$:
\begin{equation}
    f_\nu (r) = 2 \pi \int_0^r I_\nu (r') r' {\rm d}r'.
\end{equation}
The dust disk sizes are defined to be a radius that encloses 68\% and 90\% of total fluxes, consistent with choices from previous studies \citep[see, for example, the review by][]{miotello2023}. The continuum fluxes, sizes, and their uncertainties are then computed from the last 5000 steps of the chains, as shown in Table \ref{tab:uvfit_results}. Most adopted models reproduce lower fluxes than those seen at short baselines (left column in Figure \ref{fig:model_compare}), likely due to the fact that either large scale emission is resolved out by high angular resolution beams or faint emission buried by the noise, so that we lack extra components in our modeling which could account for the flux difference between models and observations. Future observations with deeper sensitivity and shorter baselines are needed to fix this issue.

For \obje, Nuker profiles reproduce millimeter flux better than that from Gaussian profiles. The visibility near zero-spacing baselines indicates its flux above 1 mJy and it is consistent with $1.16^{+0.09}_{-0.09}$ mJy reported in \citet{akeson2019}, specifically the Gaussian profile gives $0.60^{+0.06}_{-0.06}$ mJy, while the Nuker profile recovers a flux of $1.20^{+0.98}_{-0.51}$ mJy.

Under the assumption that the dust disk is optically thin at millimeter wavelengths, the dust disk mass $M_{\rm dust}$ is estimated following \citet{hildebrand1983}, as
\begin{equation}
    M_{\rm dust} \approx \frac{D^2F_\nu}{\kappa_\nu B_\nu(T_{\rm dust})}.
\end{equation}
where $D$ is distance from \textit{Gaia} DR3 \citep{gaiadr3}, $\kappa_\nu$ is dust opacity for which we adopt a power law of $\kappa_\nu=2.3\  {\rm cm}^2 {\rm g}^{-1} (\nu/230\ {\rm GHz})^{0.4}$ \citep{andrews2013}, and $B_\nu$ is the Planck function where the $T_{\rm dust}$ is assumed at $20\ {\rm K}$ (e.g., \citealt{ansdell2016}).

We assume 10\% uncertainty in the source flux calibration (\citealt{diaz_trigo_maria_2019_4511962}, see also \citealt{francis2020}) to calculate the dust mass uncertainty. Table \ref{tab:uvfit_results} summarizes the estimated dust masses. \objb~has the most massive disk in our sample, with a dust mass $19.31^{+1.93}_{-1.93} M_\oplus$, while \obje~has the lowest dust mass with only $0.51^{+0.05}_{-0.05} M_\oplus$. 
We note that these dust masses serve as lower limits, since part of the disk could be optically thick in reality (see Section \ref{subsec:discussion_evolution} for a brief calculation). The assumed dust temperature would also affect the estimated 
dust mass: for example, the dust mass would be higher/lower by $\sim 40\%/50\%$ if the actual averaged dust temperature is 15~K/30~K. 


\begin{deluxetable*}{cDDDDDDc}[!th]
    \tablecaption{Dust Disk Model Parameters from uv Modeling} \label{tab:uvfit_results}
    \tablehead{
    \colhead{Source Name} & \twocolhead{2M0412} & \twocolhead{2M0434}  & \twocolhead{2M0436} & \twocolhead{2M0450} & \twocolhead{CIDA 12} & \twocolhead{2M0508} & \colhead{Unit}  \\
    \colhead{ Model }     & \twocolhead{2AGR}   & \twocolhead{GP\&AGR} & \twocolhead{AGR}    & \twocolhead{GP}     & \twocolhead{GP}     & \twocolhead{AGR}    & } 
    \decimals
    \startdata
    $f_1$         & $8.63^{+0.03}_{-0.03}$     & $9.83^{+0.01}_{-0.01}$   & $9.98^{+0.03}_{-0.03}$  & $10.45^{+0.05}_{-0.05}$ & $9.54^{+0.09}_{-0.08}$    & $9.93^{+0.03}_{-0.03}$  & $\log_{10}$ Jy/sr \\
    $r_{\rm peak,1}$         & $415.10^{+8.71}_{-6.41}$   & -.-                      & $84.68^{+6.05}_{-7.35}$ & -.-                     & -.-                       & $39.80^{+6.38}_{-4.54}$ & mas \\
    $\sigma_{11}$ & $5.93^{+6.51}_{-4.21}$     & $268.38^{+3.04}_{-3.02}$ & $32.90^{+5.84}_{-6.23}$ & $16.89^{+1.44}_{-1.28}$ & $54.06^{+11.11}_{-11.15}$ & $4.69^{+5.94}_{-3.47}$  & mas \\
    $\sigma_{12}$ & $117.71^{+13.71}_{-11.45}$ & -.-                      & $6.88^{+5.34}_{-4.63}$  & -.-                     & -.-                       & $41.91^{+3.61}_{-4.01}$ & mas \\
    $f_2$         & $8.96^{+0.02}_{-0.02}$     & $9.95^{+0.03}_{-0.03}$   & -.-                     & -.-                     & -.-                       & -.-                     & $\log_{10}$ Jy/sr \\
    $r_{\rm peak,2}$         & $766.27^{+7.44}_{-14.94}$  & $67.41^{+8.91}_{-7.60}$  & -.-                     & -.-                     & -.-                       & -.-                     & mas \\
    $\sigma_{21}$ & $37.05^{+7.31}_{-12.02}$   & $10.15^{+7.94}_{-6.71}$  & -.-                     & -.-                     & -.-                       & -.-                     & mas \\
    $\sigma_{22}$ & $63.74^{+8.73}_{-5.68}$    & $56.22^{+4.81}_{-5.52}$  & -.-                     & -.-                     & -.-                       & -.-                     & mas \\[0.15cm]
    \hline 
    Inc                & $15.90^{+0.70}_{-0.82}$  & $68.54^{+0.23}_{-0.23}$ & $53.46^{+1.41}_{-1.49}$ & $43.26^{+7.84}_{-10.37}$  & $65.53^{+7.43}_{-12.65}$ & $54.34^{+1.67}_{-1.78}$ & deg \\
    PA                 & $124.14^{+2.81}_{-2.06}$ & $96.67^{+0.24}_{-0.24}$ & $46.96^{+2.03}_{-2.05}$ & $63.01^{+12.43}_{-12.87}$ & $169.94^{+5.22}_{-6.59}$ & $98.90^{+2.10}_{-2.10}$ & deg \\
    $\Delta_{\rm RA}$  & $2.77^{+1.86}_{-2.06}$   & $10.89^{+1.03}_{-1.05}$ & $7.86^{+1.12}_{-1.15}$  & $-3.13^{+0.96}_{-0.97}$   & $9.05^{+3.21}_{-3.30}$   & $12.36^{+1.38}_{-1.40}$ & mas \\
    $\Delta_{\rm Dec}$ & $-0.69^{+2.11}_{-2.12}$  & $10.37^{+0.50}_{-0.49}$ & $5.76^{+1.11}_{-1.11}$  & $-8.98^{+0.91}_{-0.92}$   & $-2.15^{+6.02}_{-6.18}$  & $5.77^{+1.10}_{-1.12}$  & mas \\[0.12cm]
    \hline
    $F_{\rm 1.3mm}$ & $17.58^{+0.33}_{-0.32}$  & $30.73^{+0.26}_{-0.24}$ & $2.71^{+0.05}_{-0.05}$  & $0.85^{+0.03}_{-0.03}$ & $0.60^{+0.06}_{-0.06}$  & $3.02^{+0.09}_{-0.09}$  & mJy \\
    $M_{\rm dust}$  & $11.50^{+1.15}_{-1.15}$  & $19.31^{+1.93}_{-1.93}$ & $1.68^{+0.17}_{-0.17}$  & $0.53^{+0.05}_{-0.05}$ & $0.51^{+0.05}_{-0.05}$  & $2.61^{+0.26}_{-0.26}$  & $M_\oplus$ \\
    $R_{68\%}$      & $118.0^{+0.5}_{-0.3}$ & $55.1^{+0.5}_{-0.5}$ & $11.8^{+0.2}_{-0.2}$ & $3.6^{+0.3}_{-0.3}$ & $13.8^{+2.8}_{-2.9}$ & $15.6^{+0.5}_{-0.5}$ & au \\
    $R_{90\%}$      & $126.0^{+0.9}_{-0.6}$ & $81.2^{+0.8}_{-0.8}$ & $13.2^{+0.4}_{-0.3}$ & $5.1^{+0.5}_{-0.4}$ & $19.6^{+4.0}_{-4.1}$ & $20.4^{+0.8}_{-0.8}$ & au \\[0.12cm]
    \enddata
    \tablecomments{Model row: AGR for asymmetric Gaussian ring, GP for Gaussian profile.}
\end{deluxetable*}

\subsection{CO emission}
\label{subsec:result_co}
Gas disk observations and comparisons to dust disks are helpful to understand disk evolution as a whole.
To compute the integrated fluxes of CO lines, extraction regions need to be determined. We first generate the cumulative flux distributions and determine the radii where CO emission no longer increases. Then we draw circular regions as large as these radii, and project the regions with the same inclination and position angle as those of dust disks. CO fluxes are then retrieved from these elliptical regions.
The uncertainties of the integrated flux are measured in non-emission channels with the same velocity range as those used to generate moment-0 images, as the standard deviation of integrated fluxes estimated from 1000 randomly distributed elliptical regions with the same shape as the above extraction regions. 
The fluxes and uncertainties of $^{12}$CO, $^{13}$CO and C$^{18}$O $J=2-1$ lines are listed in Table \ref{tab:co_flux}, as well as the $3\sigma$ upper limits for the non-detections.

We detect (SNRs of integrated fluxes above $3\sigma$) $^{12}$CO in five disks and $^{13}$CO in \obja~\&~\objb. \objd~has a marginal detection of $^{12}$CO with SNR of $2.7\sigma$. 
The gas disk sizes are measured the same way as dust disk sizes, adopting the same inclination and position angles from the fits to the dust continuum emission.
\objd~has the smallest gas disk about 26 au while \obja~has the largest gas disk about 199 au. When compared to dust disk sizes, the ratio ranges from 1.6 (\obja) to 5.1 (\objd).
Since the largest angular scale recovered by our observations is only $\sim 0.6''$, the gas emission at large radii is likely resolved out; cloud absorption could also result in underestimation of disk sizes (e.g., \citealt{long2022}); by 10\%-30\% with a typical cloud emission line width of 1-2 $\rm km \ s^{-1}$, we treat the gas radii $R_{\rm gas, 90\%}$ as lower limits. 
The measured gas disk sizes and size ratios compared to dust disks are summarized in Table \ref{tab:co_flux}.

\begin{figure*}[!th]
    \plotone{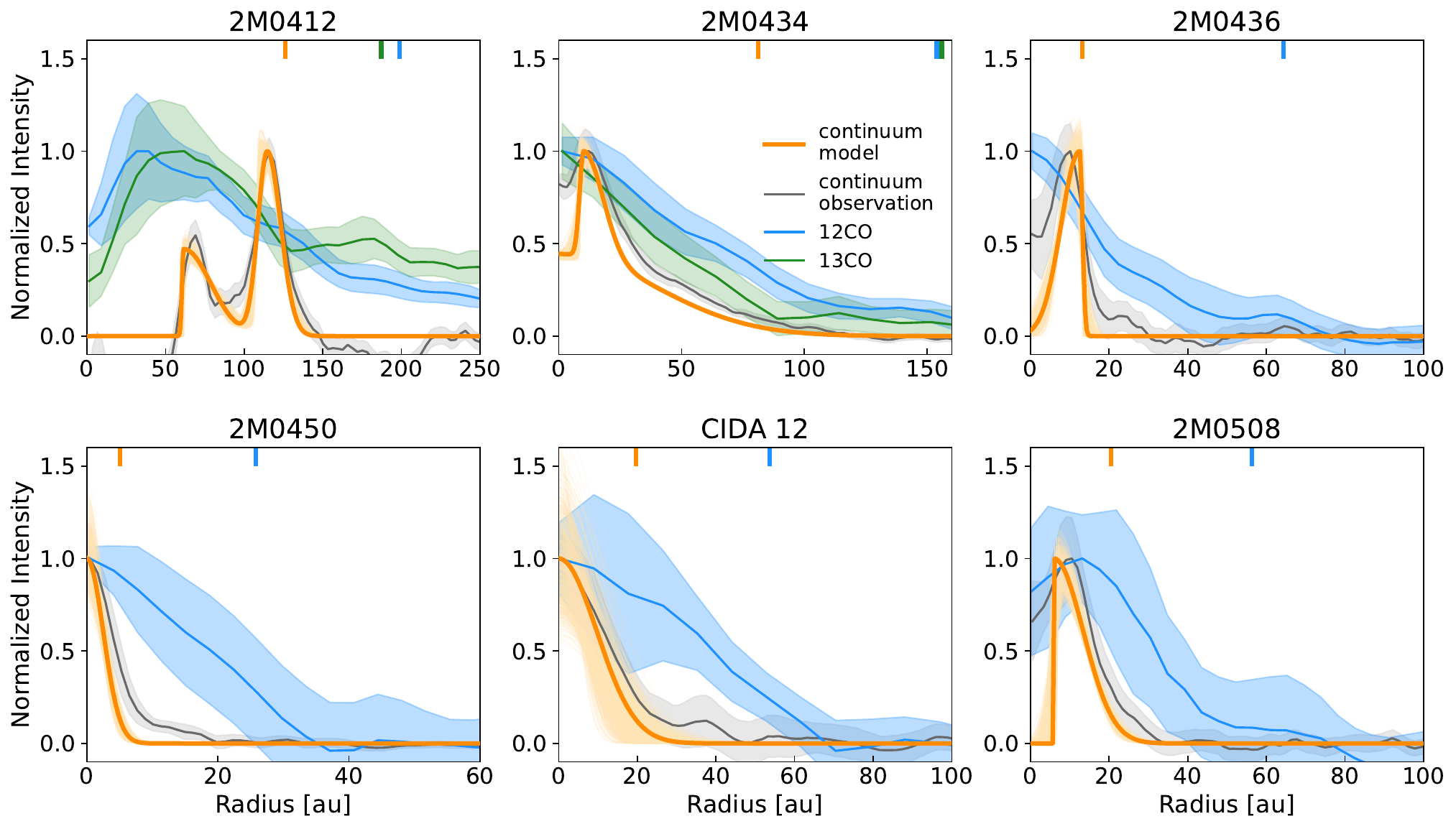}
    \caption{Normalized radial profiles of dust continuum, $^{12}$CO and $^{13}$CO (if detected). The thick orange curves are the best-fit model profiles of dust continuum (i.e. unconvolved); the thin light orange lines correspond to 1000 randomly selected model intensities from posterior sample (Section \ref{subsec:result_uvfit}). The gray lines are radial profiles of dust continuum from observations. The blue and green curves correspond to $^{12}$CO and $^{13}$CO azimuthally averaged radial profiles retrieved from moment-0 images, except for \obja~the moment-8 images are used due to severe cloud contamination. The color-filled regions represent $1\sigma$ uncertainty for each radial profile. The sticks at the top of each panel represent the radius enclosing $90\%$ emission of each profile, orange for dust continuum, blue for $^{12}$CO and green for $^{13}$CO.} 
    \label{fig:intensity}
\end{figure*}

Figure \ref{fig:intensity} shows the azimuthally averaged radial profiles of gas emission and the best-fit continuum model intensity profiles. CO emission around \obja~suffers from severe cloud absorption, so we show the profile extracted from the moment-8 map\footnote{Moment-8 map represents the maximum value along the spectrum.} instead. Among the six disks, \obja~shows a drop in emission inside the dust cavity, with peak emission located at $\sim$ 36 au for $^{12}$CO and 54 au for $^{13}$CO. The cloud absorption may affect these values. A tentative drop of $^{12}$CO is seen around \objf, with the peak location overlapping with the dust cavity size. No other clear evidence of gas emission depletion is found in the other four disks, under current $\sim 0.1''$ ($0.2''$ for \objb) resolution for gas images.

\begin{deluxetable*}{ccccccccc}
    \tablecaption{Measurement of gas emission from CO isotopes} \label{tab:co_flux}
    \tablehead{
    \colhead{Molecule} & \colhead{Integrated Flux} & \colhead{Peak Intensity} & \colhead{Velocity Range} & \colhead{Chan. Width} & \colhead{Chan. RMS} & \colhead{Beam} & \colhead{$R_{90\%}$} & \colhead{$R_{\rm gas}/R_{\rm dust}$} \\
    \colhead{ } & \colhead{[Jy $\rm km\ s^{-1}$]} & \colhead{[mJy beam$\rm ^{-1} km\ s^{-1}$]} & \colhead{[$\rm km\ s^{-1}$]} & \colhead{[$\rm km\ s^{-1}$]} & \colhead{[$\rm mJy\ beam^{-1}$]} & \colhead{[mas$\times$mas, deg]} & \colhead{[au]}
    }
    \startdata
    \multicolumn{9}{c}{2M0412} \\
           $^{12}$CO & $1.08\pm0.06$ & 71.4 & 4.0-8.0 & 0.4 & 5.5 & 215$\times$209, -3 & $\geq$198.9 & $\geq$1.6 \\
           $^{13}$CO & $0.54\pm0.06$ & 39.2 & 4.0-8.0 & 0.4 & 5.8 & 220$\times$212, -2 & $\geq$187.2 & $\geq$1.5\\
           C$^{18}$O & $<0.15$       & -    & 4.0-8.0 & 0.4 & 4.2 & 222$\times$211, 5  & - & -\\
    \hline
    \multicolumn{9}{c}{2M0434} \\
           $^{12}$CO & $1.21\pm0.11$ & 130.5 & 2.0-10.0 & 0.4 & 6.9 & 209$\times$175, -67 & $\geq$153.9 & $\geq$1.9\\
           $^{13}$CO & $0.61\pm0.09$ & 74.2  & 2.0-10.0 & 1.0 & 4.4 & 170$\times$159, -57 & $\geq$155.9 & $\geq$1.9\\
           C$^{18}$O & $<0.21$       & -     & 2.0-10.0 & 1.0 & 3.4 & 169$\times$159, -55 & - & - \\
    \hline
    \multicolumn{9}{c}{2M0436} \\
           $^{12}$CO & $0.38\pm0.07$ & 56.1 & 1.4-9.4 & 0.8 & 3.5 & 112$\times$99, -51  & $\geq$64.4 & $\geq$4.9 \\
           $^{13}$CO & $<0.24$       & -    & 1.4-9.4 & 0.8 & 3.8 & 113$\times$100, -55 & - & -\\
           C$^{18}$O & $<0.17$       & -    & 1.4-9.4 & 0.8 & 2.9 & 113$\times$101, -54 & - & -\\
    \hline
    \multicolumn{9}{c}{2M0450} \\
           $^{12}$CO & $0.10\pm0.04$ & 30.3 & 2.0-12.0 & 1.0 & 2.6 & 110$\times$97, -77 & $\geq$25.8 & $\geq$5.1\\
           $^{13}$CO & $<0.11$       & -    & 2.0-12.0 & 1.0 & 2.9 & 112$\times$98, -82 & - & -\\
           C$^{18}$O & $<0.10$       & -    & 2.0-12.0 & 1.0 & 2.2 & 111$\times$99, -85 & - & -\\
    \hline
    \multicolumn{9}{c}{CIDA 12} \\
            $^{12}$CO & $0.24\pm0.05$ & 57.6 & 2.0-12.0 & 1.0 & 3.0 & 101$\times$94, -37 & $\geq$53.7 & $\geq$2.7\\
            $^{13}$CO & $<0.15$       & -    & 2.0-12.0 & 1.0 & 3.3 & 102$\times$96, -40 & - & -\\
            C$^{18}$O & $<0.12$       & -    & 2.0-12.0 & 1.0 & 2.5 & 101$\times$97, -35 & - & -\\
    \hline
    \multicolumn{9}{c}{2M0508} \\
           $^{12}$CO & $0.19\pm0.06$ & 39.6 & 2.0-12.0 & 1.0 & 2.9 & 100$\times$93, -39 & $\geq$56.3 & $\geq$2.8\\
           $^{13}$CO & $<0.20$       & -    & 2.0-12.0 & 1.0 & 3.3 & 101$\times$96, -43 & - & -\\
           C$^{18}$O & $<0.19$       & -    & 2.0-12.0 & 1.0 & 2.5 & 101$\times$97, -38 & - & -\\
    \hline
    \enddata
\end{deluxetable*}

\section{Discussion} \label{sec:discuss}

Among the six disks, the four brighter disks show cavity and ring substructures and the two faintest disks appear as smooth emission. The cavities can be large to 60 au while the most common sizes are about 10 au. For the two smooth disks, visibility fitting suggests the possibility of them harboring  cavities down to 2 au. Comparing the gas disks with dust disks, two disks show large size ratios of about 5.
With these results, we organise our discussions as follows:
In Section \ref{subsec:discussion_evolution}, we discuss the disk global properties under disk evolution context. Section \ref{subsec:discussion_substructrue} focuses on the detected substructures at millimeter wavelength and SED hints for cavities for micron-sized dust disk. At last, we discuss the mechanisms that are possibly responsible for the observed rings and cavities in Section \ref{subsec:discussion_origins}.

\begin{figure*}[t]
    \centering
    \includegraphics[scale=0.52]{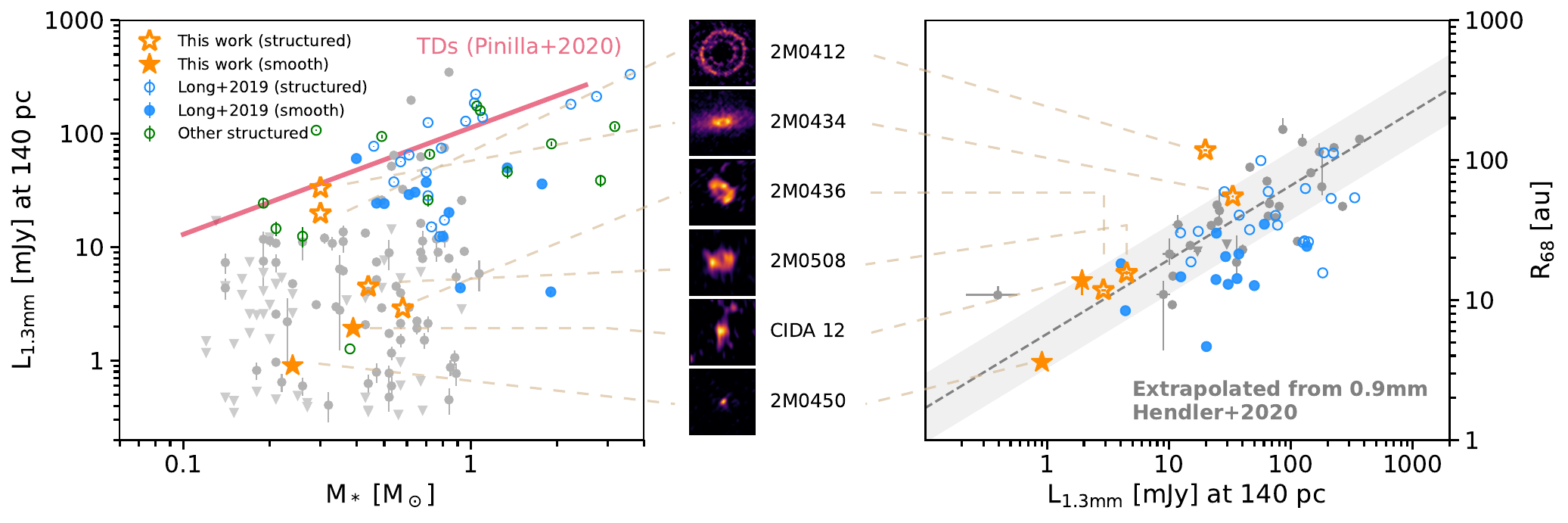}
    \caption{Scaling relations for Taurus disks. \textit{Left:} Millimeter luminosity at 1.3 mm versus stellar mass. Open orange stars are structured disks and solid orange stars are smooth disks in this work.
    Blue circles are for disks in \citet{long2019}, with open ones for structured and filled ones for smooth disks.
    Open green circles represent other structured disks in literature (see references in text).
    The gray dots (detection) and triangles (upper limits) are other Taurus disks (\citealt{manara2023}). The light pink line is the scaling relation fitted to disks with large cavities ($>$20 au) from \citet{pinilla2020}. \textit{Central:} Thumbnails of each target's continuum image, the box sizes are varied so that the targets are shown clearly. \textit{Right:} Dust disk size $R_{\rm dust,68\%}$ versus millimeter luminosity. The orange stars are the same as left panel, blue circles are from \citet{long2019} with open and filled for structured and smooth disks respectively,  and gray dots are from \citet{tripathi2017, kurtovic2021}. The scaling relation is from \citet{hendler2020}. The gray dots and the scaling relation are extrapolated from observations at ALMA Band 7 ($\sim$ 0.9mm).}
    \label{fig:bulk_property}
\end{figure*}

\subsection{Context with disk evolution} \label{subsec:discussion_evolution}

Putting disk properties together with the disk population can help constrain global disk evolution \citep{manara2023}. Previous studies have shown that the millimeter luminosity of disks scales with stellar mass \citep[e.g.,][]{andrews2013, ansdell2016, ansdell2017, barenfeld2016, pascucci2016}. The left panel of Figure \ref{fig:bulk_property} compares the six disks in this work with other Taurus disks with detected substructures in the $L_{\rm mm} - M_*$ plane (\citealt{long2018, hashimoto2021, kurtovic2021, yamaguchi2021, jennings2022, zhang2023}; review by \citealt{bae2023}),
also with other Taurus disks from \citet{manara2023}. The properties of the six disks follow the correlation for the whole Taurus population \citep{andrews2013}. The four structured disks are brighter than the two smooth disks. It could be that the two smooth disks simply formed with less dust, while it could also be the lack of substructures to trap the dust and prevent dust from drifting toward the star.
However, if the potential small cavities shown in Section \ref{subsec:result_nuker} are indeed present in \objd~and \obje, either their trapping efficiency is lower or the formation time of dust traps is later than the brighter disks.

If disks are optically thick, the low $L_{\rm mm}$ might not correspond to a low dust mass. To have a simple estimate of the optical depth for the two smooth disks, we estimate the disk midplane temperature following \citet{huang2018}: 
\begin{equation}
    T_{\rm mid}(r) = \left(\frac{\phi_{\rm fl} L_*}{8\pi r^2\sigma_{\rm SB}}\right)^{1/4}.
    \label{eq:midT}
\end{equation}
where $\sigma_{\rm SB}$ is the Stefan-Boltzmann constant, $\phi_{\rm fl}$ is the disk flaring angle. A conservative $\phi_{\rm fl}=0.02$ is used (\citealt{huang2018}). Then the optical depth $\tau_\nu$ is calculated using:
\begin{equation}
    I_\nu(r) = B_\nu(T_{\rm mid}(r)) (1-\exp(-\tau_\nu(r))).
\end{equation}
where $B_\nu$ is the Planck's law for black-body radiation.
The conservatively estimated peak $\tau$ is $\sim 0.5$ for \objd~and $0.15$ for \obje~so that \objd~is partially optically thick and \obje~is nearly optically thin. We expect the above arguments regarding dust trapping to hold for \obje~at least.

The $L_{\rm mm} - M_*$ relation has been suggested to be flatter for disks with inner cavities \citep[pink line in left panel of Figure \ref{fig:bulk_property}]{pinilla2018, pinilla2020}.
While those studies lack data for the very low mass stars and disks with small cavities ($<$20 au). Furthermore, the flatter relation is not observed in any individual star-forming region.
The two fainter disks with small inner cavities (\objc~and \objf) deviate significantly from this flatter relation. Conservative estimates of their optical depth yield maximum values of 0.6 for \objc~and 0.5 for \objf, suggesting the two disks are partially optically thick. Boulder formation could also lower the observable millimeter flux (\citealt{pinilla2020}). A combination of optical thickness and boulder formation in these disks might explain their deviation. 

In addition to stellar mass, the millimeter luminosity also scales with disk size \citep{tripathi2017, andrews2018scaling}. The right panel of Figure \ref{fig:bulk_property} compares the six disks in this work with other Taurus disks \citep{tripathi2017, long2019, kurtovic2021} in the $R_{\rm eff} - L_{\rm mm}$ plane with the best-fit scaling relation $R_{\rm eff} \propto L_{\rm mm}^{0.53\pm0.12}$ from \citet{hendler2020}.\footnote{Fluxes of sources in \citet{kurtovic2021} and the scaling relation in \citet{hendler2020} are extrapolated from 0.9 mm to 1.3 mm using a spectral index of 2.2 \citep{andrews2020}, the slopes of the scaling relation at these two wavelength are found to be identical \citep{tazzari2021}.} 
Numerical simulations have shown that disks fall along the observed relation $L_{\rm mm}\propto R^2_{\rm eff}$ if disks are in the radial drift dominated regime, while for disks with strong substructures they follow a relation of $L_{\rm mm}\propto R^{5/4}_{\rm eff}$ \citep{rosotti2019,zormpas2022}. 
The two relations intersect at the top end of the size-luminosity relation (i.e. for bright and large disks, \citealt{zormpas2022}).

Most of our six disks follow the observed relation $L_{\rm mm}\propto R^2_{\rm eff}$ except \obja, whose large size stands out at its luminosity regime, suggesting that a pressure bump was built early at large disk radii of \obja~\citep{long2023}. 
\objc, \objd, \obje~and \objf~fall at the faint end of $L_{\rm mm}\propto R^2_{\rm eff}$ relation hence are likely drift dominated. For the structured \objc~and \objf, either their substructures are too weak to retain the dust in their rings or their rings formed late when the disks have already evolved to the faint end of the $L_{\rm mm}\propto R^2_{\rm eff}$ relation.

The gas disk is universally found to be more extended than the dust disk (e.g., \citealt{barenfeld2017, ansdell2018, long2022}). In our six disks, we find that the two largest dust disks (\obja~and \objb) have a $R_{\rm 90,gas}$/$R_{\rm 90,dust}$  of $\sim$ 1.5 and 1.9 respectively. \obje~and \objf~have a ratio of $\sim$ 2.7. The two smallest dust disks (\objc~and \objd) have the most extreme ratios around 5, which are also among the largest values for disk population (\citealt{long2022}).
Early formation of pressure bumps in the two largest dust disks \obja~and \objb~may explain their lower size ratios. \obja~show a large ring at large radii of 114 au, while \objb's ring peaks at about 10 au which is far smaller than its size of 81 au, indicating its extended emission outside the ring holds unresolved substructures.
\citet{trapman2019} suggests that disks with $R_{90,\rm gas}/R_{90,\rm dust} > 4$ can only be explained with dust evolution and radial drift. This is likely what \objc~and \objd~have undergone due to their such extreme size ratios, which is also consistent with more efficient dust radial drift expected in lower mass stars \citep[]{pinilla2013}. 
These size ratios should be considered as lower limits since they are affected by cloud absorption and the observations lack short-spacing data (see Figures in Appendix \ref{sec:app_co}), hence the gas emission is probably not recovered as well as dust emission. 

\subsection{Substructures in disks around mid-to-late M stars} \label{subsec:discussion_substructrue}

Current studies of substructures mostly come from disks around solar-like stars (e.g., \citealt{andrews2018, long2018, cieza2021}).
\citet{vdMarel2021} found that substructures are less common around M stars, which was mostly based on studies with intermediate spatial resolution ($\sim$ 25 au in radius) and biased towards larger disks, while detecting substructures around M stars needs higher resolution since they are smaller. Currently only a few disks around M stars (especially mid-to-late type) have been imaged at high spatial resolution.
In Taurus, \citet{kurtovic2021} surveyed at a resolution of $\sim 0.1''$ (14 au) for a sample of six disks around M4-M5 stars and found two disks with a cavity and one ringed disk. The sample in \citet{kurtovic2021} is biased to the brightest disks in the corresponding stellar regime hence favorable for substructure detection. \citet{hashimoto2021} reported an asymmetric dust ring around a cavity around the M4.5 star ZZ Tau IRS with resolution $\sim 0.2''$. Other studies toward Lupus and Ophiuchus regions have also detected a few structured disks around mid-to-late M stars with resolutions down to $\sim 4$ au \citep{gonzalez-ruilova2020, cieza2021, vdMarel2022}.

\begin{figure*}[t!]
    \centering
    \plotone{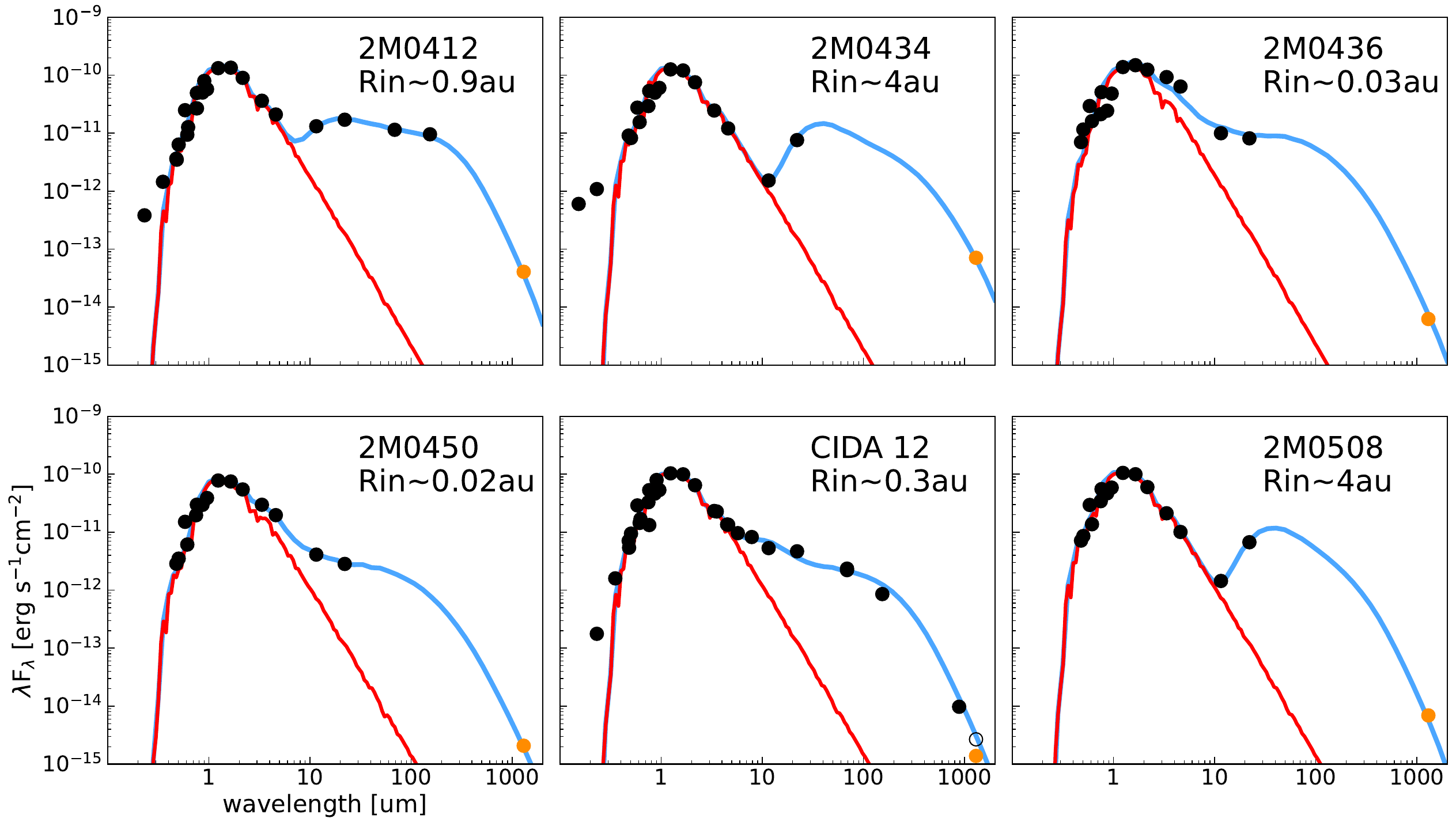}
    \caption{Spectral energy distributions from observations (black dots) and radiative transfer modeling (blue lines). The red lines are the stellar photospheric emission applied with extinction. The orange dots are measurement at 1.3mm in this work, for CIDA 12 the open black circle represents the 1.3mm flux from \citet{akeson2019}. $R_{\rm in}$ at upper right of each panel are cavity sizes from radiative transfer modeling for micron-sized dust grains.}
    \label{fig:sed}
\end{figure*}

Cavity+ring seem to be the most common substructure around low mass stars. In our sample, substructures are detected in four (\obja, \objb, \objc, and \objf) of six disks. 
The substructures we identify are all cavities surrounded by rings, with varying numbers in individual disks. 
Combined with structured disks around very low-mass stars collected in \citet{pinilla2022}, 
the cavity frequency seems higher than that around solar-like stars. 
Taking disks in Taurus as an example (\citealt{hashimoto2021, kurtovic2021}; this work), high spatial resolution imaging revealed 7 disks with cavities out of 13 disks around M3-M6 stars. In comparison to stars earlier than M3 in Taurus, only 4 disks out of 32 disks  analyzed by \citet{long2019} show cavities (5/32 considering the reanalysis in \citealt{zhang2023}). 
However, the sample around mid-to-late M stars used here might be more biased than that in \citet{long2019}. 
Based on SED modeling (see below and Figure \ref{fig:sed}), two disks out of the six disks in this work have cavities $>1$ au, which is higher than the transition disk fraction $\sim 8\%$ within the full Class II disks (\citealt{vdMarel2016}, similar fraction is obtained when limiting to M3-M6 stars).
Furthermore, disks in this work have been imaged with $\sim 0.05''$ compared to $0.1''$ in \citet{long2019}. A complete sample around mid-to-late M stars and higher resolution imaging of the inner regions of disks around solar-like stars are needed to evaluate any difference in their cavity occurrence rate in millimeter wavelength.

Gap+ring pairs are only detected in the large double ring disk \obja. \objb~has a ring located at about 10 au while its dust disk size is about 80 au, the very extended emission in between may hold shallow rings and gaps inside. 
Rings around \objc~and \objf~have radial widths comparable to the beam size hence are not resolved -- at higher resolution they might be resolved into multiple narrower rings (e.g., \citealt{facchini2020, perez2020}). It is also possible that fainter rings at large radii are not recovered with current sensitivity. Observations with a spatial resolution that is smaller than the pressure scale height and with deeper sensitivity are needed to have complete characterization of substructure type and occurrence rate in disks around mid-to-late M star.

Smooth disks \objd~and \obje~are candidates to host small cavities (Section \ref{subsec:result_nuker}). Though their visibility profiles do not show clear null points (Figure \ref{fig:model_compare}), small cavities below $5$ au could remain hidden at our current angular resolution, especially for \obje, which has an inclination of $\sim 65^\circ$ \citep[see analysis of high inclination disks by][]{vdMarel2022}.

ALMA revealed a central depression of large grains in four structured disks in our sample.  The SED complements ALMA imaging by providing information on small grains, since deficits of near infrared emission indicate an inner disk clearing of small dust grains. Figure \ref{fig:sed} shows the collected photometry and the modelled SEDs with modeling processes given in Appendix \ref{sec:app_sed}. In short, the modeling uses a continuous disk component with a cavity sized $R_{\rm\mu m,in}$ and an outer radius adopting $R_{\rm 90\%,dust}$ retrieved from millimeter images (Table \ref{tab:uvfit_results}).
The goal is not to have a stringent constraint on disk parameters since the available infrared photometry are limited, but to have a rough sense of the cavity sizes of small dust grains and how they compare to the millimeter-sized grains observed by ALMA. Since we lack knowledge of other parameters describing disk density profiles (e.g., power-law index of surface density and pressure scale height), the cavity size is degenerated with other parameters, including the power index for dust mass density and the pressure scale height. Through experiments, the constrained cavity sizes of small dust grains can differ by a few au under different combinations of other parameters. However, the uncertainty on cavity sizes does not affect the discussions below.

For the four disks with cavities in millimeter images, three disks show evidence of inner dust clearing from their SEDs. \obja's SED is consistent with $R_{\rm\mu m,in} \sim$ 0.9 au for small dust grains, which is far smaller than its $\sim 60$ au millimeter cavity. \objb~and \objf~are consistent with $R_{\rm\mu m,in} \sim$ 4 au cavities. This test of simple SED fitting suggests that micron-sized dust grains are not as depleted as mm-sized grains in the cavity, consistent with a morphology that is also seen around larger and more massive disks with both scattered light and ALMA images \citep{villenave2019}.
$R_{\rm\mu m,in}$ for \objc~is around 0.03 au, consistent with dust sublimation radius,
so there is no evidence for clearing of small grains. There is evidence of an inner disk around \objc~both from its brightness profile (Figure \ref{fig:bulk_property}) and visibility fitting result (Section \ref{subsec:result_uvfit}). Models with the same inclination as that from \objc's ALMA image have a bit of difficulty in reproducing the high flux at near infrared. These suggest that \objc~may host a puffed up or misaligned inner disk (e.g., \citealt{dullemond2001, vdMarel2018, vdMarel2022}).
For the two smooth disks, \objd~show no clear small grain cavity with $R_{\rm\mu m,in}$ consistent with dust sublimation radius, while \obje~show evidence of a 0.3 au cavity. 
Further evaluation of these potential cavities will require follow-up at much higher spatial resolution and sensitivity.  Scattered light imaging would likely require the next-generation of ground-based telescopes, since these disks are too faint to observe with current ground-based telescopes, and the cavities are too small for JWST imaging.

\subsection{Origins of small cavities} \label{subsec:discussion_origins}
Rings are the most common substructures in our sample, similar to surveys on disks around early-type stars as DSHARP (\citealt{andrews2018, huang2018}) and the Taurus survey (\citealt{long2018}).
With the current resolution and sensitivity of our observations, the detected rings all appear axis-symmetric, without significant azimuthally asymmetric features. Most rings encircle cavities, except for the outer ring of \obja~(see \citealt{long2023} for a detailed analysis of the origins of substructures of the \obja\ disk). In short, planet-disk interaction is preferred over other mechanisms in \obja, with Saturn-mass planets capable of shaping the observed disk morphology. However, a combination of dead zones and photoevaporation cannot be ruled out. Below we focus on discussion of possible origins\footnote{We exclude the dead zone outer edge as a possible origin of radial dust traps. In the up-to-date picture where all the three non-ideal MHD effects are considered, the level of MRI turbulence is damped smoothly in the outer disk (no abrupt change). As a result, a dead zone outer edge will not develop (see, e.g., Figure 2 in \citealt{bai2016}).} for substructures of \objb, \objc, and \objf, which show small cavities around 10 au.

\paragraph{Condensation fronts. }
Major volatiles in protoplanetary disks freeze onto dust grains from the gas phase as the disk temperature decreases outward. Across these regions, which are referred as condensation fronts or icelines, the dust opacity and critical fragmentation velocity are expected to change and further ring/gap substructures can be produced (e.g., \citealt{zhangke2015, okuzumi2016}).

We consider an irradiated flared disk with a disk midplane temperature formulated as Equation \ref{eq:midT},
where the flaring angle $\phi_{\rm fl}$ is assumed to be 0.02 (e.g., \citealt{huang2018, dullemond2018}). 
The iceline location of a volatile can be calculated from a stellar luminosity and the freezing point of the molecule. 
Since icelines of abundant species like $\rm H_2O$ and $\rm NH_3$ are close to the host star ($<$ 2 au) and are not resolvable by our observations, we focus on species with lower condensation temperatures in \citet{zhangke2015}, which are clathrate-hydrated CO and $\rm N_2$ (41-46 K), CO (23-28 K) and $\rm N_2$ (12-15 K). 

\begin{figure*}[!t]
    \centering
    \includegraphics[scale=0.35]{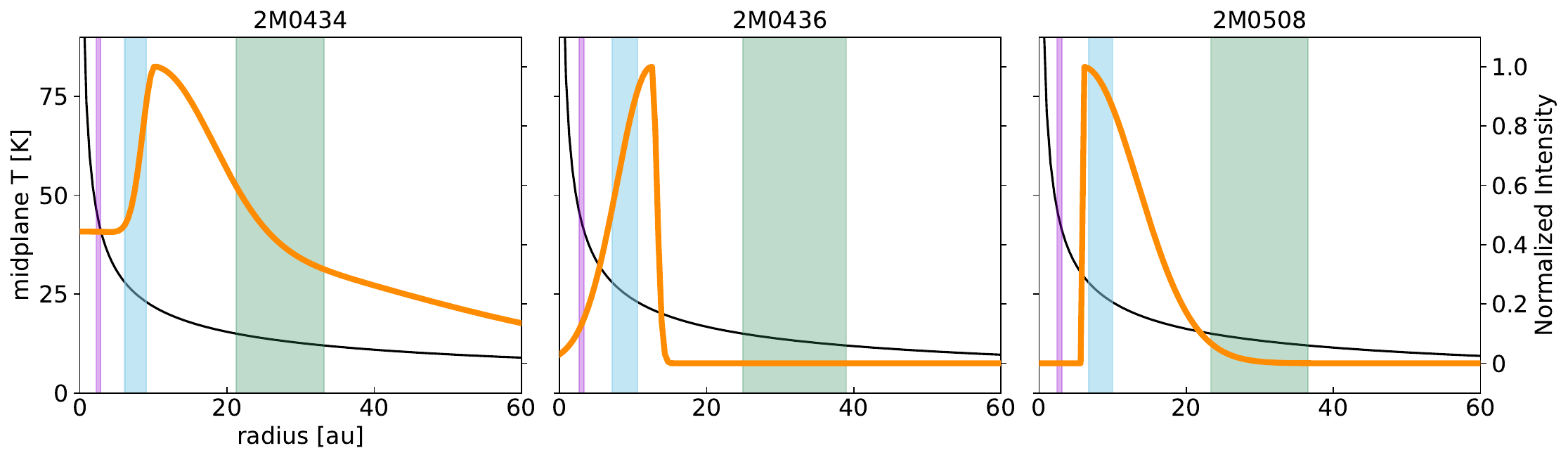}
    \includegraphics[scale=0.34]{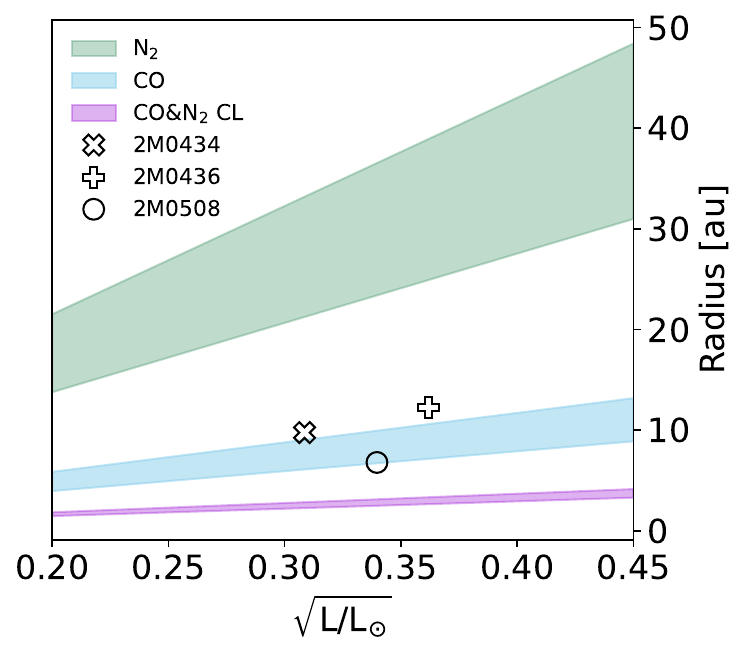}
    \caption{\textit{Left panels} (1)-(3): disk midplane temperature profiles calculated using Eq. \ref{eq:midT} are shown as black curves. Orange curves show the normalized radial profile of best-fit models from visibility fitting. Shaded regions are the iceline locations for $\rm N_2$ (light green), CO (light blue) and clathrate-hydrated CO and $\rm N_2$ (light purple). \textit{Right panel:} Disk radius versus the square root of stellar luminosity. The peak model intensity locations are shown for \objb~(cross), \objc~(plus) and \objf~(circle). The shaded regions for icelines are the same as the left panels.}
    \label{fig:iceline}
\end{figure*}

The comparison between iceline locations and brightness profiles from best-fit models is shown in Figure \ref{fig:iceline}. Interestingly, the peak emission radii are all close to the CO icelines. However, inferring the iceline radius is a challenging task. Besides the uncertainties in the estimation of disk midplane temperature (e.g., \citealt{liu2021iceline}), icelines can be thermally unstable and dynamically evolve on time scales from 1000 to 10000 years \citep{owen2020}. In simulations, icelines have not been found to carve out cavities (see \citealt{pinilla2017iceline} for a simulation around a Herbig AeBe star). Deeper observations of CO isotopologues are needed to better constrain disk mid-plane temperature, and future simulations tailored to M stars could together help determine the contribution of icelines to the cavities/rings detected.

\paragraph{Photoevaporation and MHD wind. }

When the mass loss rate by photoevaporation surpasses the accretion rate through the disk, a gas and dust gap will open in the disk. 
Recent new photoevaporation models 
by \citet{picogna2019} predict transition disks with accretion rates $\leq 10^{-9} M_\odot\ {\rm yr}^{-1}$ and maximum cavity sizes about 30 au when surrounding solar-type stars. 
The lower mass loss rate at large radii around very low-mass stars \citep{picogna2021} will likely lead to a smaller maximu cavity size,
which we expect to be above 10 au.
This is because the old photoevaporation models around $0.1 M_\odot$ stars by \citet{owen2012} have cavities $\leq 10$ au and new mass loss profiles by \citet{picogna2019} are more efficient at removing material at larger disc radii, which will lead to larger cavities. We therefore expect the maximum cavity size around very low mass stars will fall between 10 and 30 au. 
\objb's 10 au cavity, \objc's 12 au cavity and their accretion rates about a few $\times 10^{-10} M_\odot\ {\rm yr}^{-1}$ fall in the predicted parameter space.

Further evaluating the role of photoevaporation in the creation of cavities will require additional observations.  Possible discriminators include high angular resolution imaging of the gas -- in this paper, we have limited sensitivity that allows us to only tentatively identify a cavity around one source, \objf.  In addition, empirical measurements of photoevaporation rates through lines of \ion{O}{1} and \ion{Ne}{2} \citep{pascucci2011}  but challenging to interpret \citep[e.g.,][]{banzatti2019,fang23,rab2023}, given the possibility that these lines may form in a magnetothermal wind \citep[e.g.,][]{wang2019mhd}.
                                      
Furthermore, accreting transition disks could be sustained by other mechanisms. \citet{garate2021} showed that a disk with X-ray photoevaporation in combination with dead zones can reproduce both the accretion rates and gap sizes observed in transition disks. Their models predict a long-lived compact inner disk, which is not seen in our ALMA images under current resolution and sensitivity. MHD wind as another scenario, if the large-scale magnetic field distribution is not significantly modified, could generate a positive radial slope of the gas surface density (e.g., \citealt{suzuki2016}) which inhibits the inward drift of pebbles, and sustain the accretion rates similar to classic T Tauri stars (e.g., \citealt{lesur2021, martel2022}).



\paragraph{Embedded planets}
\label{subsubsec:planets}
Giant planets can open gaps in the gas disks and form gas pressure bumps that trap dust particles outside its orbit. Dust cavities will then form after the inner dust disk materials are accreted onto the star. By applying the analytical criterion for opening a gap in the gas \citep{crida2006}, the estimated masses of planets needed to open the small cavities in our disk sample are: for viscous $\alpha\sim10^{-4}$ and $10^{-3}$, 0.1 \& 0.3 $M_{\rm Jup}$ at $\sim 8$ au for \objb, 0.1 \& 0.3 $M_{\rm Jup}$ at $\sim 10$ au for \objc, and 0.1 \& 0.2 $M_{\rm Jup}$ at $\sim 6$ au for \objf. The locations of planets are assumed to be five times the Hill radius $r_H = r_p (M_p/3M_*)^{1/3}$ away from the ring peak locations (e.g., \citealt{dodson-robinson2011}). Since only \objf~shows a tentative gas cavity, these planet masses should be treated as upper limits. Moreover, if the disk mass and angular momentum transport are dominated by MHD wind rather than turbulent viscosity, planets can open wider and deeper gaps more easily (e.g., \citealt{elbakyan2022, aoyama2023, wafflard-fernandez2023}). As a result, the masses of embedded planets could be a factor of a few to ten smaller than estimated above (\citealt{elbakyan2022}). 

Simulations of planet-disk interaction predict the segregation between millimeter-sized dust continuum emission peak and gas emission peak (or micron-sized dust, e.g., \citealt{deJuanOvelar2013, facchini2018}). 
We find hints of such segregation between micron-sized and millimeter-sized dust grains from SEDs (Section \ref{subsec:discussion_substructrue}). While for gas emission, only \objf~shows tentative inner gas depletion and the gas peak location overlap with the ring location. Deeper sensitivity and higher angular resolution of molecular line mapping are needed to resolve the gas depletion in the inner disk if present. 

Taking disk dust masses in Section \ref{subsec:result_flux_mass_size} and assuming a gas-to-dust mass ratio of 100, the estimated current disk masses are 6.1 $M_{\rm Jup}$ for \objb, 0.5 $M_{\rm Jup}$ for \objc~and 0.8 $M_{\rm Jup}$ for \objf. If we assume a constant accretion rate after disk formation, then the total mass accreted onto the star from the disk is 1.5 $M_{\rm Jup}$ for \objb\ and 4.5 $M_{\rm Jup}$ for \objc\ (using stellar ages with 50\% spot coverage).
The planet masses needed to open cavities for \objb~and \objc~ are within 10\% of their estimated total disk masses at initial stages. If \objf~has an accretion rate $\sim 3\times10^{-10}M_\odot\ {\rm yr}^{-1}$, similar to those in our sample (Table \ref{tab:host_properties}), and that accretion rate has remained constant (most likely, the accretion rate at earlier stages of formation was higher, e.g., \citealt{fischer17, fang23usco}), its potential planet mass is also within 10\% of disk mass. Hence from the perspective of disk mass budget, the three small cavity disks had enough materials to form those Saturn-mass planets (e.g., \citealt{boss2006, lin2018}). 

If the observed cavities are carved out by giant planets, how does the current detection rate of cavities around mid-to-late M stars compared to the occurrence rate of exoplanets?
The detection rate of cavities around mid-to-late M stars in Taurus is around $24\%$, given the fraction of 7/13 for cavities in disks (Section \ref{subsec:discussion_substructrue}) and accounting for the disk fraction around $45\%$ for M3-M6 stars in Taurus (\citealt{esplin2019}; this latter correction applies only if the age spread of the sample is much less than the disk survival timescale). However, this is likely higher than the real cavity rate, since the sample of 13 disks is slightly biased to the brighter end of disks around M3-M6 stars. 
As for exoplanets around M-dwarfs, 
the CARMENES survey obtained occurrence rates of giant planets with $M_{\rm p}\sin i$ > $100M_\oplus$ and periods of 1-1000 days being $0.021^{+0.018}_{-0.011}$ planets per star and $0.045^{+0.021}_{-0.016}$ planets per star for stars less and more massive than $0.337 M_\odot$ respectively (\citealt{ribas2023}). 
Wide-orbit (>$10$ au) giant planets massive than Jupiter are found to be rare around M-dwarfs (e.g., \citealt{bowler2015}). Currently, the constraints on sub-Jovian giants on orbits of a few au around M-dwarfs are limited. 
Measurement around Sun-like stars show giant planet occurrence is enhanced by a factor of four beyond 1 au compared to within 1 au (e.g., \citealt{fernandes2019, fulton2021}), while it is unclear whether the result could be extrapolated to M-dwarfs.
Samples from both disk and exoplanet need to be developed to make a robust comparison.

\paragraph{Binary companions}
Binary stars could also carve out inner cavities in circumbinary disks, with the cavity sizes expected to be 3-5 times the binary semimajor axis (e.g., \citealt{artymowicz1994, miranda2017}). Hence stellar companions within 4 au could be responsible for the 7-12 au cavities in the three disks.
To our knowledge, no evidence of stellar multiplicity have been found in our sample (\citealt{kraus2011, kraus2012}), although the scales of a few au have yet to be explored.
If the companions are sufficiently massive (mass ratios $>$ 0.1-0.2), the binary would produce eccentric cavities (0.05-0.35 for circular binaries. \citealt{miranda2017, ragusa2020}) and cause a shift between the stellar mass center and the geometric center of the cavity. For the $\sim10$ au cavities in our sample, resolutions better than the binary separation are needed to detect the potential shift.

\vspace{0.2cm}
Among the mechanisms described above, photoevaporation and planets are more preferred origins of small cavities (around 7 and 10 au) in our sample. To distinguish between these possibilities, more deep and higher angular resolution observations of molecular lines are needed. Furthermore, multi-wavelength analysis of the spectral index at the cavity edge might help to distinguish photoevaporation and planet scenarios (\citealt{picogna2023}), where a cavity created by photoevaporation is expected to have a higher spectral index due to its lower dust filtering efficiency.

\section{Summary}
\label{sec:summary}
This paper presents high angular resolution ($\sim$50 mas, 8 au) ALMA Band 6 observations of six disks around mid-to-late type M stars in Taurus. We characterize the disk continuum emission by fitting parametric models in the visibility plane. We explore these disks' global properties and the possible origins of their substructures. The main findings are summarized as follows:
\vspace{0.2cm}

1. We detect all six disks in millimeter continuum emission. $^{12}$CO is detected in five disks with a marginal detection in \objd, and $^{13}$CO is detected in \obja~and \objb.
Dust substructures are detected in 4 disks: \obja, \objb, \objc~and \objf, which are also the four brightest disks in our sample. We perform fitting in the visibility plane, with rings modeled as radially asymmetric Gaussian rings.
With our current sensitivity ($\sim$ 40 mJy/beam) and spatial resolution, \obja~show a large cavity surrounded by a ring at 60 au, followed by a gap and an outer dust ring at 110 au; the other 3 disks all display a cavity surrounded by a dust ring at 10 au for \objb, 12 au for \objc~and 7 au for \objf. 

2. The Nuker profile fitting on the two compact smooth disks \objd~and \obje~show that they may hold small cavities, with sizes being $\sim$ 1.7 au for \objd~and $\sim$ 5.7 au for \obje. Current spatial resolution and sensitivity may hide these small and shallow cavities. Nuker profiles can reproduce the flux for \obje~better than Gaussian profile, with the flux from the Nuker profile being twice as much as Gaussian's flux. Higher spatial resolution is needed to resolve the potential cavities.

3. The structured disks are brighter than the smooth disks in our sample, the presence of dust trapping by substructures could account for this. \objc~and \objf~deviate from the flatter $L_{\rm mm}-M_*$ relation for structured disks, which could be due to a combination of partially optical thickness and boulder formation (\citealt{pinilla2018, pinilla2020}). Disks around \objc, \objd, \obje~and \objf~fall at the faint end of $L_{\rm mm}\propto R^2_{\rm eff}$ relation, consistent with being radial drift dominated. After measuring the ratios between gas disk size $R_{\rm gas,90\%}$ and the dust disk size $R_{\rm dust,90\%}$, \objc~and \objd~show large ratios $R_{\rm gas,90\%}/R_{\rm dust,90\%} \sim 5$, indicating very efficient dust radial drift in these two disks.

4. All structured disks in our sample show a central cavity, indicating the clearing of millimeter-sized dust grains. Radiative transfer fits to SEDs providing a rough constraint on the clearing of micron-sized dust particles. Of the 4 structured disks, only \objc~does not show evidence of clearing of micron dust, with the modeled cavity size being the dust sublimation radius. \obja~show a 0.9 au cavity size far smaller than its millimeter cavity. \objb~and \objf~show 4 au cavities for micron dust. For the 2 smooth disks, \obje~show a 0.3 au cavity consistent with its near-IR deficit. These SED-derived cavity sizes are smaller than their counterpart in millimeter images.

5. Various mechanisms could be responsible for the observed substructures. For \obja, \citet{long2023} shows that Saturn-mass planets are able to carve out its cavity and gap. The large cavity alone can also be shaped by a combination of dead zone and photoevaporation. \obja's large rings are unlikely to be related to icelines. For the other three disks, Saturn-mass planets or photoevaporation could be responsible for their cavities around 10 au. Icelines could also play a role but their locations are highly uncertain.
High resolution mapping of gas emission and multi-wavelength analysis of the spectral index at the edge of cavities, could help to distinguish different mechanisms.
\vspace{0.2cm}

Our sample covers a wide range of millimeter luminosity, the four cavity disks together with the other two harboring potential small cavities suggest that substructures are likely ubiquitous in disks around mid-to-late M stars. Current exoplanet statistics does not rule out that all these observed cavities are created by giant planets.

High spatial resolution imaging of disks around mid-to-late M stars are still very limited in number. The more frequent small cavities around 10 au around those disks may originate from various mechanisms, which will make them great laboratories for testing planet formation and disk physics.


%

\section{Acknowledgments}

G.J.H and Y.S. are supported by the National Key R\&D Program of China No. 2019YFA0405100 and by grant 12173003 from the National Natural Science Foundation of China.
Support for F.L. was provided by NASA through the NASA Hubble Fellowship grant \#HST-HF2-51512.001-A awarded by the Space Telescope Science Institute, which is operated by the Association of Universities for Research in Astronomy, Incorporated, under NASA contract NAS5-26555. 
D. H. is supported by Center for Informatics and Computation in Astronomy (CICA) grant and grant number 110J0353I9 from the Ministry of Education of Taiwan. D. H. also acknowledges support from the National Science and Technology Council of Taiwan through grant number 111B3005191.
P. P. acknowledges support from the UK Research and Innovation (UKRI) under the UK government’s Horizon Europe funding guarantee from ERC (under grant agreement No 101076489).
E.R. acknowledges financial support from the European Union's Horizon Europe research and innovation programme under the Marie Sk\l{}odowska-Curie grant agreement No. 101102964 (ORBIT-D).
E.R. also acknowledges financial support from the European Research Council (ERC) under the European Union’s Horizon 2020 research and innovation programme (grant agreement No. 864965, PODCAST).
D.J.\ is supported by NRC Canada and by an NSERC Discovery Grant.
C.F.M. is Funded by the European Union (ERC, WANDA, 101039452). Views and opinions expressed are however those of the author(s) only and do not necessarily reflect those of the European Union or the European Research Council Executive Agency. Neither the European Union nor the granting authority can be held responsible for them.
G.D.M. acknowledges support from FONDECYT project 11221206, from ANID --- Millennium Science Initiative --- ICN12\_009, and the ANID BASAL project FB210003.
L.A.C. acknowledges support from FONDECYT project 1211656 from ANID and the Millennium Science Initiative, Center Code NCN2021\_080.

This paper makes use of the following ALMA data: ADS/JAO.ALMA\#2019.1.00566.S. ALMA is a partnership of ESO (representing its member states), NSF (USA) and NINS (Japan), together with NRC (Canada), MOST and ASIAA (Taiwan), and KASI (Republic of Korea), in cooperation with the Republic of Chile. The Joint ALMA Observatory is operated by ESO, AUI/NRAO and NAOJ. The National Radio Astronomy Observatory is a facility of the National Science Foundation operated under cooperative agreement by Associated Universities, Inc.
\vspace{5mm}
\facilities{ALMA}


\software{CASA \citep{casa2022},
        GALARIO \citep{tazzari2018},  
          emcee \citep{foremanMackey2013},
          RADMC-3D \citep{dullemond2012}
          }



\appendix

\restartappendixnumbering
\section{CO Channel maps and moment maps}
\label{sec:app_co}
\begin{figure*}[!h]
    \centering
    \plotone{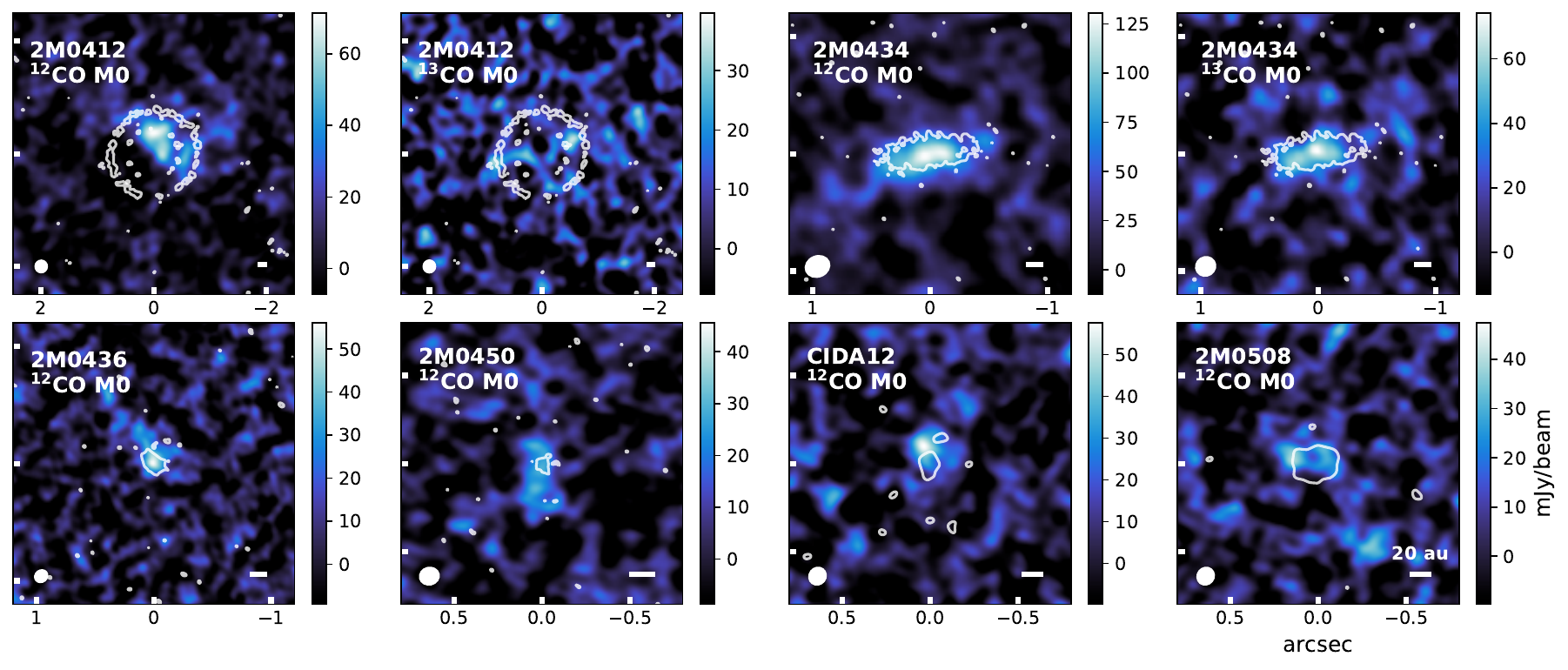}
    \caption{Moment 0 and 8 maps for (marginally) detected $^{12}$CO and $^{13}$CO. The $3\sigma$ white contours from dust continuum images are overlaid. For each panel, beams are shown as white ellipses at lower left and 20 au white scale bars are shown at lower right. }
    \label{fig:co_mommap}
\end{figure*}

\begin{figure*}[!h]
    \centering
    \plotone{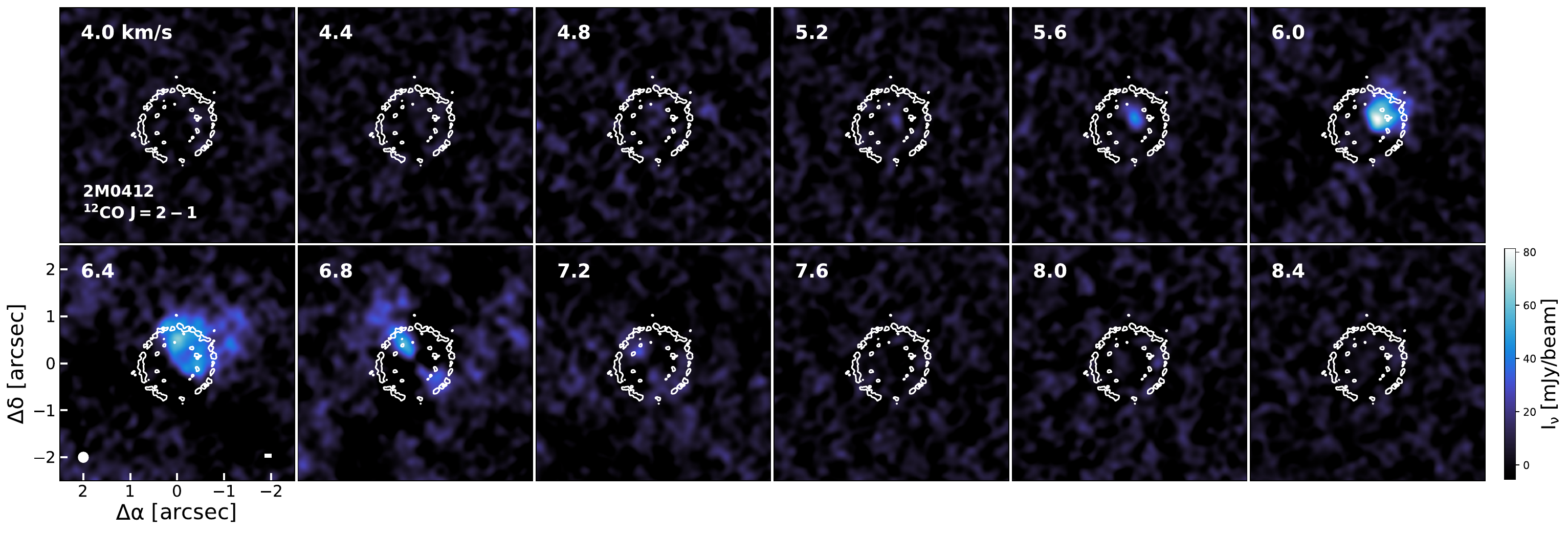} \\
    \plotone{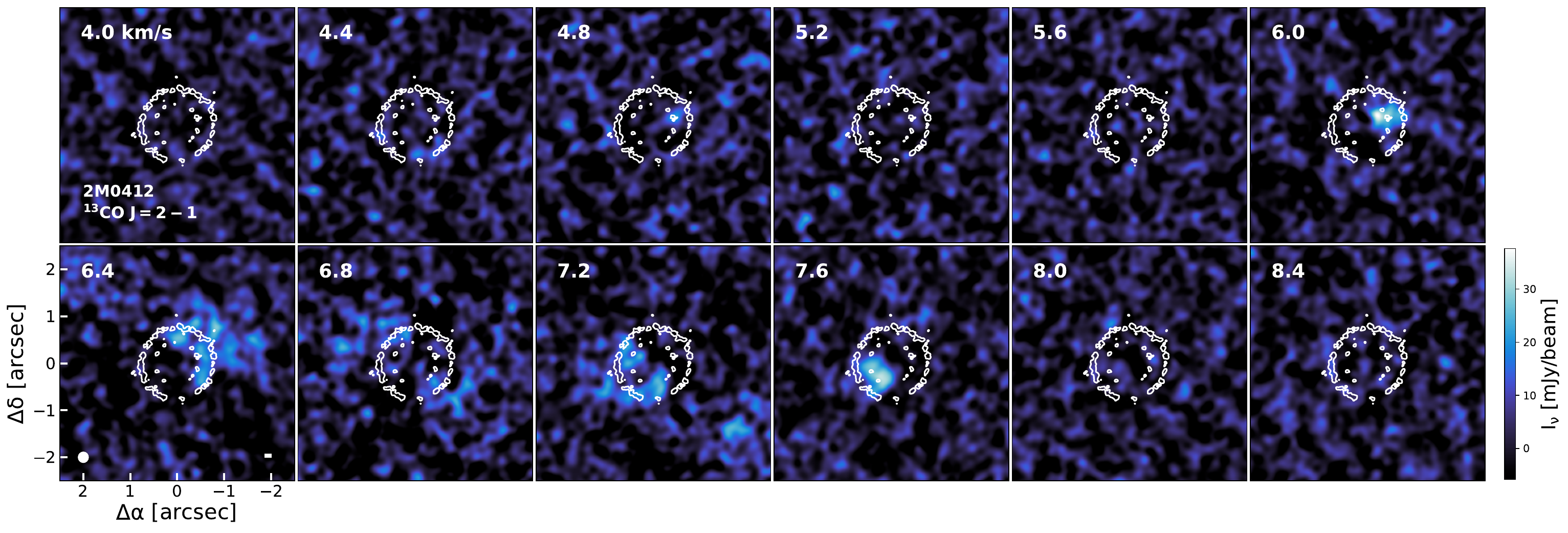} 
    \caption{Channel maps of $^{12}$CO (top) and $^{13}$CO (bottom) for \obja. Central velocities are given in the upper left corner of each channel in unit of $\rm km\ s^{-1}$. The $3\sigma$ white contours from dust continuum images are overlaid.
    Beams as white ellipses and 20 au white scale bars are shown at lower left channel, at lower left and lower right corner correspondingly.}
    \label{fig:co_channelmap1}
\end{figure*}

\begin{figure*}[!h]
    \centering
    \plotone{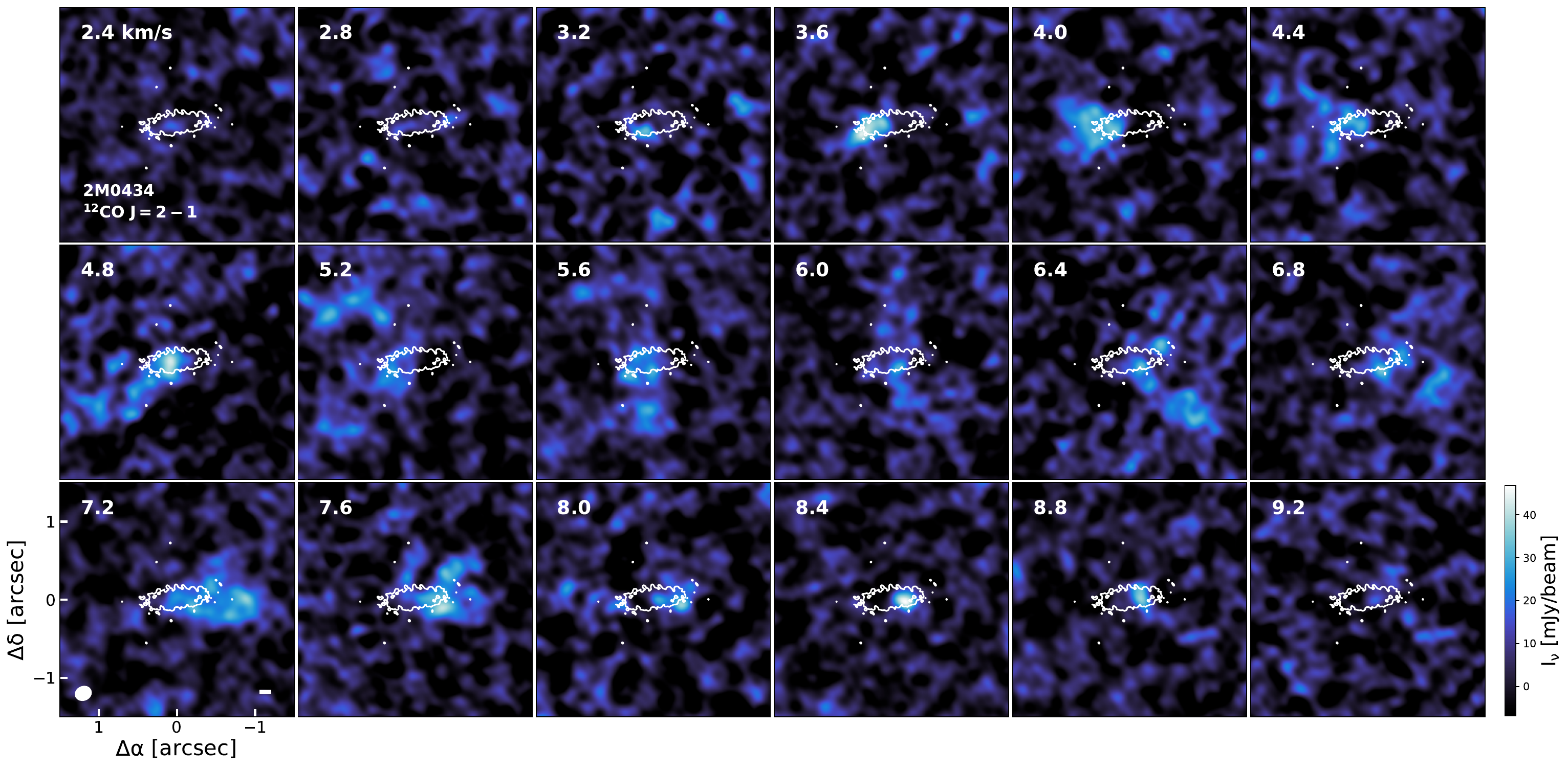} \\
    \plotone{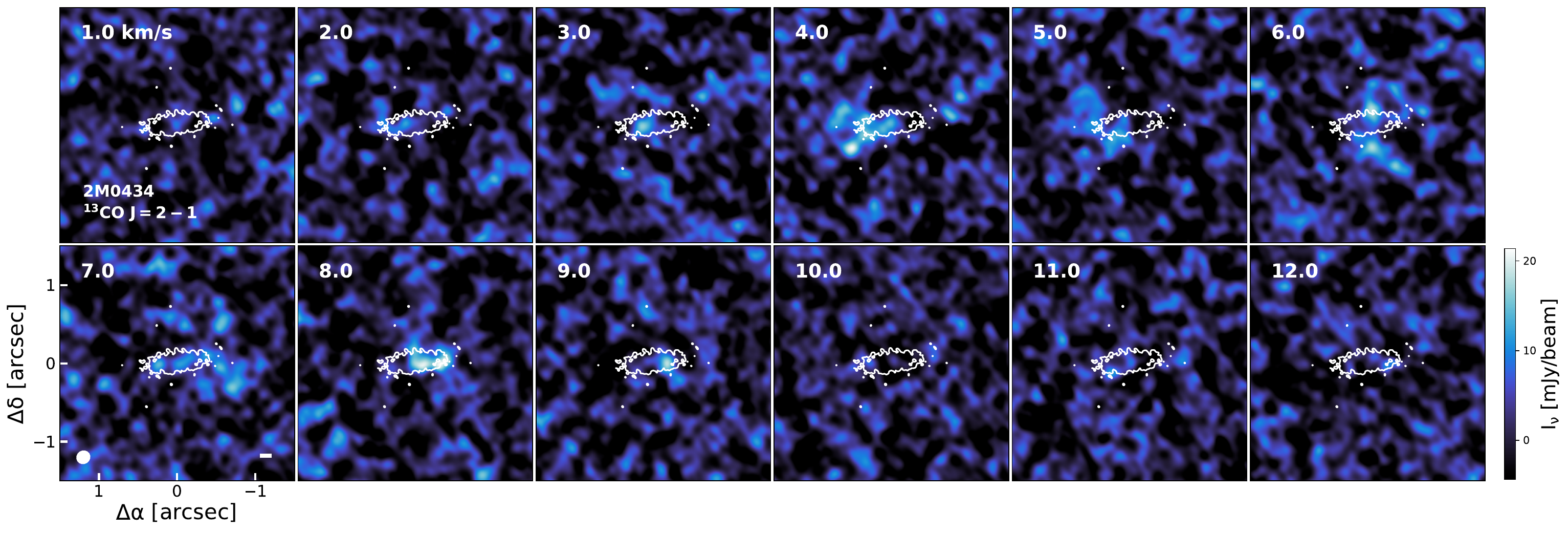} 
    \caption{Same as Figure \ref{fig:co_channelmap1} for \objb.}
    \label{fig:co_channelmap2}
\end{figure*}

\begin{figure*}[!h]
    \centering
    \plotone{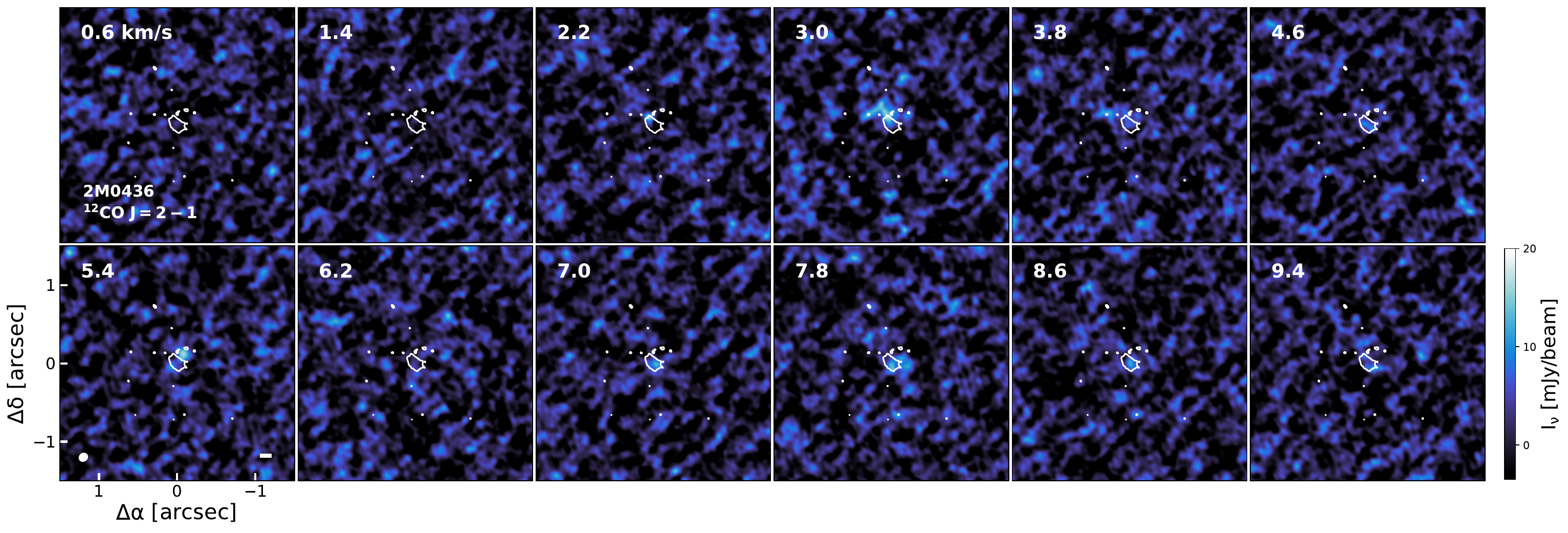} 
    \caption{Same as Figure \ref{fig:co_channelmap1} for \objc.}
    \label{fig:co_channelmap3}
\end{figure*}

\begin{figure*}[!h]
    \centering
    \plotone{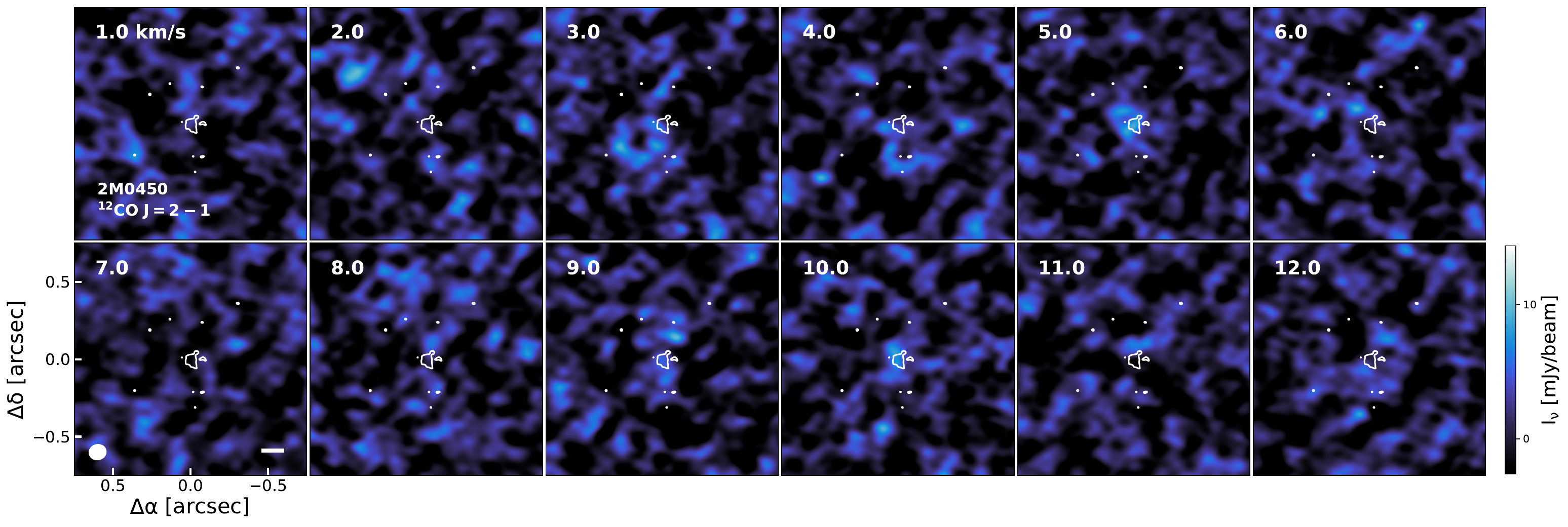} 
    \caption{Same as Figure \ref{fig:co_channelmap1} for \objd.}
    \label{fig:co_channelmap4}
\end{figure*}

\begin{figure*}[!h]
    \centering
    \plotone{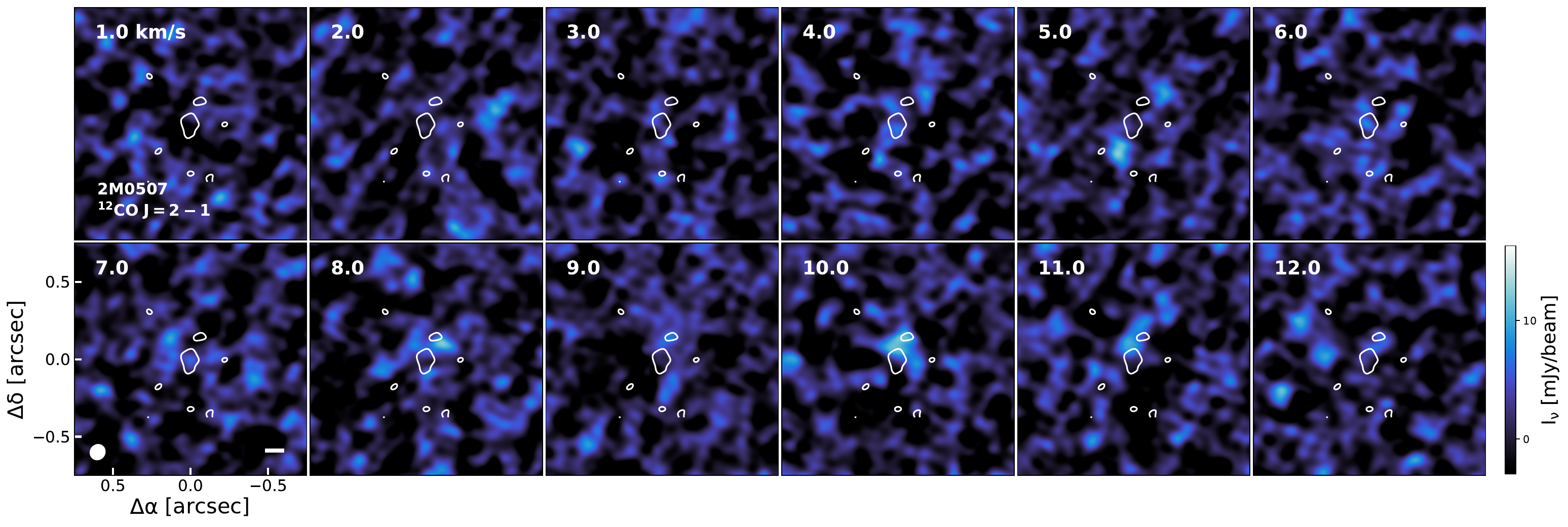} 
    \caption{Same as Figure \ref{fig:co_channelmap1} for \obje.}
    \label{fig:co_channelmap5}
\end{figure*}

\begin{figure*}[!h]
    \centering
    \plotone{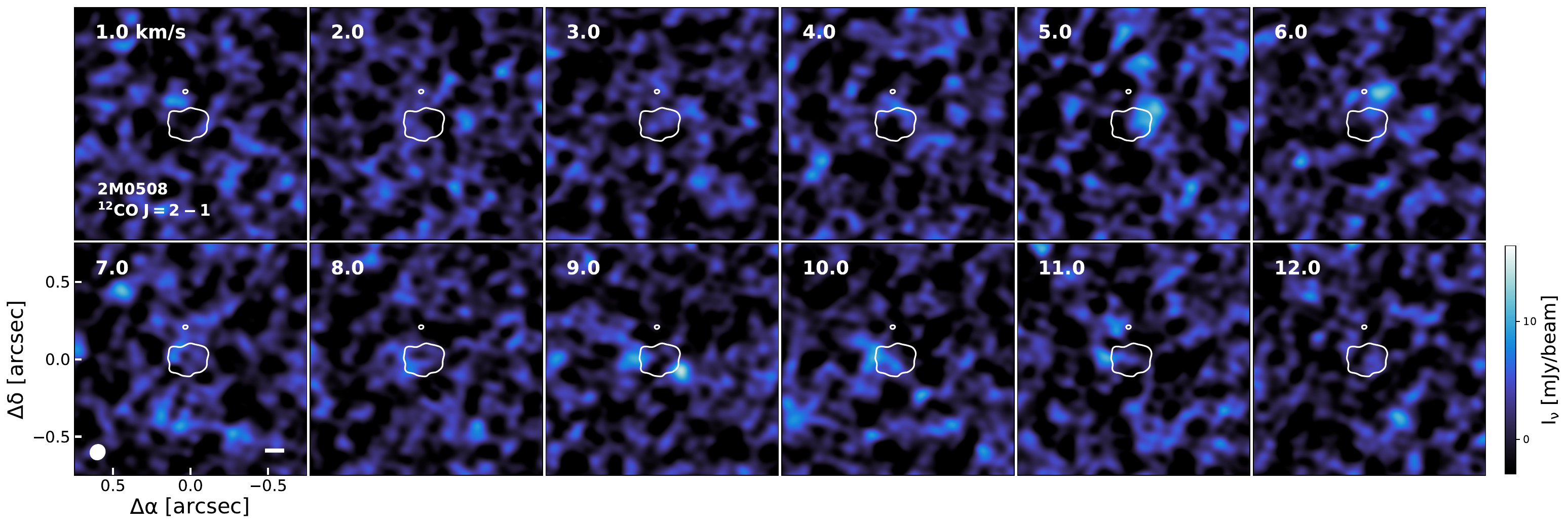} 
    \caption{Same as Figure \ref{fig:co_channelmap1} for \objf.}
    \label{fig:co_channelmap6}
\end{figure*}

Moment maps of $^{12}\rm CO$ and $^{13}\rm CO$ (if detected) are shown in Figure \ref{fig:co_mommap} and corresponding channel maps are shown in Figures \ref{fig:co_channelmap1}-\ref{fig:co_channelmap6}.


\restartappendixnumbering
\section{Central emission in 2M0436?}
\label{sec:app_central_2m0436}
\begin{figure*}[!ht]
    \centering
    \includegraphics[scale=0.5]{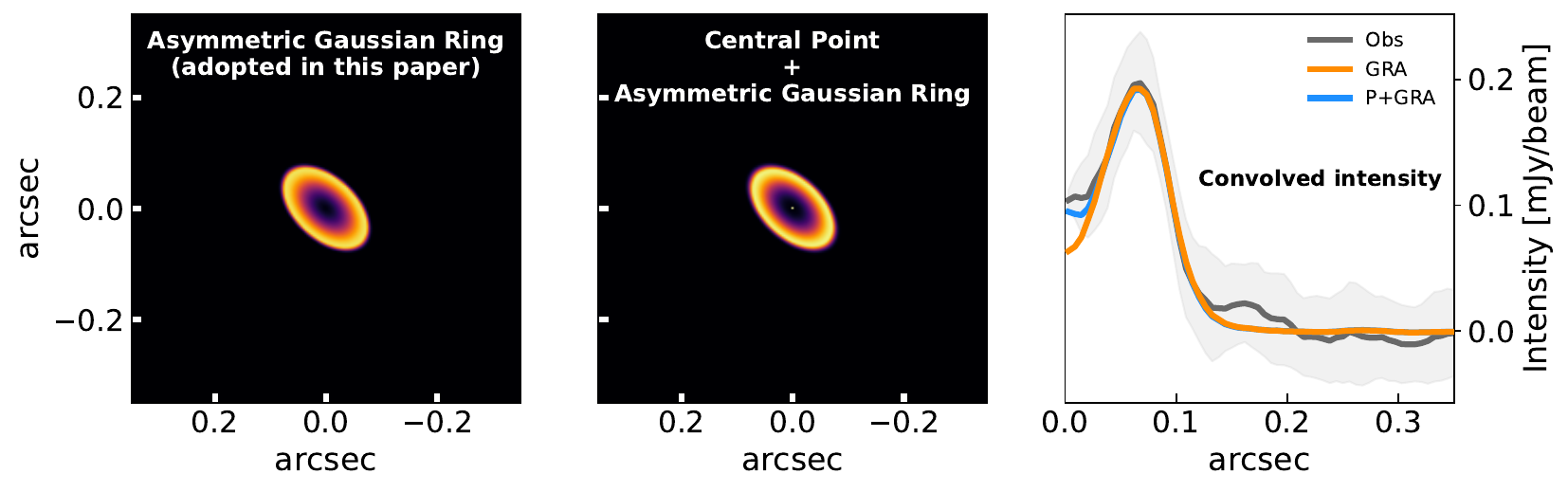}
    \caption{Two different models for \objc. \textit{From left to right:} (1) Best-fit image from the model with only an asymmetric Gaussian ring. (2) Best-fit image from the model with a central point and an asymmetric Gaussian ring. (3) Azimuthall averaged radial profiles retrieved from beam convolved images.}
    \label{fig:2M0436_models}
\end{figure*}

\begin{figure}
    \centering
    \includegraphics[scale=0.45]{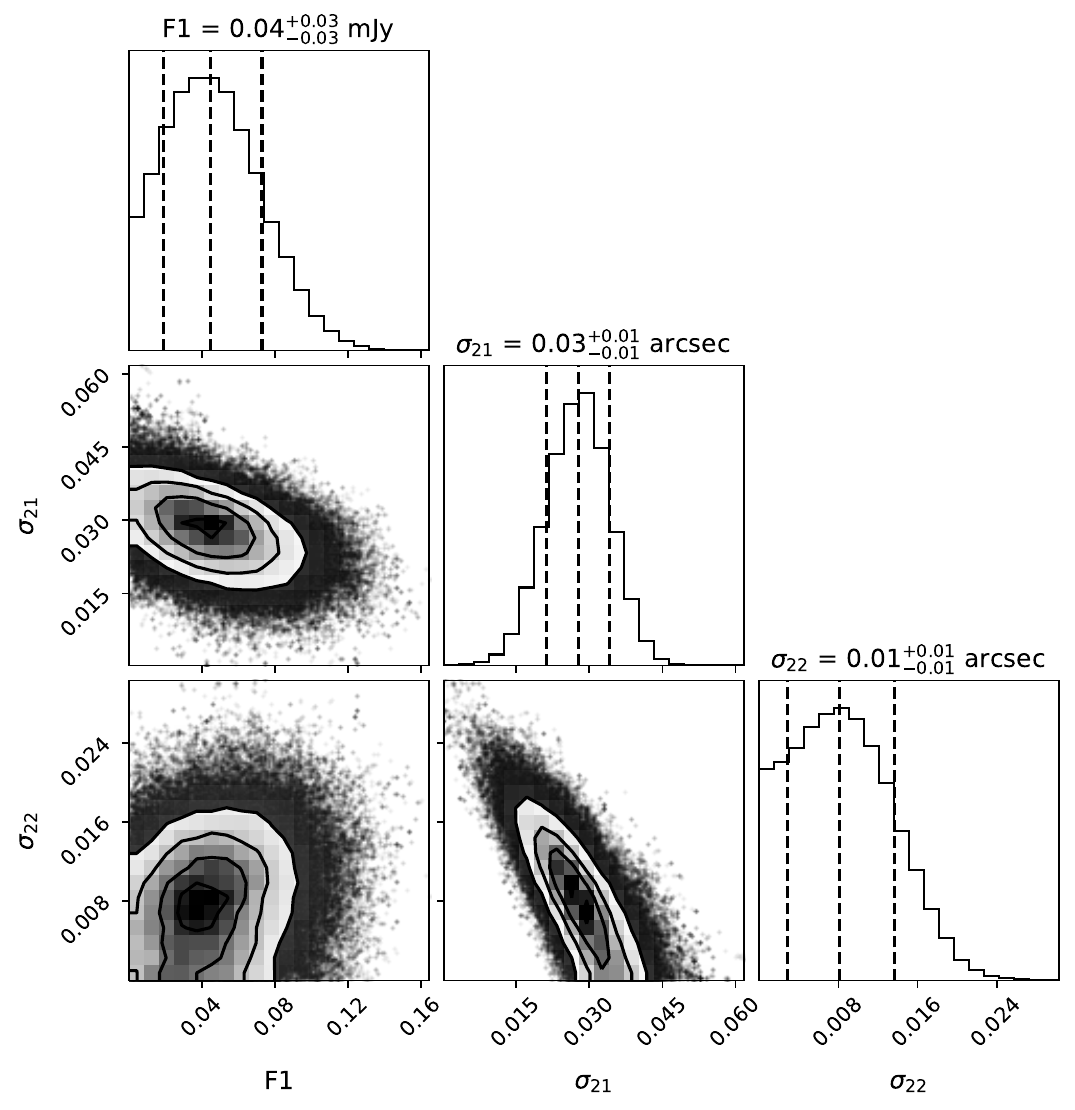}
    \caption{Cornerplots of key parameters of the model includes a central point and an asymmetric Gaussian ring. From left to right, the three parameters are the flux of point emission, the inner width of the ring and the outer width of the ring.}
    \label{fig:2M0436_cornerplot}
\end{figure}

In this section, we test whether \objc~has unresolved central emission (Section \ref{subsec:result_uvfit}).  A central point emission is added to the model of a radially asymmetric Gaussian ring, with point source intensity formulated as $F_1\delta(r)$.
$F_1$ is set to distributed uniformly between 0 and 1 mJy.
Figure \ref{fig:2M0436_models} shows the best-fit model images and intensity profiles for the two models. The central point emission model reaches a convergence with the point flux being $0.04^{+0.03}_{-0.03}$ mJy (Figure \ref{fig:2M0436_cornerplot}), indicating a non-detection (detection at 1$\sigma$) of central emission. Section \ref{subsec:result_uvfit} mentioned that it could be the unresolved central emission blends with the outer ring emission, resulting in a ring with its inner width than outer width which is contrary to dust trapping in pressure bump. However, in our attempts to add central emission component the outer ring still shows wider inner width and the ratio between the two widths is similar to that from ring-only model (Section \ref{subsec:result_substructure}). 

\restartappendixnumbering
\section{Nuker fitting results on \objd~\& \obje}
\label{sec:app_nuker}
\begin{figure*}
    \centering
    \includegraphics[scale=0.65]{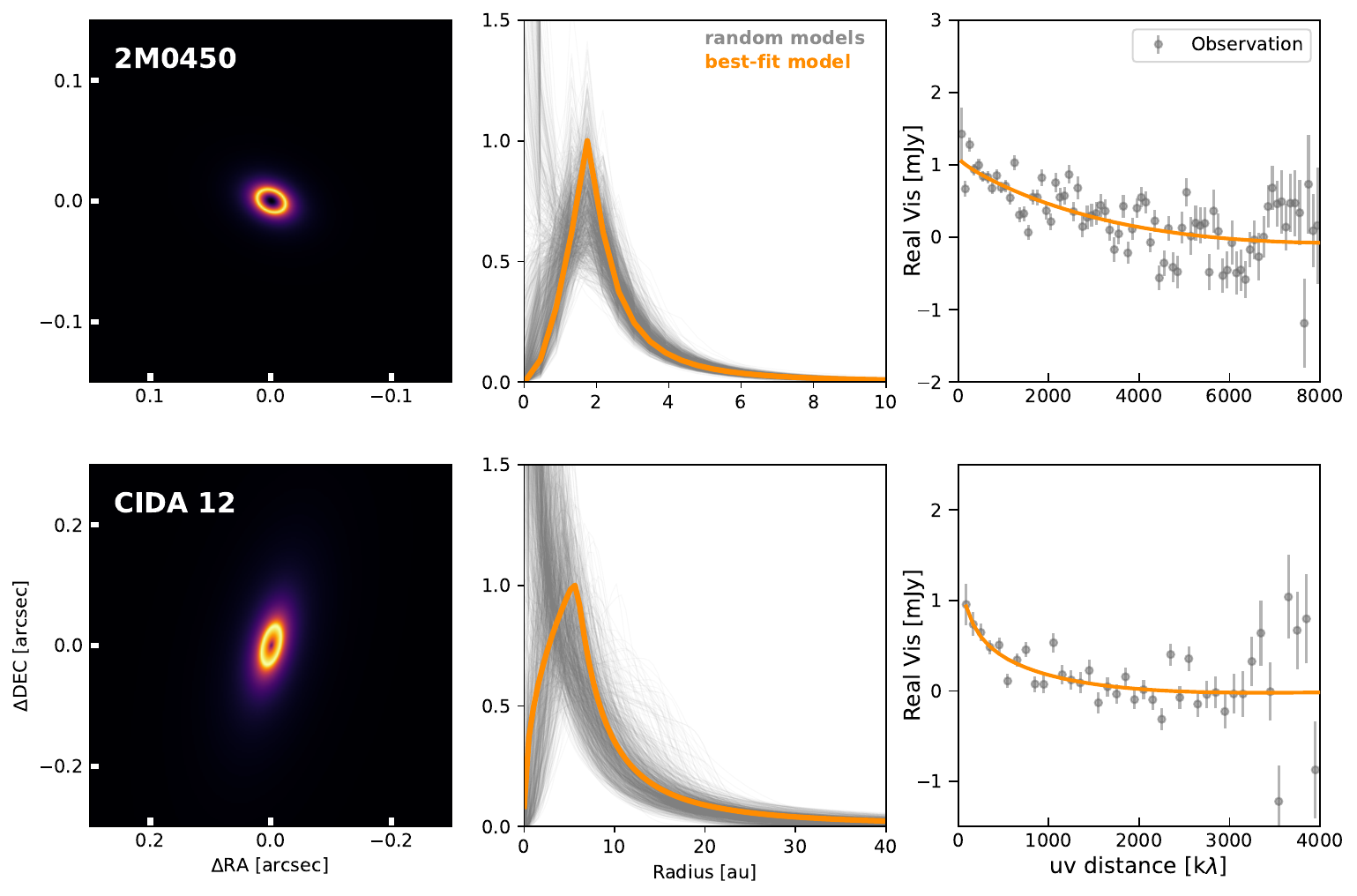}
    \caption{Nuker fitting results of disks that seem to lack structures in the image plane. \textit{Columns from left to right:} (1) unconvolved best-fit Nuker model images, (2) radial profiles retrieved from best-fit (orange) and 1000 randomly selected (gray) Nuker models, (3) real part of deprojected and binned visibilities from observations and best-fit models.}
    \label{fig:nukerimage}
\end{figure*}

The modeling procedure follows Section \ref{subsec:result_uvfit}, with Nuker parameters' priors following \citet{tripathi2017}: $p(\log_{10}\alpha)=\mathcal{U}(0,2),\ p(\beta)=\mathcal{U}(0,10)$ and $\gamma$ sampled approximated uniform in $(-3,2)$:
\begin{equation}
    p(\gamma) \propto \frac{1}{1+e^{-5(\gamma+3)}} - \frac{1}{1+e^{-15(\gamma-2)}}.
\end{equation}
Figure \ref{fig:nukerimage} shows the best-fit Nuker model images, visibilities and random radial profiles retrieved from posterior distributions.

\restartappendixnumbering
\section{SED Fitting}
\label{sec:app_sed}

The SED data points are collected from GALEX GR6+7 \citep{bianchi2017}, SDSS DR17 \citep{abdurrouf2022}, PAN-STARRS \citep{flewelling2020}, GAIA DR3 \citep{gaia2022}, 2MASS \citep{skrutskie2006}, WISE \citep{wright2010}, Spitzer \citep{luhman2010}, Herschel \citep{marton2017}, and our measurements at 1.3mm (Literature value at 1.3mm for CIDA 12 in \citealt{akeson2019}).

We fit the photospheric emission of each host star using the BT-Settl models (\citealt{allard2011}), with the temperature and extinction in Table \ref{tab:host_properties} and assuming a gravity of $\log g=4.0$. With the fitted spectrum and stellar luminosity as inputs, we then use the \texttt{RADMC-3D} code \citep{dullemond2012} to perform self-consistent radiative transfer to obtain dust temperature profiles and then simulate the SEDs.
For a flared disk model with well-mixed dust and gas (for small grains this is approximately correct), the disk mass density profile is given as: 
\begin{equation}
    \rho(r, z) = \frac{\Sigma(r)}{\sqrt{2\pi}H_p} \exp{\left(-\frac{z^2}{2H_p^2}\right)}.
\end{equation}
where $r$ is the distance to the central star measured in the disk mid-plane, $z$ is the distance to the disk mid-plane, $H_p$ is the pressure scale height and $\Sigma$ is the surface density integrated over vertical direction. We adopt simple power-law for $\Sigma$ and $H_p$:
\begin{align}
    \Sigma(r) &= \Sigma_0\left(\frac{r}{r_{\rm out}}\right)^p. \\
    H_p(r) &= H_{100} \left(\frac{r}{100 {\rm\ au}}\right)^\beta.
\end{align}
where $H_{100}$ is the scale height of small dust grains at 100 au and $p, \beta$ are the slope.

Six parameters are needed to describe the simple flared disk model: $R_{\rm in}, R_{\rm out}, p, \beta, H_{100}, M_{\rm dust}$. We fix $p$ to -1 and $R_{\rm out}$ to the $R_{90\%}$ in millimeter images (Section \ref{subsec:result_flux_mass_size} and Table \ref{tab:uvfit_results}). $M_{\rm dust}$ is first set to the estimated value in Table \ref{tab:uvfit_results} and adjusted to match the $F_{\rm1.3mm}$. We note that $M_{\rm dust}$ has negligible effect on the infrared SEDs (hence the parameter $R_{\rm in}$) since dust is optically thick at that wavelength. We test each combination of the other three parameters with $R_{\rm in}$ between dust sublimation radius and mm cavity size, $\beta$ in [1.0, 1.2] and $H_{100}$ in [5,15]. 
As for the dust prescription, we adopt the model in \citep{birnstiel2018} with grain sizes ranging from 0.01 $\mu \rm m$ to 1000 $\mu \rm m$.
The parameters of our best-fit models are listed in Table \ref{tab:sedfit}, and Figure \ref{fig:sed} shows the comparison between the models and observations.

\begin{deluxetable*}{cccccccc}
    \label{tab:sedfit}
    \tablecaption{Best-fit models for SED}
    \tablehead{\colhead{Parameters} & \colhead{2M0412} & \colhead{2M0434} & \colhead{2M0436} & \colhead{2M0450} & \colhead{CIDA 12} & \colhead{2M0508} & \colhead{Unit}
    }
    \startdata
    $R_{\rm in}$ & 0.90 & 4 & 0.03 & 0.02 & 0.3 & 4 & au \\
    $R_{\rm out}$ & 126 & 82 & 14 & 6 & 20 & 21 & au \\
    $p$ & -1 & -1 & -1 & -1 & -1 & -1 & -- \\
    $\beta$ & 1.15 & 1.20 & 1.15 & 1.15 & 1.20 & 1.15 & -- \\
    $H_{100}$ & 10.0 & 7.5 & 12.5 & 10.0 & 6.3 & 7.5 & au \\
    \enddata
\end{deluxetable*}


\bibliography{sample631}

\begin{thebibliography}{}
\expandafter\ifx\csname natexlab\endcsname\relax\def\natexlab#1{#1}\fi
\providecommand{\url}[1]{\href{#1}{#1}}
\providecommand{\dodoi}[1]{doi:~\href{http://doi.org/#1}{\nolinkurl{#1}}}
\providecommand{\doeprint}[1]{\href{http://ascl.net/#1}{\nolinkurl{http://ascl.net/#1}}}
\providecommand{\doarXiv}[1]{\href{https://arxiv.org/abs/#1}{\nolinkurl{https://arxiv.org/abs/#1}}}

\bibitem[{{Abdurro'uf} {et~al.}(2022){Abdurro'uf}, {Accetta}, {Aerts}, {Silva Aguirre}, {Ahumada}, {Ajgaonkar}, {Filiz Ak}, {Alam}, {Allende Prieto}, {Almeida}, {Anders}, {Anderson}, {Andrews}, {Anguiano}, {Aquino-Ort{\'\i}z}, {Arag{\'o}n-Salamanca}, {Argudo-Fern{\'a}ndez}, {Ata}, {Aubert}, {Avila-Reese}, {Badenes}, {Barb{\'a}}, {Barger}, {Barrera-Ballesteros}, {Beaton}, {Beers}, {Belfiore}, {Bender}, {Bernardi}, {Bershady}, {Beutler}, {Bidin}, {Bird}, {Bizyaev}, {Blanc}, {Blanton}, {Boardman}, {Bolton}, {Boquien}, {Borissova}, {Bovy}, {Brandt}, {Brown}, {Brownstein}, {Brusa}, {Buchner}, {Bundy}, {Burchett}, {Bureau}, {Burgasser}, {Cabang}, {Campbell}, {Cappellari}, {Carlberg}, {Wanderley}, {Carrera}, {Cash}, {Chen}, {Chen}, {Cherinka}, {Chiappini}, {Choi}, {Chojnowski}, {Chung}, {Clerc}, {Cohen}, {Comerford}, {Comparat}, {da Costa}, {Covey}, {Crane}, {Cruz-Gonzalez}, {Culhane}, {Cunha}, {Dai}, {Damke}, {Darling}, {Davidson}, {Davies}, {Dawson}, {De Lee}, {Diamond-Stanic}, {Cano-D{\'\i}az}, {S{\'a}nchez},
  {Donor}, {Duckworth}, {Dwelly}, {Eisenstein}, {Elsworth}, {Emsellem}, {Eracleous}, {Escoffier}, {Fan}, {Farr}, {Feng}, {Fern{\'a}ndez-Trincado}, {Feuillet}, {Filipp}, {Fillingham}, {Frinchaboy}, {Fromenteau}, {Galbany}, {Garc{\'\i}a}, {Garc{\'\i}a-Hern{\'a}ndez}, {Ge}, {Geisler}, {Gelfand}, {G{\'e}ron}, {Gibson}, {Goddy}, {Godoy-Rivera}, {Grabowski}, {Green}, {Greener}, {Grier}, {Griffith}, {Guo}, {Guy}, {Hadjara}, {Harding}, {Hasselquist}, {Hayes}, {Hearty}, {Hern{\'a}ndez}, {Hill}, {Hogg}, {Holtzman}, {Horta}, {Hsieh}, {Hsu}, {Hsu}, {Huber}, {Huertas-Company}, {Hutchinson}, {Hwang}, {Ibarra-Medel}, {Chitham}, {Ilha}, {Imig}, {Jaekle}, {Jayasinghe}, {Ji}, {Johnson}, {Jones}, {J{\"o}nsson}, {Katkov}, {Khalatyan}, {Kinemuchi}, {Kisku}, {Knapen}, {Kneib}, {Kollmeier}, {Kong}, {Kounkel}, {Kreckel}, {Krishnarao}, {Lacerna}, {Lane}, {Langgin}, {Lavender}, {Law}, {Lazarz}, {Leung}, {Leung}, {Lewis}, {Li}, {Li}, {Lian}, {Liang}, {Lin}, {Lin}, {Lin}, {Lintott}, {Long}, {Longa-Pe{\~n}a}, {L{\'o}pez-Cob{\'a}}, {Lu},
  {Lundgren}, {Luo}, {Mackereth}, {de la Macorra}, {Mahadevan}, {Majewski}, {Manchado}, {Mandeville}, {Maraston}, {Margalef-Bentabol}, {Masseron}, {Masters}, {Mathur}, {McDermid}, {Mckay}, {Merloni}, {Merrifield}, {Meszaros}, {Miglio}, {Di Mille}, {Minniti}, {Minsley}, {Monachesi}, {Moon}, {Mosser}, {Mulchaey}, {Muna}, {Mu{\~n}oz}, {Myers}, {Myers}, {Nadathur}, {Nair}, {Nandra}, {Neumann}, {Newman}, {Nidever}, {Nikakhtar}, {Nitschelm}, {O'Connell}, {Garma-Oehmichen}, {Luan Souza de Oliveira}, {Olney}, {Oravetz}, {Ortigoza-Urdaneta}, {Osorio}, {Otter}, {Pace}, {Padilla}, {Pan}, {Pan}, {Parikh}, {Parker}, {Peirani}, {Pe{\~n}a Ram{\'\i}rez}, {Penny}, {Percival}, {Perez-Fournon}, {Pinsonneault}, {Poidevin}, {Poovelil}, {Price-Whelan}, {B{\'a}rbara de Andrade Queiroz}, {Raddick}, {Ray}, {Rembold}, {Riddle}, {Riffel}, {Riffel}, {Rix}, {Robin}, {Rodr{\'\i}guez-Puebla}, {Roman-Lopes}, {Rom{\'a}n-Z{\'u}{\~n}iga}, {Rose}, {Ross}, {Rossi}, {Rubin}, {Salvato}, {S{\'a}nchez}, {S{\'a}nchez-Gallego}, {Sanderson}, {Santana
  Rojas}, {Sarceno}, {Sarmiento}, {Sayres}, {Sazonova}, {Schaefer}, {Schiavon}, {Schlegel}, {Schneider}, {Schultheis}, {Schwope}, {Serenelli}, {Serna}, {Shao}, {Shapiro}, {Sharma}, {Shen}, {Shetrone}, {Shu}, {Simon}, {Skrutskie}, {Smethurst}, {Smith}, {Sobeck}, {Spoo}, {Sprague}, {Stark}, {Stassun}, {Steinmetz}, {Stello}, {Stone-Martinez}, {Storchi-Bergmann}, {Stringfellow}, {Stutz}, {Su}, {Taghizadeh-Popp}, {Talbot}, {Tayar}, {Telles}, {Teske}, {Thakar}, {Theissen}, {Tkachenko}, {Thomas}, {Tojeiro}, {Hernandez Toledo}, {Troup}, {Trump}, {Trussler}, {Turner}, {Tuttle}, {Unda-Sanzana}, {V{\'a}zquez-Mata}, {Valentini}, {Valenzuela}, {Vargas-Gonz{\'a}lez}, {Vargas-Maga{\~n}a}, {Alfaro}, {Villanova}, {Vincenzo}, {Wake}, {Warfield}, {Washington}, {Weaver}, {Weijmans}, {Weinberg}, {Weiss}, {Westfall}, {Wild}, {Wilde}, {Wilson}, {Wilson}, {Wilson}, {Wolf}, {Wood-Vasey}, {Yan}, {Zamora}, {Zasowski}, {Zhang}, {Zhao}, {Zheng}, {Zheng}, \& {Zhu}}]{abdurrouf2022}
{Abdurro'uf}, {Accetta}, K., {Aerts}, C., {et~al.} 2022, \apjs, 259, 35, \dodoi{10.3847/1538-4365/ac4414}

\bibitem[{{Akeson} {et~al.}(2019){Akeson}, {Jensen}, {Carpenter}, {Ricci}, {Laos}, {Nogueira}, \& {Suen-Lewis}}]{akeson2019}
{Akeson}, R.~L., {Jensen}, E. L.~N., {Carpenter}, J., {et~al.} 2019, \apj, 872, 158, \dodoi{10.3847/1538-4357/aaff6a}

\bibitem[{{Allard} {et~al.}(2011){Allard}, {Homeier}, \& {Freytag}}]{allard2011}
{Allard}, F., {Homeier}, D., \& {Freytag}, B. 2011, in Astronomical Society of the Pacific Conference Series, Vol. 448, 16th Cambridge Workshop on Cool Stars, Stellar Systems, and the Sun, ed. C.~{Johns-Krull}, M.~K. {Browning}, \& A.~A. {West}, 91, \dodoi{10.48550/arXiv.1011.5405}

\bibitem[{{ALMA Partnership} {et~al.}(2015){ALMA Partnership}, {Brogan}, {P{\'e}rez}, {Hunter}, {Dent}, {Hales}, {Hills}, {Corder}, {Fomalont}, {Vlahakis}, {Asaki}, {Barkats}, {Hirota}, {Hodge}, {Impellizzeri}, {Kneissl}, {Liuzzo}, {Lucas}, {Marcelino}, {Matsushita}, {Nakanishi}, {Phillips}, {Richards}, {Toledo}, {Aladro}, {Broguiere}, {Cortes}, {Cortes}, {Espada}, {Galarza}, {Garcia-Appadoo}, {Guzman-Ramirez}, {Humphreys}, {Jung}, {Kameno}, {Laing}, {Leon}, {Marconi}, {Mignano}, {Nikolic}, {Nyman}, {Radiszcz}, {Remijan}, {Rod{\'o}n}, {Sawada}, {Takahashi}, {Tilanus}, {Vila Vilaro}, {Watson}, {Wiklind}, {Akiyama}, {Chapillon}, {de Gregorio-Monsalvo}, {Di Francesco}, {Gueth}, {Kawamura}, {Lee}, {Nguyen Luong}, {Mangum}, {Pietu}, {Sanhueza}, {Saigo}, {Takakuwa}, {Ubach}, {van Kempen}, {Wootten}, {Castro-Carrizo}, {Francke}, {Gallardo}, {Garcia}, {Gonzalez}, {Hill}, {Kaminski}, {Kurono}, {Liu}, {Lopez}, {Morales}, {Plarre}, {Schieven}, {Testi}, {Videla}, {Villard}, {Andreani}, {Hibbard}, \&
  {Tatematsu}}]{almapartner2015}
{ALMA Partnership}, {Brogan}, C.~L., {P{\'e}rez}, L.~M., {et~al.} 2015, \apjl, 808, L3, \dodoi{10.1088/2041-8205/808/1/L3}

\bibitem[{{Andrews}(2020)}]{andrews2020}
{Andrews}, S.~M. 2020, \araa, 58, 483, \dodoi{10.1146/annurev-astro-031220-010302}

\bibitem[{{Andrews} {et~al.}(2013){Andrews}, {Rosenfeld}, {Kraus}, \& {Wilner}}]{andrews2013}
{Andrews}, S.~M., {Rosenfeld}, K.~A., {Kraus}, A.~L., \& {Wilner}, D.~J. 2013, \apj, 771, 129, \dodoi{10.1088/0004-637X/771/2/129}

\bibitem[{{Andrews} {et~al.}(2018{\natexlab{a}}){Andrews}, {Terrell}, {Tripathi}, {Ansdell}, {Williams}, \& {Wilner}}]{andrews2018scaling}
{Andrews}, S.~M., {Terrell}, M., {Tripathi}, A., {et~al.} 2018{\natexlab{a}}, \apj, 865, 157, \dodoi{10.3847/1538-4357/aadd9f}

\bibitem[{{Andrews} {et~al.}(2018{\natexlab{b}}){Andrews}, {Huang}, {P{\'e}rez}, {Isella}, {Dullemond}, {Kurtovic}, {Guzm{\'a}n}, {Carpenter}, {Wilner}, {Zhang}, {Zhu}, {Birnstiel}, {Bai}, {Benisty}, {Hughes}, {{\"O}berg}, \& {Ricci}}]{andrews2018}
{Andrews}, S.~M., {Huang}, J., {P{\'e}rez}, L.~M., {et~al.} 2018{\natexlab{b}}, \apjl, 869, L41, \dodoi{10.3847/2041-8213/aaf741}

\bibitem[{{Ansdell} {et~al.}(2017){Ansdell}, {Williams}, {Manara}, {Miotello}, {Facchini}, {van der Marel}, {Testi}, \& {van Dishoeck}}]{ansdell2017}
{Ansdell}, M., {Williams}, J.~P., {Manara}, C.~F., {et~al.} 2017, \aj, 153, 240, \dodoi{10.3847/1538-3881/aa69c0}

\bibitem[{{Ansdell} {et~al.}(2016){Ansdell}, {Williams}, {van der Marel}, {Carpenter}, {Guidi}, {Hogerheijde}, {Mathews}, {Manara}, {Miotello}, {Natta}, {Oliveira}, {Tazzari}, {Testi}, {van Dishoeck}, \& {van Terwisga}}]{ansdell2016}
{Ansdell}, M., {Williams}, J.~P., {van der Marel}, N., {et~al.} 2016, \apj, 828, 46, \dodoi{10.3847/0004-637X/828/1/46}

\bibitem[{{Ansdell} {et~al.}(2018){Ansdell}, {Williams}, {Trapman}, {van Terwisga}, {Facchini}, {Manara}, {van der Marel}, {Miotello}, {Tazzari}, {Hogerheijde}, {Guidi}, {Testi}, \& {van Dishoeck}}]{ansdell2018}
{Ansdell}, M., {Williams}, J.~P., {Trapman}, L., {et~al.} 2018, \apj, 859, 21, \dodoi{10.3847/1538-4357/aab890}

\bibitem[{{Aoyama} \& {Bai}(2023)}]{aoyama2023}
{Aoyama}, Y., \& {Bai}, X.-N. 2023, \apj, 946, 5, \dodoi{10.3847/1538-4357/acb81f}

\bibitem[{{Artymowicz} \& {Lubow}(1994)}]{artymowicz1994}
{Artymowicz}, P., \& {Lubow}, S.~H. 1994, \apj, 421, 651, \dodoi{10.1086/173679}

\bibitem[{{Bae} {et~al.}(2023){Bae}, {Isella}, {Zhu}, {Martin}, {Okuzumi}, \& {Suriano}}]{bae2023}
{Bae}, J., {Isella}, A., {Zhu}, Z., {et~al.} 2023, in Astronomical Society of the Pacific Conference Series, Vol. 534, Astronomical Society of the Pacific Conference Series, ed. S.~{Inutsuka}, Y.~{Aikawa}, T.~{Muto}, K.~{Tomida}, \& M.~{Tamura}, 423

\bibitem[{{Bai}(2016)}]{bai2016}
{Bai}, X.-N. 2016, \apj, 821, 80, \dodoi{10.3847/0004-637X/821/2/80}

\bibitem[{{Banzatti} {et~al.}(2019){Banzatti}, {Pascucci}, {Edwards}, {Fang}, {Gorti}, \& {Flock}}]{banzatti2019}
{Banzatti}, A., {Pascucci}, I., {Edwards}, S., {et~al.} 2019, \apj, 870, 76, \dodoi{10.3847/1538-4357/aaf1aa}

\bibitem[{{Barenfeld} {et~al.}(2016){Barenfeld}, {Carpenter}, {Ricci}, \& {Isella}}]{barenfeld2016}
{Barenfeld}, S.~A., {Carpenter}, J.~M., {Ricci}, L., \& {Isella}, A. 2016, \apj, 827, 142, \dodoi{10.3847/0004-637X/827/2/142}

\bibitem[{{Barenfeld} {et~al.}(2017){Barenfeld}, {Carpenter}, {Sargent}, {Isella}, \& {Ricci}}]{barenfeld2017}
{Barenfeld}, S.~A., {Carpenter}, J.~M., {Sargent}, A.~I., {Isella}, A., \& {Ricci}, L. 2017, \apj, 851, 85, \dodoi{10.3847/1538-4357/aa989d}

\bibitem[{{Bianchi} {et~al.}(2017){Bianchi}, {Shiao}, \& {Thilker}}]{bianchi2017}
{Bianchi}, L., {Shiao}, B., \& {Thilker}, D. 2017, \apjs, 230, 24, \dodoi{10.3847/1538-4365/aa7053}

\bibitem[{{Birnstiel} {et~al.}(2018){Birnstiel}, {Dullemond}, {Zhu}, {Andrews}, {Bai}, {Wilner}, {Carpenter}, {Huang}, {Isella}, {Benisty}, {P{\'e}rez}, \& {Zhang}}]{birnstiel2018}
{Birnstiel}, T., {Dullemond}, C.~P., {Zhu}, Z., {et~al.} 2018, \apjl, 869, L45, \dodoi{10.3847/2041-8213/aaf743}

\bibitem[{{Boss}(1997)}]{boss1997}
{Boss}, A.~P. 1997, Science, 276, 1836, \dodoi{10.1126/science.276.5320.1836}

\bibitem[{{Boss}(2006)}]{boss2006}
---. 2006, \apj, 643, 501, \dodoi{10.1086/501522}

\bibitem[{{Boss} \& {Kanodia}(2023)}]{boss2023}
{Boss}, A.~P., \& {Kanodia}, S. 2023, \apj, 956, 4, \dodoi{10.3847/1538-4357/acf373}

\bibitem[{{Bowler} {et~al.}(2015){Bowler}, {Liu}, {Shkolnik}, \& {Tamura}}]{bowler2015}
{Bowler}, B.~P., {Liu}, M.~C., {Shkolnik}, E.~L., \& {Tamura}, M. 2015, \apjs, 216, 7, \dodoi{10.1088/0067-0049/216/1/7}

\bibitem[{{Bryant} {et~al.}(2023){Bryant}, {Bayliss}, \& {Van Eylen}}]{bryant2023}
{Bryant}, E.~M., {Bayliss}, D., \& {Van Eylen}, V. 2023, \mnras, 521, 3663, \dodoi{10.1093/mnras/stad626}

\bibitem[{{Burn} {et~al.}(2021){Burn}, {Schlecker}, {Mordasini}, {Emsenhuber}, {Alibert}, {Henning}, {Klahr}, \& {Benz}}]{burn2021}
{Burn}, R., {Schlecker}, M., {Mordasini}, C., {et~al.} 2021, \aap, 656, A72, \dodoi{10.1051/0004-6361/202140390}

\bibitem[{{CASA Team} {et~al.}(2022){CASA Team}, {Bean}, {Bhatnagar}, {Castro}, {Donovan Meyer}, {Emonts}, {Garcia}, {Garwood}, {Golap}, {Gonzalez Villalba}, {Harris}, {Hayashi}, {Hoskins}, {Hsieh}, {Jagannathan}, {Kawasaki}, {Keimpema}, {Kettenis}, {Lopez}, {Marvil}, {Masters}, {McNichols}, {Mehringer}, {Miel}, {Moellenbrock}, {Montesino}, {Nakazato}, {Ott}, {Petry}, {Pokorny}, {Raba}, {Rau}, {Schiebel}, {Schweighart}, {Sekhar}, {Shimada}, {Small}, {Steeb}, {Sugimoto}, {Suoranta}, {Tsutsumi}, {van Bemmel}, {Verkouter}, {Wells}, {Xiong}, {Szomoru}, {Griffith}, {Glendenning}, \& {Kern}}]{casa2022}
{CASA Team}, {Bean}, B., {Bhatnagar}, S., {et~al.} 2022, \pasp, 134, 114501, \dodoi{10.1088/1538-3873/ac9642}

\bibitem[{{Cieza} {et~al.}(2019){Cieza}, {Ru{\'\i}z-Rodr{\'\i}guez}, {Hales}, {Casassus}, {P{\'e}rez}, {Gonzalez-Ruilova}, {C{\'a}novas}, {Williams}, {Zurlo}, {Ansdell}, {Avenhaus}, {Bayo}, {Bertrang}, {Christiaens}, {Dent}, {Ferrero}, {Gamen}, {Olofsson}, {Orcajo}, {Pe{\~n}a Ram{\'\i}rez}, {Principe}, {Schreiber}, \& {van der Plas}}]{cieza2019}
{Cieza}, L.~A., {Ru{\'\i}z-Rodr{\'\i}guez}, D., {Hales}, A., {et~al.} 2019, \mnras, 482, 698, \dodoi{10.1093/mnras/sty2653}

\bibitem[{{Cieza} {et~al.}(2021){Cieza}, {Gonz{\'a}lez-Ruilova}, {Hales}, {Pinilla}, {Ru{\'\i}z-Rodr{\'\i}guez}, {Zurlo}, {Casassus}, {P{\'e}rez}, {C{\'a}novas}, {Arce-Tord}, {Flock}, {Kurtovic}, {Marino}, {Nogueira}, {Perez}, {Price}, {Principe}, \& {Williams}}]{cieza2021}
{Cieza}, L.~A., {Gonz{\'a}lez-Ruilova}, C., {Hales}, A.~S., {et~al.} 2021, \mnras, 501, 2934, \dodoi{10.1093/mnras/staa3787}

\bibitem[{{Crida} {et~al.}(2006){Crida}, {Morbidelli}, \& {Masset}}]{crida2006}
{Crida}, A., {Morbidelli}, A., \& {Masset}, F. 2006, \icarus, 181, 587, \dodoi{10.1016/j.icarus.2005.10.007}

\bibitem[{{Curone} {et~al.}(2022){Curone}, {Izquierdo}, {Testi}, {Lodato}, {Facchini}, {Natta}, {Pinilla}, {Kurtovic}, {Toci}, {Benisty}, {Tazzari}, {Borsa}, {Lombardi}, {Manara}, {Sanchis}, \& {Ricci}}]{curone2022}
{Curone}, P., {Izquierdo}, A.~F., {Testi}, L., {et~al.} 2022, \aap, 665, A25, \dodoi{10.1051/0004-6361/202142748}

\bibitem[{{Cutri} {et~al.}(2003){Cutri}, {Skrutskie}, {van Dyk}, {Beichman}, {Carpenter}, {Chester}, {Cambresy}, {Evans}, {Fowler}, {Gizis}, {Howard}, {Huchra}, {Jarrett}, {Kopan}, {Kirkpatrick}, {Light}, {Marsh}, {McCallon}, {Schneider}, {Stiening}, {Sykes}, {Weinberg}, {Wheaton}, {Wheelock}, \& {Zacarias}}]{cutri03}
{Cutri}, R.~M., {Skrutskie}, M.~F., {van Dyk}, S., {et~al.} 2003, {2MASS All Sky Catalog of point sources.}

\bibitem[{{Daemgen} {et~al.}(2015){Daemgen}, {Bonavita}, {Jayawardhana}, {Lafreni{\`e}re}, \& {Janson}}]{daemgen2015}
{Daemgen}, S., {Bonavita}, M., {Jayawardhana}, R., {Lafreni{\`e}re}, D., \& {Janson}, M. 2015, \apj, 799, 155, \dodoi{10.1088/0004-637X/799/2/155}

\bibitem[{{de Juan Ovelar} {et~al.}(2013){de Juan Ovelar}, {Min}, {Dominik}, {Thalmann}, {Pinilla}, {Benisty}, \& {Birnstiel}}]{deJuanOvelar2013}
{de Juan Ovelar}, M., {Min}, M., {Dominik}, C., {et~al.} 2013, \aap, 560, A111, \dodoi{10.1051/0004-6361/201322218}

\bibitem[{Diaz~Trigo {et~al.}(2019)Diaz~Trigo, Carpenter, Maude, Miura, \& Plunkett}]{diaz_trigo_maria_2019_4511962}
Diaz~Trigo, M., Carpenter, J., Maude, L., Miura, R., \& Plunkett, A. 2019, ALMA Cycle 7 Proposer's Guide, \dodoi{10.5281/zenodo.4511962}

\bibitem[{{Dodson-Robinson} \& {Salyk}(2011)}]{dodson-robinson2011}
{Dodson-Robinson}, S.~E., \& {Salyk}, C. 2011, \apj, 738, 131, \dodoi{10.1088/0004-637X/738/2/131}

\bibitem[{{Dong} {et~al.}(2018){Dong}, {Liu}, {Eisner}, {Andrews}, {Fung}, {Zhu}, {Chiang}, {Hashimoto}, {Liu}, {Casassus}, {Esposito}, {Hasegawa}, {Muto}, {Pavlyuchenkov}, {Wilner}, {Akiyama}, {Tamura}, \& {Wisniewski}}]{dong2018}
{Dong}, R., {Liu}, S.-y., {Eisner}, J., {et~al.} 2018, \apj, 860, 124, \dodoi{10.3847/1538-4357/aac6cb}

\bibitem[{{Dullemond} {et~al.}(2001){Dullemond}, {Dominik}, \& {Natta}}]{dullemond2001}
{Dullemond}, C.~P., {Dominik}, C., \& {Natta}, A. 2001, \apj, 560, 957, \dodoi{10.1086/323057}

\bibitem[{{Dullemond} {et~al.}(2012){Dullemond}, {Juhasz}, {Pohl}, {Sereshti}, {Shetty}, {Peters}, {Commercon}, \& {Flock}}]{dullemond2012}
{Dullemond}, C.~P., {Juhasz}, A., {Pohl}, A., {et~al.} 2012, {RADMC-3D: A multi-purpose radiative transfer tool}, Astrophysics Source Code Library, record ascl:1202.015.
\newblock \doeprint{1202.015}

\bibitem[{{Dullemond} {et~al.}(2018){Dullemond}, {Birnstiel}, {Huang}, {Kurtovic}, {Andrews}, {Guzm{\'a}n}, {P{\'e}rez}, {Isella}, {Zhu}, {Benisty}, {Wilner}, {Bai}, {Carpenter}, {Zhang}, \& {Ricci}}]{dullemond2018}
{Dullemond}, C.~P., {Birnstiel}, T., {Huang}, J., {et~al.} 2018, \apjl, 869, L46, \dodoi{10.3847/2041-8213/aaf742}

\bibitem[{{Elbakyan} {et~al.}(2022){Elbakyan}, {Wu}, {Nayakshin}, \& {Rosotti}}]{elbakyan2022}
{Elbakyan}, V., {Wu}, Y., {Nayakshin}, S., \& {Rosotti}, G. 2022, \mnras, 515, 3113, \dodoi{10.1093/mnras/stac1774}

\bibitem[{{Esplin} \& {Luhman}(2019)}]{esplin2019}
{Esplin}, T.~L., \& {Luhman}, K.~L. 2019, \aj, 158, 54, \dodoi{10.3847/1538-3881/ab2594}

\bibitem[{{Facchini} {et~al.}(2018){Facchini}, {Pinilla}, {van Dishoeck}, \& {de Juan Ovelar}}]{facchini2018}
{Facchini}, S., {Pinilla}, P., {van Dishoeck}, E.~F., \& {de Juan Ovelar}, M. 2018, \aap, 612, A104, \dodoi{10.1051/0004-6361/201731390}

\bibitem[{{Facchini} {et~al.}(2020){Facchini}, {Benisty}, {Bae}, {Loomis}, {Perez}, {Ansdell}, {Mayama}, {Pinilla}, {Teague}, {Isella}, \& {Mann}}]{facchini2020}
{Facchini}, S., {Benisty}, M., {Bae}, J., {et~al.} 2020, \aap, 639, A121, \dodoi{10.1051/0004-6361/202038027}

\bibitem[{{Fang} {et~al.}(2023{\natexlab{a}}){Fang}, {Pascucci}, {Edwards}, {Gorti}, {Hillenbrand}, \& {Carpenter}}]{fang23usco}
{Fang}, M., {Pascucci}, I., {Edwards}, S., {et~al.} 2023{\natexlab{a}}, \apj, 945, 112, \dodoi{10.3847/1538-4357/acb2c9}

\bibitem[{{Fang} {et~al.}(2023{\natexlab{b}}){Fang}, {Wang}, {Herczeg}, {Hashimoto}, {Xu}, {Nemer}, {Pascucci}, {Haffert}, \& {Aoyama}}]{fang23}
{Fang}, M., {Wang}, L., {Herczeg}, G.~J., {et~al.} 2023{\natexlab{b}}, Nature Astronomy, 7, 905, \dodoi{10.1038/s41550-023-02004-x}

\bibitem[{{Fernandes} {et~al.}(2019){Fernandes}, {Mulders}, {Pascucci}, {Mordasini}, \& {Emsenhuber}}]{fernandes2019}
{Fernandes}, R.~B., {Mulders}, G.~D., {Pascucci}, I., {Mordasini}, C., \& {Emsenhuber}, A. 2019, \apj, 874, 81, \dodoi{10.3847/1538-4357/ab0300}

\bibitem[{{Fischer} {et~al.}(2017){Fischer}, {Megeath}, {Furlan}, {Ali}, {Stutz}, {Tobin}, {Osorio}, {Stanke}, {Manoj}, {Poteet}, {Booker}, {Hartmann}, {Wilson}, {Myers}, \& {Watson}}]{fischer17}
{Fischer}, W.~J., {Megeath}, S.~T., {Furlan}, E., {et~al.} 2017, \apj, 840, 69, \dodoi{10.3847/1538-4357/aa6d69}

\bibitem[{{Flewelling} {et~al.}(2020){Flewelling}, {Magnier}, {Chambers}, {Heasley}, {Holmberg}, {Huber}, {Sweeney}, {Waters}, {Calamida}, {Casertano}, {Chen}, {Farrow}, {Hasinger}, {Henderson}, {Long}, {Metcalfe}, {Narayan}, {Nieto-Santisteban}, {Norberg}, {Rest}, {Saglia}, {Szalay}, {Thakar}, {Tonry}, {Valenti}, {Werner}, {White}, {Denneau}, {Draper}, {Hodapp}, {Jedicke}, {Kaiser}, {Kudritzki}, {Price}, {Wainscoat}, {Chastel}, {McLean}, {Postman}, \& {Shiao}}]{flewelling2020}
{Flewelling}, H.~A., {Magnier}, E.~A., {Chambers}, K.~C., {et~al.} 2020, \apjs, 251, 7, \dodoi{10.3847/1538-4365/abb82d}

\bibitem[{{Foreman-Mackey} {et~al.}(2013){Foreman-Mackey}, {Hogg}, {Lang}, \& {Goodman}}]{foremanMackey2013}
{Foreman-Mackey}, D., {Hogg}, D.~W., {Lang}, D., \& {Goodman}, J. 2013, \pasp, 125, 306, \dodoi{10.1086/670067}

\bibitem[{{Francis} {et~al.}(2020){Francis}, {Johnstone}, {Herczeg}, {Hunter}, \& {Harsono}}]{francis2020}
{Francis}, L., {Johnstone}, D., {Herczeg}, G., {Hunter}, T.~R., \& {Harsono}, D. 2020, \aj, 160, 270, \dodoi{10.3847/1538-3881/abbe1a}

\bibitem[{{Fulton} {et~al.}(2021){Fulton}, {Rosenthal}, {Hirsch}, {Isaacson}, {Howard}, {Dedrick}, {Sherstyuk}, {Blunt}, {Petigura}, {Knutson}, {Behmard}, {Chontos}, {Crepp}, {Crossfield}, {Dalba}, {Fischer}, {Henry}, {Kane}, {Kosiarek}, {Marcy}, {Rubenzahl}, {Weiss}, \& {Wright}}]{fulton2021}
{Fulton}, B.~J., {Rosenthal}, L.~J., {Hirsch}, L.~A., {et~al.} 2021, \apjs, 255, 14, \dodoi{10.3847/1538-4365/abfcc1}

\bibitem[{{Gaia Collaboration} {et~al.}(2022){Gaia Collaboration}, {Vallenari}, {Brown}, {Prusti}, {de Bruijne}, {Arenou}, {Babusiaux}, {Biermann}, {Creevey}, {Ducourant}, {Evans}, {Eyer}, {Guerra}, {Hutton}, {Jordi}, {Klioner}, {Lammers}, {Lindegren}, {Luri}, {Mignard}, {Panem}, {Pourbaix}, {Randich}, {Sartoretti}, {Soubiran}, {Tanga}, {Walton}, {Bailer-Jones}, {Bastian}, {Drimmel}, {Jansen}, {Katz}, {Lattanzi}, {van Leeuwen}, {Bakker}, {Cacciari}, {Casta{\~n}eda}, {De Angeli}, {Fabricius}, {Fouesneau}, {Fr{\'e}mat}, {Galluccio}, {Guerrier}, {Heiter}, {Masana}, {Messineo}, {Mowlavi}, {Nicolas}, {Nienartowicz}, {Pailler}, {Panuzzo}, {Riclet}, {Roux}, {Seabroke}, {Sordo{\o}rcit}, {Th{\'e}venin}, {Gracia-Abril}, {Portell}, {Teyssier}, {Altmann}, {Andrae}, {Audard}, {Bellas-Velidis}, {Benson}, {Berthier}, {Blomme}, {Burgess}, {Busonero}, {Busso}, {C{\'a}novas}, {Carry}, {Cellino}, {Cheek}, {Clementini}, {Damerdji}, {Davidson}, {de Teodoro}, {Nu{\~n}ez Campos}, {Delchambre}, {Dell'Oro}, {Esquej},
  {Fern{\'a}ndez-Hern{\'a}ndez}, {Fraile}, {Garabato}, {Garc{\'\i}a-Lario}, {Gosset}, {Haigron}, {Halbwachs}, {Hambly}, {Harrison}, {Hern{\'a}ndez}, {Hestroffer}, {Hodgkin}, {Holl}, {Jan{\ss}en}, {Jevardat de Fombelle}, {Jordan}, {Krone-Martins}, {Lanzafame}, {L{\"o}ffler}, {Marchal}, {Marrese}, {Moitinho}, {Muinonen}, {Osborne}, {Pancino}, {Pauwels}, {Recio-Blanco}, {Reyl{\'e}}, {Riello}, {Rimoldini}, {Roegiers}, {Rybizki}, {Sarro}, {Siopis}, {Smith}, {Sozzetti}, {Utrilla}, {van Leeuwen}, {Abbas}, {{\'A}brah{\'a}m}, {Abreu Aramburu}, {Aerts}, {Aguado}, {Ajaj}, {Aldea-Montero}, {Altavilla}, {{\'A}lvarez}, {Alves}, {Anders}, {Anderson}, {Anglada Varela}, {Antoja}, {Baines}, {Baker}, {Balaguer-N{\'u}{\~n}ez}, {Balbinot}, {Balog}, {Barache}, {Barbato}, {Barros}, {Barstow}, {Bartolom{\'e}}, {Bassilana}, {Bauchet}, {Becciani}, {Bellazzini}, {Berihuete}, {Bernet}, {Bertone}, {Bianchi}, {Binnenfeld}, {Blanco-Cuaresma}, {Blazere}, {Boch}, {Bombrun}, {Bossini}, {Bouquillon}, {Bragaglia}, {Bramante}, {Breedt},
  {Bressan}, {Brouillet}, {Brugaletta}, {Bucciarelli}, {Burlacu}, {Butkevich}, {Buzzi}, {Caffau}, {Cancelliere}, {Cantat-Gaudin}, {Carballo}, {Carlucci}, {Carnerero}, {Carrasco}, {Casamiquela}, {Castellani}, {Castro-Ginard}, {Chaoul}, {Charlot}, {Chemin}, {Chiaramida}, {Chiavassa}, {Chornay}, {Comoretto}, {Contursi}, {Cooper}, {Cornez}, {Cowell}, {Crifo}, {Cropper}, {Crosta}, {Crowley}, {Dafonte}, {Dapergolas}, {David}, {David}, {de Laverny}, {De Luise}, {De March}, {De Ridder}, {de Souza}, {de Torres}, {del Peloso}, {del Pozo}, {Delbo}, {Delgado}, {Delisle}, {Demouchy}, {Dharmawardena}, {Di Matteo}, {Diakite}, {Diener}, {Distefano}, {Dolding}, {Edvardsson}, {Enke}, {Fabre}, {Fabrizio}, {Faigler}, {Fedorets}, {Fernique}, {Fienga}, {Figueras}, {Fournier}, {Fouron}, {Fragkoudi}, {Gai}, {Garcia-Gutierrez}, {Garcia-Reinaldos}, {Garc{\'\i}a-Torres}, {Garofalo}, {Gavel}, {Gavras}, {Gerlach}, {Geyer}, {Giacobbe}, {Gilmore}, {Girona}, {Giuffrida}, {Gomel}, {Gomez}, {Gonz{\'a}lez-N{\'u}{\~n}ez},
  {Gonz{\'a}lez-Santamar{\'\i}a}, {Gonz{\'a}lez-Vidal}, {Granvik}, {Guillout}, {Guiraud}, {Guti{\'e}rrez-S{\'a}nchez}, {Guy}, {Hatzidimitriou}, {Hauser}, {Haywood}, {Helmer}, {Helmi}, {Sarmiento}, {Hidalgo}, {Hilger}, {H{\l}adczuk}, {Hobbs}, {Holland}, {Huckle}, {Jardine}, {Jasniewicz}, {Jean-Antoine Piccolo}, {Jim{\'e}nez-Arranz}, {Jorissen}, {Juaristi Campillo}, {Julbe}, {Karbevska}, {Kervella}, {Khanna}, {Kontizas}, {Kordopatis}, {Korn}, {K{\'o}sp{\'a}l}, {Kostrzewa-Rutkowska}, {Kruszy{\'n}ska}, {Kun}, {Laizeau}, {Lambert}, {Lanza}, {Lasne}, {Le Campion}, {Lebreton}, {Lebzelter}, {Leccia}, {Leclerc}, {Lecoeur-Taibi}, {Liao}, {Licata}, {Lindstr{\o}m}, {Lister}, {Livanou}, {Lobel}, {Lorca}, {Loup}, {Madrero Pardo}, {Magdaleno Romeo}, {Managau}, {Mann}, {Manteiga}, {Marchant}, {Marconi}, {Marcos}, {Marcos Santos}, {Mar{\'\i}n Pina}, {Marinoni}, {Marocco}, {Marshall}, {Polo}, {Mart{\'\i}n-Fleitas}, {Marton}, {Mary}, {Masip}, {Massari}, {Mastrobuono-Battisti}, {Mazeh}, {McMillan}, {Messina}, {Michalik},
  {Millar}, {Mints}, {Molina}, {Molinaro}, {Moln{\'a}r}, {Monari}, {Mongui{\'o}}, {Montegriffo}, {Montero}, {Mor}, {Mora}, {Morbidelli}, {Morel}, {Morris}, {Muraveva}, {Murphy}, {Musella}, {Nagy}, {Noval}, {Oca{\~n}a}, {Ogden}, {Ordenovic}, {Osinde}, {Pagani}, {Pagano}, {Palaversa}, {Palicio}, {Pallas-Quintela}, {Panahi}, {Payne-Wardenaar}, {Pe{\~n}alosa Esteller}, {Penttil{\"a}}, {Pichon}, {Piersimoni}, {Pineau}, {Plachy}, {Plum}, {Poggio}, {Pr{\v{s}}a}, {Pulone}, {Racero}, {Ragaini}, {Rainer}, {Raiteri}, {Rambaux}, {Ramos}, {Ramos-Lerate}, {Re Fiorentin}, {Regibo}, {Richards}, {Rios Diaz}, {Ripepi}, {Riva}, {Rix}, {Rixon}, {Robichon}, {Robin}, {Robin}, {Roelens}, {Rogues}, {Rohrbasser}, {Romero-G{\'o}mez}, {Rowell}, {Royer}, {Ruz Mieres}, {Rybicki}, {Sadowski}, {S{\'a}ez N{\'u}{\~n}ez}, {Sagrist{\`a} Sell{\'e}s}, {Sahlmann}, {Salguero}, {Samaras}, {Sanchez Gimenez}, {Sanna}, {Santove{\~n}a}, {Sarasso}, {Schultheis}, {Sciacca}, {Segol}, {Segovia}, {S{\'e}gransan}, {Semeux}, {Shahaf}, {Siddiqui}, {Siebert},
  {Siltala}, {Silvelo}, {Slezak}, {Slezak}, {Smart}, {Snaith}, {Solano}, {Solitro}, {Souami}, {Souchay}, {Spagna}, {Spina}, {Spoto}, {Steele}, {Steidelm{\"u}ller}, {Stephenson}, {S{\"u}veges}, {Surdej}, {Szabados}, {Szegedi-Elek}, {Taris}, {Taylo}, {Teixeira}, {Tolomei}, {Tonello}, {Torra}, {Torra}, {Torralba Elipe}, {Trabucchi}, {Tsounis}, {Turon}, {Ulla}, {Unger}, {Vaillant}, {van Dillen}, {van Reeven}, {Vanel}, {Vecchiato}, {Viala}, {Vicente}, {Voutsinas}, {Weiler}, {Wevers}, {Wyrzykowski}, {Yoldas}, {Yvard}, {Zhao}, {Zorec}, {Zucker}, \& {Zwitter}}]{gaia2022}
{Gaia Collaboration}, {Vallenari}, A., {Brown}, A.~G.~A., {et~al.} 2022, arXiv e-prints, arXiv:2208.00211, \dodoi{10.48550/arXiv.2208.00211}

\bibitem[{{Gaia Collaboration} {et~al.}(2023){Gaia Collaboration}, {Vallenari}, {Brown}, {Prusti}, {de Bruijne}, {Arenou}, {Babusiaux}, {Biermann}, {Creevey}, {Ducourant}, {Evans}, {Eyer}, {Guerra}, {Hutton}, {Jordi}, {Klioner}, {Lammers}, {Lindegren}, {Luri}, {Mignard}, {Panem}, {Pourbaix}, {Randich}, {Sartoretti}, {Soubiran}, {Tanga}, {Walton}, {Bailer-Jones}, {Bastian}, {Drimmel}, {Jansen}, {Katz}, {Lattanzi}, {van Leeuwen}, {Bakker}, {Cacciari}, {Casta{\~n}eda}, {De Angeli}, {Fabricius}, {Fouesneau}, {Fr{\'e}mat}, {Galluccio}, {Guerrier}, {Heiter}, {Masana}, {Messineo}, {Mowlavi}, {Nicolas}, {Nienartowicz}, {Pailler}, {Panuzzo}, {Riclet}, {Roux}, {Seabroke}, {Sordo}, {Th{\'e}venin}, {Gracia-Abril}, {Portell}, {Teyssier}, {Altmann}, {Andrae}, {Audard}, {Bellas-Velidis}, {Benson}, {Berthier}, {Blomme}, {Burgess}, {Busonero}, {Busso}, {C{\'a}novas}, {Carry}, {Cellino}, {Cheek}, {Clementini}, {Damerdji}, {Davidson}, {de Teodoro}, {Nu{\~n}ez Campos}, {Delchambre}, {Dell'Oro}, {Esquej},
  {Fern{\'a}ndez-Hern{\'a}ndez}, {Fraile}, {Garabato}, {Garc{\'\i}a-Lario}, {Gosset}, {Haigron}, {Halbwachs}, {Hambly}, {Harrison}, {Hern{\'a}ndez}, {Hestroffer}, {Hodgkin}, {Holl}, {Jan{\ss}en}, {Jevardat de Fombelle}, {Jordan}, {Krone-Martins}, {Lanzafame}, {L{\"o}ffler}, {Marchal}, {Marrese}, {Moitinho}, {Muinonen}, {Osborne}, {Pancino}, {Pauwels}, {Recio-Blanco}, {Reyl{\'e}}, {Riello}, {Rimoldini}, {Roegiers}, {Rybizki}, {Sarro}, {Siopis}, {Smith}, {Sozzetti}, {Utrilla}, {van Leeuwen}, {Abbas}, {{\'A}brah{\'a}m}, {Abreu Aramburu}, {Aerts}, {Aguado}, {Ajaj}, {Aldea-Montero}, {Altavilla}, {{\'A}lvarez}, {Alves}, {Anders}, {Anderson}, {Anglada Varela}, {Antoja}, {Baines}, {Baker}, {Balaguer-N{\'u}{\~n}ez}, {Balbinot}, {Balog}, {Barache}, {Barbato}, {Barros}, {Barstow}, {Bartolom{\'e}}, {Bassilana}, {Bauchet}, {Becciani}, {Bellazzini}, {Berihuete}, {Bernet}, {Bertone}, {Bianchi}, {Binnenfeld}, {Blanco-Cuaresma}, {Blazere}, {Boch}, {Bombrun}, {Bossini}, {Bouquillon}, {Bragaglia}, {Bramante}, {Breedt},
  {Bressan}, {Brouillet}, {Brugaletta}, {Bucciarelli}, {Burlacu}, {Butkevich}, {Buzzi}, {Caffau}, {Cancelliere}, {Cantat-Gaudin}, {Carballo}, {Carlucci}, {Carnerero}, {Carrasco}, {Casamiquela}, {Castellani}, {Castro-Ginard}, {Chaoul}, {Charlot}, {Chemin}, {Chiaramida}, {Chiavassa}, {Chornay}, {Comoretto}, {Contursi}, {Cooper}, {Cornez}, {Cowell}, {Crifo}, {Cropper}, {Crosta}, {Crowley}, {Dafonte}, {Dapergolas}, {David}, {David}, {de Laverny}, {De Luise}, {De March}, {De Ridder}, {de Souza}, {de Torres}, {del Peloso}, {del Pozo}, {Delbo}, {Delgado}, {Delisle}, {Demouchy}, {Dharmawardena}, {Di Matteo}, {Diakite}, {Diener}, {Distefano}, {Dolding}, {Edvardsson}, {Enke}, {Fabre}, {Fabrizio}, {Faigler}, {Fedorets}, {Fernique}, {Fienga}, {Figueras}, {Fournier}, {Fouron}, {Fragkoudi}, {Gai}, {Garcia-Gutierrez}, {Garcia-Reinaldos}, {Garc{\'\i}a-Torres}, {Garofalo}, {Gavel}, {Gavras}, {Gerlach}, {Geyer}, {Giacobbe}, {Gilmore}, {Girona}, {Giuffrida}, {Gomel}, {Gomez}, {Gonz{\'a}lez-N{\'u}{\~n}ez},
  {Gonz{\'a}lez-Santamar{\'\i}a}, {Gonz{\'a}lez-Vidal}, {Granvik}, {Guillout}, {Guiraud}, {Guti{\'e}rrez-S{\'a}nchez}, {Guy}, {Hatzidimitriou}, {Hauser}, {Haywood}, {Helmer}, {Helmi}, {Sarmiento}, {Hidalgo}, {Hilger}, {H{\l}adczuk}, {Hobbs}, {Holland}, {Huckle}, {Jardine}, {Jasniewicz}, {Jean-Antoine Piccolo}, {Jim{\'e}nez-Arranz}, {Jorissen}, {Juaristi Campillo}, {Julbe}, {Karbevska}, {Kervella}, {Khanna}, {Kontizas}, {Kordopatis}, {Korn}, {K{\'o}sp{\'a}l}, {Kostrzewa-Rutkowska}, {Kruszy{\'n}ska}, {Kun}, {Laizeau}, {Lambert}, {Lanza}, {Lasne}, {Le Campion}, {Lebreton}, {Lebzelter}, {Leccia}, {Leclerc}, {Lecoeur-Taibi}, {Liao}, {Licata}, {Lindstr{\o}m}, {Lister}, {Livanou}, {Lobel}, {Lorca}, {Loup}, {Madrero Pardo}, {Magdaleno Romeo}, {Managau}, {Mann}, {Manteiga}, {Marchant}, {Marconi}, {Marcos}, {Marcos Santos}, {Mar{\'\i}n Pina}, {Marinoni}, {Marocco}, {Marshall}, {Martin Polo}, {Mart{\'\i}n-Fleitas}, {Marton}, {Mary}, {Masip}, {Massari}, {Mastrobuono-Battisti}, {Mazeh}, {McMillan}, {Messina}, {Michalik},
  {Millar}, {Mints}, {Molina}, {Molinaro}, {Moln{\'a}r}, {Monari}, {Mongui{\'o}}, {Montegriffo}, {Montero}, {Mor}, {Mora}, {Morbidelli}, {Morel}, {Morris}, {Muraveva}, {Murphy}, {Musella}, {Nagy}, {Noval}, {Oca{\~n}a}, {Ogden}, {Ordenovic}, {Osinde}, {Pagani}, {Pagano}, {Palaversa}, {Palicio}, {Pallas-Quintela}, {Panahi}, {Payne-Wardenaar}, {Pe{\~n}alosa Esteller}, {Penttil{\"a}}, {Pichon}, {Piersimoni}, {Pineau}, {Plachy}, {Plum}, {Poggio}, {Pr{\v{s}}a}, {Pulone}, {Racero}, {Ragaini}, {Rainer}, {Raiteri}, {Rambaux}, {Ramos}, {Ramos-Lerate}, {Re Fiorentin}, {Regibo}, {Richards}, {Rios Diaz}, {Ripepi}, {Riva}, {Rix}, {Rixon}, {Robichon}, {Robin}, {Robin}, {Roelens}, {Rogues}, {Rohrbasser}, {Romero-G{\'o}mez}, {Rowell}, {Royer}, {Ruz Mieres}, {Rybicki}, {Sadowski}, {S{\'a}ez N{\'u}{\~n}ez}, {Sagrist{\`a} Sell{\'e}s}, {Sahlmann}, {Salguero}, {Samaras}, {Sanchez Gimenez}, {Sanna}, {Santove{\~n}a}, {Sarasso}, {Schultheis}, {Sciacca}, {Segol}, {Segovia}, {S{\'e}gransan}, {Semeux}, {Shahaf}, {Siddiqui}, {Siebert},
  {Siltala}, {Silvelo}, {Slezak}, {Slezak}, {Smart}, {Snaith}, {Solano}, {Solitro}, {Souami}, {Souchay}, {Spagna}, {Spina}, {Spoto}, {Steele}, {Steidelm{\"u}ller}, {Stephenson}, {S{\"u}veges}, {Surdej}, {Szabados}, {Szegedi-Elek}, {Taris}, {Taylor}, {Teixeira}, {Tolomei}, {Tonello}, {Torra}, {Torra}, {Torralba Elipe}, {Trabucchi}, {Tsounis}, {Turon}, {Ulla}, {Unger}, {Vaillant}, {van Dillen}, {van Reeven}, {Vanel}, {Vecchiato}, {Viala}, {Vicente}, {Voutsinas}, {Weiler}, {Wevers}, {Wyrzykowski}, {Yoldas}, {Yvard}, {Zhao}, {Zorec}, {Zucker}, \& {Zwitter}}]{gaiadr3}
---. 2023, \aap, 674, A1, \dodoi{10.1051/0004-6361/202243940}

\bibitem[{{G{\'a}rate} {et~al.}(2021){G{\'a}rate}, {Delage}, {Stadler}, {Pinilla}, {Birnstiel}, {Stammler}, {Picogna}, {Ercolano}, {Franz}, \& {Lenz}}]{garate2021}
{G{\'a}rate}, M., {Delage}, T.~N., {Stadler}, J., {et~al.} 2021, \aap, 655, A18, \dodoi{10.1051/0004-6361/202141444}

\bibitem[{{Gonz{\'a}lez-Ruilova} {et~al.}(2020){Gonz{\'a}lez-Ruilova}, {Cieza}, {Hales}, {P{\'e}rez}, {Zurlo}, {Arce-Tord}, {Casassus}, {C{\'a}novas}, {Flock}, {Herczeg}, {Pinilla}, {Price}, {Principe}, {Ru{\'\i}z-Rodr{\'\i}guez}, \& {Williams}}]{gonzalez-ruilova2020}
{Gonz{\'a}lez-Ruilova}, C., {Cieza}, L.~A., {Hales}, A.~S., {et~al.} 2020, \apjl, 902, L33, \dodoi{10.3847/2041-8213/abbcce}

\bibitem[{{Guo} {et~al.}(2018){Guo}, {Herczeg}, {Jose}, {Fu}, {Chiang}, {Grankin}, {Michel}, {Kesh Yadav}, {Liu}, {Chen}, {Li}, {Xue}, {Niu}, {Subramaniam}, {Sharma}, {Prasert}, {Flores-Fajardo}, {Castro}, \& {Altamirano}}]{guo18}
{Guo}, Z., {Herczeg}, G.~J., {Jose}, J., {et~al.} 2018, \apj, 852, 56, \dodoi{10.3847/1538-4357/aa9e52}

\bibitem[{{Hardegree-Ullman} {et~al.}(2019){Hardegree-Ullman}, {Cushing}, {Muirhead}, \& {Christiansen}}]{hardegree-ullman2019}
{Hardegree-Ullman}, K.~K., {Cushing}, M.~C., {Muirhead}, P.~S., \& {Christiansen}, J.~L. 2019, \aj, 158, 75, \dodoi{10.3847/1538-3881/ab21d2}

\bibitem[{{Hashimoto} {et~al.}(2021){Hashimoto}, {Dong}, \& {Muto}}]{hashimoto2021}
{Hashimoto}, J., {Dong}, R., \& {Muto}, T. 2021, \aj, 161, 264, \dodoi{10.3847/1538-3881/abf431}

\bibitem[{{Hendler} {et~al.}(2020){Hendler}, {Pascucci}, {Pinilla}, {Tazzari}, {Carpenter}, {Malhotra}, \& {Testi}}]{hendler2020}
{Hendler}, N., {Pascucci}, I., {Pinilla}, P., {et~al.} 2020, \apj, 895, 126, \dodoi{10.3847/1538-4357/ab70ba}

\bibitem[{{Herczeg} \& {Hillenbrand}(2008)}]{herczeg08}
{Herczeg}, G.~J., \& {Hillenbrand}, L.~A. 2008, \apj, 681, 594, \dodoi{10.1086/586728}

\bibitem[{{Herczeg} \& {Hillenbrand}(2014)}]{herczeg14}
---. 2014, \apj, 786, 97, \dodoi{10.1088/0004-637X/786/2/97}

\bibitem[{{Hildebrand}(1983)}]{hildebrand1983}
{Hildebrand}, R.~H. 1983, \qjras, 24, 267

\bibitem[{{Huang} {et~al.}(2018{\natexlab{a}}){Huang}, {Andrews}, {P{\'e}rez}, {Zhu}, {Dullemond}, {Isella}, {Benisty}, {Bai}, {Birnstiel}, {Carpenter}, {Guzm{\'a}n}, {Hughes}, {{\"O}berg}, {Ricci}, {Wilner}, \& {Zhang}}]{huang2018_spiral}
{Huang}, J., {Andrews}, S.~M., {P{\'e}rez}, L.~M., {et~al.} 2018{\natexlab{a}}, \apjl, 869, L43, \dodoi{10.3847/2041-8213/aaf7a0}

\bibitem[{{Huang} {et~al.}(2018{\natexlab{b}}){Huang}, {Andrews}, {Dullemond}, {Isella}, {P{\'e}rez}, {Guzm{\'a}n}, {{\"O}berg}, {Zhu}, {Zhang}, {Bai}, {Benisty}, {Birnstiel}, {Carpenter}, {Hughes}, {Ricci}, {Weaver}, \& {Wilner}}]{huang2018}
{Huang}, J., {Andrews}, S.~M., {Dullemond}, C.~P., {et~al.} 2018{\natexlab{b}}, \apjl, 869, L42, \dodoi{10.3847/2041-8213/aaf740}

\bibitem[{{Huang} {et~al.}(2020){Huang}, {Andrews}, {Dullemond}, {{\"O}berg}, {Qi}, {Zhu}, {Birnstiel}, {Carpenter}, {Isella}, {Mac{\'\i}as}, {McClure}, {P{\'e}rez}, {Teague}, {Wilner}, \& {Zhang}}]{huang2020}
---. 2020, \apj, 891, 48, \dodoi{10.3847/1538-4357/ab711e}

\bibitem[{{Jennings} {et~al.}(2022){Jennings}, {Tazzari}, {Clarke}, {Booth}, \& {Rosotti}}]{jennings2022}
{Jennings}, J., {Tazzari}, M., {Clarke}, C.~J., {Booth}, R.~A., \& {Rosotti}, G.~P. 2022, \mnras, 514, 6053, \dodoi{10.1093/mnras/stac1770}

\bibitem[{{Kraus} \& {Hillenbrand}(2012)}]{kraus2012}
{Kraus}, A.~L., \& {Hillenbrand}, L.~A. 2012, \apj, 757, 141, \dodoi{10.1088/0004-637X/757/2/141}

\bibitem[{{Kraus} {et~al.}(2011){Kraus}, {Ireland}, {Martinache}, \& {Hillenbrand}}]{kraus2011}
{Kraus}, A.~L., {Ireland}, M.~J., {Martinache}, F., \& {Hillenbrand}, L.~A. 2011, \apj, 731, 8, \dodoi{10.1088/0004-637X/731/1/8}

\bibitem[{{Kurtovic} {et~al.}(2021){Kurtovic}, {Pinilla}, {Long}, {Benisty}, {Manara}, {Natta}, {Pascucci}, {Ricci}, {Scholz}, \& {Testi}}]{kurtovic2021}
{Kurtovic}, N.~T., {Pinilla}, P., {Long}, F., {et~al.} 2021, \aap, 645, A139, \dodoi{10.1051/0004-6361/202038983}

\bibitem[{{Lantz} {et~al.}(2004){Lantz}, {Aldering}, {Antilogus}, {Bonnaud}, {Capoani}, {Castera}, {Copin}, {Dubet}, {Gangler}, {Henault}, {Lemonnier}, {Pain}, {Pecontal}, {Pecontal}, \& {Smadja}}]{lantz04}
{Lantz}, B., {Aldering}, G., {Antilogus}, P., {et~al.} 2004, in Society of Photo-Optical Instrumentation Engineers (SPIE) Conference Series, Vol. 5249, Optical Design and Engineering, ed. L.~{Mazuray}, P.~J. {Rogers}, \& R.~{Wartmann}, 146--155, \dodoi{10.1117/12.512493}

\bibitem[{{Lauer} {et~al.}(1995){Lauer}, {Ajhar}, {Byun}, {Dressler}, {Faber}, {Grillmair}, {Kormendy}, {Richstone}, \& {Tremaine}}]{lauer1995}
{Lauer}, T.~R., {Ajhar}, E.~A., {Byun}, Y.~I., {et~al.} 1995, \aj, 110, 2622, \dodoi{10.1086/117719}

\bibitem[{{Lesur}(2021)}]{lesur2021}
{Lesur}, G. R.~J. 2021, \aap, 650, A35, \dodoi{10.1051/0004-6361/202040109}

\bibitem[{{Lin} {et~al.}(2018){Lin}, {Lee}, \& {Chiang}}]{lin2018}
{Lin}, J.~W., {Lee}, E.~J., \& {Chiang}, E. 2018, \mnras, 480, 4338, \dodoi{10.1093/mnras/sty2159}

\bibitem[{{Liu} {et~al.}(2019){Liu}, {Lambrechts}, {Johansen}, \& {Liu}}]{liu2019}
{Liu}, B., {Lambrechts}, M., {Johansen}, A., \& {Liu}, F. 2019, \aap, 632, A7, \dodoi{10.1051/0004-6361/201936309}

\bibitem[{{Liu}(2021)}]{liu2021iceline}
{Liu}, Y. 2021, Research in Astronomy and Astrophysics, 21, 164, \dodoi{10.1088/1674-4527/21/7/164}

\bibitem[{{Long} {et~al.}(2018){Long}, {Pinilla}, {Herczeg}, {Harsono}, {Dipierro}, {Pascucci}, {Hendler}, {Tazzari}, {Ragusa}, {Salyk}, {Edwards}, {Lodato}, {van de Plas}, {Johnstone}, {Liu}, {Boehler}, {Cabrit}, {Manara}, {Menard}, {Mulders}, {Nisini}, {Fischer}, {Rigliaco}, {Banzatti}, {Avenhaus}, \& {Gully-Santiago}}]{long2018}
{Long}, F., {Pinilla}, P., {Herczeg}, G.~J., {et~al.} 2018, \apj, 869, 17, \dodoi{10.3847/1538-4357/aae8e1}

\bibitem[{{Long} {et~al.}(2019){Long}, {Herczeg}, {Harsono}, {Pinilla}, {Tazzari}, {Manara}, {Pascucci}, {Cabrit}, {Nisini}, {Johnstone}, {Edwards}, {Salyk}, {Menard}, {Lodato}, {Boehler}, {Mace}, {Liu}, {Mulders}, {Hendler}, {Ragusa}, {Fischer}, {Banzatti}, {Rigliaco}, {van de Plas}, {Dipierro}, {Gully-Santiago}, \& {Lopez-Valdivia}}]{long2019}
{Long}, F., {Herczeg}, G.~J., {Harsono}, D., {et~al.} 2019, \apj, 882, 49, \dodoi{10.3847/1538-4357/ab2d2d}

\bibitem[{{Long} {et~al.}(2022){Long}, {Andrews}, {Rosotti}, {Harsono}, {Pinilla}, {Wilner}, {{\"O}berg}, {Teague}, {Trapman}, \& {Tabone}}]{long2022}
{Long}, F., {Andrews}, S.~M., {Rosotti}, G., {et~al.} 2022, \apj, 931, 6, \dodoi{10.3847/1538-4357/ac634e}

\bibitem[{{Long} {et~al.}(2023){Long}, {Ren}, {Wallack}, {Harsono}, {Herczeg}, {Pinilla}, {Mawet}, {Liu}, {Andrews}, {Bai}, {Cabrit}, {Cieza}, {Johnstone}, {Leisenring}, {Lodato}, {Liu}, {Manara}, {Mulders}, {Ragusa}, {Sallum}, {Shi}, {Tazzari}, {Uyama}, {Wagner}, {Wilner}, \& {Xuan}}]{long2023}
{Long}, F., {Ren}, B.~B., {Wallack}, N.~L., {et~al.} 2023, \apj, 949, 27, \dodoi{10.3847/1538-4357/acc843}

\bibitem[{{Longarini} {et~al.}(2023){Longarini}, {Lodato}, {Bertin}, \& {Armitage}}]{longarini2023}
{Longarini}, C., {Lodato}, G., {Bertin}, G., \& {Armitage}, P.~J. 2023, \mnras, 519, 2017, \dodoi{10.1093/mnras/stac3653}

\bibitem[{{Luhman}(2018)}]{luhman18}
{Luhman}, K.~L. 2018, \aj, 156, 271, \dodoi{10.3847/1538-3881/aae831}

\bibitem[{{Luhman} {et~al.}(2010){Luhman}, {Allen}, {Espaillat}, {Hartmann}, \& {Calvet}}]{luhman2010}
{Luhman}, K.~L., {Allen}, P.~R., {Espaillat}, C., {Hartmann}, L., \& {Calvet}, N. 2010, \apjs, 186, 111, \dodoi{10.1088/0067-0049/186/1/111}

\bibitem[{{Luhman} {et~al.}(2017){Luhman}, {Mamajek}, {Shukla}, \& {Loutrel}}]{luhman2017}
{Luhman}, K.~L., {Mamajek}, E.~E., {Shukla}, S.~J., \& {Loutrel}, N.~P. 2017, \aj, 153, 46, \dodoi{10.3847/1538-3881/153/1/46}

\bibitem[{{MacGregor} {et~al.}(2018){MacGregor}, {Weinberger}, {Wilner}, {Kowalski}, \& {Cranmer}}]{macgregor2018}
{MacGregor}, M.~A., {Weinberger}, A.~J., {Wilner}, D.~J., {Kowalski}, A.~F., \& {Cranmer}, S.~R. 2018, \apjl, 855, L2, \dodoi{10.3847/2041-8213/aaad6b}

\bibitem[{{Mairs} {et~al.}(2019){Mairs}, {Lalchand}, {Bower}, {Forbrich}, {Bell}, {Herczeg}, {Johnstone}, {Chen}, {Lee}, \& {Hacar}}]{mairs2019}
{Mairs}, S., {Lalchand}, B., {Bower}, G.~C., {et~al.} 2019, \apj, 871, 72, \dodoi{10.3847/1538-4357/aaf3b1}

\bibitem[{{Manara} {et~al.}(2023){Manara}, {Ansdell}, {Rosotti}, {Hughes}, {Armitage}, {Lodato}, \& {Williams}}]{manara2023}
{Manara}, C.~F., {Ansdell}, M., {Rosotti}, G.~P., {et~al.} 2023, in Astronomical Society of the Pacific Conference Series, Vol. 534, Protostars and Planets VII, ed. S.~{Inutsuka}, Y.~{Aikawa}, T.~{Muto}, K.~{Tomida}, \& M.~{Tamura}, 539, \dodoi{10.48550/arXiv.2203.09930}

\bibitem[{{Martel} \& {Lesur}(2022)}]{martel2022}
{Martel}, {\'E}., \& {Lesur}, G. 2022, \aap, 667, A17, \dodoi{10.1051/0004-6361/202142946}

\bibitem[{{Marton} {et~al.}(2017){Marton}, {Calzoletti}, {Perez Garcia}, {Kiss}, {Paladini}, {Altieri}, {Sanchez Portal}, {Kidger}, \& {the Herschel Point Source Catalogue Working Group}}]{marton2017}
{Marton}, G., {Calzoletti}, L., {Perez Garcia}, A.~M., {et~al.} 2017, arXiv e-prints, arXiv:1705.05693, \dodoi{10.48550/arXiv.1705.05693}

\bibitem[{{Mercer} \& {Stamatellos}(2020)}]{mercer2020}
{Mercer}, A., \& {Stamatellos}, D. 2020, \aap, 633, A116, \dodoi{10.1051/0004-6361/201936954}

\bibitem[{{Miguel} {et~al.}(2020){Miguel}, {Cridland}, {Ormel}, {Fortney}, \& {Ida}}]{miguel2020}
{Miguel}, Y., {Cridland}, A., {Ormel}, C.~W., {Fortney}, J.~J., \& {Ida}, S. 2020, \mnras, 491, 1998, \dodoi{10.1093/mnras/stz3007}

\bibitem[{{Miotello} {et~al.}(2023){Miotello}, {Kamp}, {Birnstiel}, {Cleeves}, \& {Kataoka}}]{miotello2023}
{Miotello}, A., {Kamp}, I., {Birnstiel}, T., {Cleeves}, L.~C., \& {Kataoka}, A. 2023, in Astronomical Society of the Pacific Conference Series, Vol. 534, Protostars and Planets VII, ed. S.~{Inutsuka}, Y.~{Aikawa}, T.~{Muto}, K.~{Tomida}, \& M.~{Tamura}, 501, \dodoi{10.48550/arXiv.2203.09818}

\bibitem[{{Miranda} {et~al.}(2017){Miranda}, {Mu{\~n}oz}, \& {Lai}}]{miranda2017}
{Miranda}, R., {Mu{\~n}oz}, D.~J., \& {Lai}, D. 2017, \mnras, 466, 1170, \dodoi{10.1093/mnras/stw3189}

\bibitem[{{Morales} {et~al.}(2019){Morales}, {Mustill}, {Ribas}, {Davies}, {Reiners}, {Bauer}, {Kossakowski}, {Herrero}, {Rodr{\'\i}guez}, {L{\'o}pez-Gonz{\'a}lez}, {Rodr{\'\i}guez-L{\'o}pez}, {B{\'e}jar}, {Gonz{\'a}lez-Cuesta}, {Luque}, {Pall{\'e}}, {Perger}, {Baroch}, {Johansen}, {Klahr}, {Mordasini}, {Anglada-Escud{\'e}}, {Caballero}, {Cort{\'e}s-Contreras}, {Dreizler}, {Lafarga}, {Nagel}, {Passegger}, {Reffert}, {Rosich}, {Schweitzer}, {Tal-Or}, {Trifonov}, {Zechmeister}, {Quirrenbach}, {Amado}, {Guenther}, {Hagen}, {Henning}, {Jeffers}, {Kaminski}, {K{\"u}rster}, {Montes}, {Seifert}, {Abell{\'a}n}, {Abril}, {Aceituno}, {Aceituno}, {Alonso-Floriano}, {Ammler-von Eiff}, {Antona}, {Arroyo-Torres}, {Azzaro}, {Barrado}, {Becerril-Jarque}, {Ben{\'\i}tez}, {Berdi{\~n}as}, {Bergond}, {Brinkm{\"o}ller}, {del Burgo}, {Burn}, {Calvo-Ortega}, {Cano}, {C{\'a}rdenas}, {Cardona Guill{\'e}n}, {Carro}, {Casal}, {Casanova}, {Casasayas-Barris}, {Chaturvedi}, {Cifuentes}, {Claret}, {Colom{\'e}}, {Czesla},
  {D{\'\i}ez-Alonso}, {Dorda}, {Emsenhuber}, {Fern{\'a}ndez}, {Fern{\'a}ndez-Mart{\'\i}n}, {Ferro}, {Fuhrmeister}, {Galad{\'\i}-Enr{\'\i}quez}, {Gallardo Cava}, {Garc{\'\i}a Vargas}, {Garcia-Piquer}, {Gesa}, {Gonz{\'a}lez-{\'A}lvarez}, {Gonz{\'a}lez Hern{\'a}ndez}, {Gonz{\'a}lez-Peinado}, {Gu{\`a}rdia}, {Guijarro}, {de Guindos}, {Hatzes}, {Hauschildt}, {Hedrosa}, {Hermelo}, {Hern{\'a}ndez Arabi}, {Hern{\'a}ndez Otero}, {Hintz}, {Holgado}, {Huber}, {Huke}, {Johnson}, {de Juan}, {Kehr}, {Kemmer}, {Kim}, {Kl{\"u}ter}, {Klutsch}, {Labarga}, {Labiche}, {Lalitha}, {Lamp{\'o}n}, {Lara}, {Launhardt}, {L{\'a}zaro}, {Lizon}, {Llamas}, {Lodieu}, {L{\'o}pez del Fresno}, {L{\'o}pez Salas}, {L{\'o}pez-Santiago}, {Mag{\'a}n Madinabeitia}, {Mall}, {Mancini}, {Mandel}, {Marfil}, {Mar{\'\i}n Molina}, {Mart{\'\i}n}, {Mart{\'\i}n-Fern{\'a}ndez}, {Mart{\'\i}n-Ruiz}, {Mart{\'\i}nez-Rodr{\'\i}guez}, {Marvin}, {Mirabet}, {Moya}, {Naranjo}, {Nelson}, {Nortmann}, {Nowak}, {Ofir}, {Pascual}, {Pavlov}, {Pedraz}, {P{\'e}rez Medialdea},
  {P{\'e}rez-Calpena}, {Perryman}, {Rabaza}, {Ram{\'o}n Ballesta}, {Rebolo}, {Redondo}, {Rix}, {Rodler}, {Rodr{\'\i}guez Trinidad}, {Sabotta}, {Sadegi}, {Salz}, {S{\'a}nchez-Blanco}, {S{\'a}nchez Carrasco}, {S{\'a}nchez-L{\'o}pez}, {Sanz-Forcada}, {Sarkis}, {Sarmiento}, {Sch{\"a}fer}, {Schlecker}, {Schmitt}, {Sch{\"o}fer}, {Solano}, {Sota}, {Stahl}, {Stock}, {Stuber}, {St{\"u}rmer}, {Su{\'a}rez}, {Tabernero}, {Tulloch}, {Veredas}, {Vico-Linares}, {Vilardell}, {Wagner}, {Winkler}, {Wolthoff}, {Yan}, \& {Zapatero Osorio}}]{morales2019}
{Morales}, J.~C., {Mustill}, A.~J., {Ribas}, I., {et~al.} 2019, Science, 365, 1441, \dodoi{10.1126/science.aax3198}

\bibitem[{{Mulders} {et~al.}(2021){Mulders}, {Dr{\k{a}}{\.z}kowska}, {van der Marel}, {Ciesla}, \& {Pascucci}}]{mulders2021}
{Mulders}, G.~D., {Dr{\k{a}}{\.z}kowska}, J., {van der Marel}, N., {Ciesla}, F.~J., \& {Pascucci}, I. 2021, \apjl, 920, L1, \dodoi{10.3847/2041-8213/ac2947}

\bibitem[{{Mulders} {et~al.}(2015){Mulders}, {Pascucci}, \& {Apai}}]{mulders2015}
{Mulders}, G.~D., {Pascucci}, I., \& {Apai}, D. 2015, \apj, 798, 112, \dodoi{10.1088/0004-637X/798/2/112}

\bibitem[{{Oke} \& {Gunn}(1982)}]{oke82}
{Oke}, J.~B., \& {Gunn}, J.~E. 1982, \pasp, 94, 586, \dodoi{10.1086/131027}

\bibitem[{{Okuzumi} {et~al.}(2016){Okuzumi}, {Momose}, {Sirono}, {Kobayashi}, \& {Tanaka}}]{okuzumi2016}
{Okuzumi}, S., {Momose}, M., {Sirono}, S.-i., {Kobayashi}, H., \& {Tanaka}, H. 2016, \apj, 821, 82, \dodoi{10.3847/0004-637X/821/2/82}

\bibitem[{{Owen}(2020)}]{owen2020}
{Owen}, J.~E. 2020, \mnras, 495, 3160, \dodoi{10.1093/mnras/staa1309}

\bibitem[{{Owen} {et~al.}(2012){Owen}, {Clarke}, \& {Ercolano}}]{owen2012}
{Owen}, J.~E., {Clarke}, C.~J., \& {Ercolano}, B. 2012, \mnras, 422, 1880, \dodoi{10.1111/j.1365-2966.2011.20337.x}

\bibitem[{{Paneque-Carre{\~n}o} {et~al.}(2021){Paneque-Carre{\~n}o}, {P{\'e}rez}, {Benisty}, {Hall}, {Veronesi}, {Lodato}, {Sierra}, {Carpenter}, {Andrews}, {Bae}, {Henning}, {Kwon}, {Linz}, {Loinard}, {Pinte}, {Ricci}, {Tazzari}, {Testi}, \& {Wilner}}]{paneque-carreno2021}
{Paneque-Carre{\~n}o}, T., {P{\'e}rez}, L.~M., {Benisty}, M., {et~al.} 2021, \apj, 914, 88, \dodoi{10.3847/1538-4357/abf243}

\bibitem[{{Pascucci} {et~al.}(2011){Pascucci}, {Sterzik}, {Alexander}, {Alencar}, {Gorti}, {Hollenbach}, {Owen}, {Ercolano}, \& {Edwards}}]{pascucci2011}
{Pascucci}, I., {Sterzik}, M., {Alexander}, R.~D., {et~al.} 2011, \apj, 736, 13, \dodoi{10.1088/0004-637X/736/1/13}

\bibitem[{{Pascucci} {et~al.}(2016){Pascucci}, {Testi}, {Herczeg}, {Long}, {Manara}, {Hendler}, {Mulders}, {Krijt}, {Ciesla}, {Henning}, {Mohanty}, {Drabek-Maunder}, {Apai}, {Sz{\H{u}}cs}, {Sacco}, \& {Olofsson}}]{pascucci2016}
{Pascucci}, I., {Testi}, L., {Herczeg}, G.~J., {et~al.} 2016, \apj, 831, 125, \dodoi{10.3847/0004-637X/831/2/125}

\bibitem[{{Pecaut} \& {Mamajek}(2013)}]{pecaut13}
{Pecaut}, M.~J., \& {Mamajek}, E.~E. 2013, \apjs, 208, 9, \dodoi{10.1088/0067-0049/208/1/9}

\bibitem[{{P{\'e}rez} {et~al.}(2016){P{\'e}rez}, {Carpenter}, {Andrews}, {Ricci}, {Isella}, {Linz}, {Sargent}, {Wilner}, {Henning}, {Deller}, {Chandler}, {Dullemond}, {Lazio}, {Menten}, {Corder}, {Storm}, {Testi}, {Tazzari}, {Kwon}, {Calvet}, {Greaves}, {Harris}, \& {Mundy}}]{perez2016}
{P{\'e}rez}, L.~M., {Carpenter}, J.~M., {Andrews}, S.~M., {et~al.} 2016, Science, 353, 1519, \dodoi{10.1126/science.aaf8296}

\bibitem[{{P{\'e}rez} {et~al.}(2020){P{\'e}rez}, {Casassus}, {Hales}, {Marino}, {Cheetham}, {Zurlo}, {Cieza}, {Dong}, {Alarc{\'o}n}, {Ben{\'\i}tez-Llambay}, {Fomalont}, \& {Avenhaus}}]{perez2020}
{P{\'e}rez}, S., {Casassus}, S., {Hales}, A., {et~al.} 2020, \apjl, 889, L24, \dodoi{10.3847/2041-8213/ab6b2b}

\bibitem[{{Picogna} {et~al.}(2021){Picogna}, {Ercolano}, \& {Espaillat}}]{picogna2021}
{Picogna}, G., {Ercolano}, B., \& {Espaillat}, C.~C. 2021, \mnras, 508, 3611, \dodoi{10.1093/mnras/stab2883}

\bibitem[{{Picogna} {et~al.}(2019){Picogna}, {Ercolano}, {Owen}, \& {Weber}}]{picogna2019}
{Picogna}, G., {Ercolano}, B., {Owen}, J.~E., \& {Weber}, M.~L. 2019, \mnras, 487, 691, \dodoi{10.1093/mnras/stz1166}

\bibitem[{{Picogna} {et~al.}(2023){Picogna}, {Sch{\"a}fer}, {Ercolano}, {Rab}, {Franz}, \& {G{\'a}rate}}]{picogna2023}
{Picogna}, G., {Sch{\"a}fer}, C., {Ercolano}, B., {et~al.} 2023, \mnras, 523, 3318, \dodoi{10.1093/mnras/stad1504}

\bibitem[{{Pinilla}(2022)}]{pinilla2022}
{Pinilla}, P. 2022, European Physical Journal Plus, 137, 1206, \dodoi{10.1140/epjp/s13360-022-03384-1}

\bibitem[{{Pinilla} {et~al.}(2013){Pinilla}, {Birnstiel}, {Benisty}, {Ricci}, {Natta}, {Dullemond}, {Dominik}, \& {Testi}}]{pinilla2013}
{Pinilla}, P., {Birnstiel}, T., {Benisty}, M., {et~al.} 2013, \aap, 554, A95, \dodoi{10.1051/0004-6361/201220875}

\bibitem[{{Pinilla} {et~al.}(2012){Pinilla}, {Birnstiel}, {Ricci}, {Dullemond}, {Uribe}, {Testi}, \& {Natta}}]{pinilla2012}
{Pinilla}, P., {Birnstiel}, T., {Ricci}, L., {et~al.} 2012, \aap, 538, A114, \dodoi{10.1051/0004-6361/201118204}

\bibitem[{{Pinilla} {et~al.}(2020){Pinilla}, {Pascucci}, \& {Marino}}]{pinilla2020}
{Pinilla}, P., {Pascucci}, I., \& {Marino}, S. 2020, \aap, 635, A105, \dodoi{10.1051/0004-6361/201937003}

\bibitem[{{Pinilla} {et~al.}(2017{\natexlab{a}}){Pinilla}, {Pohl}, {Stammler}, \& {Birnstiel}}]{pinilla2017iceline}
{Pinilla}, P., {Pohl}, A., {Stammler}, S.~M., \& {Birnstiel}, T. 2017{\natexlab{a}}, \apj, 845, 68, \dodoi{10.3847/1538-4357/aa7edb}

\bibitem[{{Pinilla} {et~al.}(2015){Pinilla}, {van der Marel}, {P{\'e}rez}, {van Dishoeck}, {Andrews}, {Birnstiel}, {Herczeg}, {Pontoppidan}, \& {van Kempen}}]{pinilla2015}
{Pinilla}, P., {van der Marel}, N., {P{\'e}rez}, L.~M., {et~al.} 2015, \aap, 584, A16, \dodoi{10.1051/0004-6361/201526655}

\bibitem[{{Pinilla} {et~al.}(2017{\natexlab{b}}){Pinilla}, {P{\'e}rez}, {Andrews}, {van der Marel}, {van Dishoeck}, {Ataiee}, {Benisty}, {Birnstiel}, {Juh{\'a}sz}, {Natta}, {Ricci}, \& {Testi}}]{pinilla2017}
{Pinilla}, P., {P{\'e}rez}, L.~M., {Andrews}, S., {et~al.} 2017{\natexlab{b}}, \apj, 839, 99, \dodoi{10.3847/1538-4357/aa6973}

\bibitem[{{Pinilla} {et~al.}(2018){Pinilla}, {Tazzari}, {Pascucci}, {Youdin}, {Garufi}, {Manara}, {Testi}, {van der Plas}, {Barenfeld}, {Canovas}, {Cox}, {Hendler}, {P{\'e}rez}, \& {van der Marel}}]{pinilla2018}
{Pinilla}, P., {Tazzari}, M., {Pascucci}, I., {et~al.} 2018, \apj, 859, 32, \dodoi{10.3847/1538-4357/aabf94}

\bibitem[{{Pinilla} {et~al.}(2021){Pinilla}, {Kurtovic}, {Benisty}, {Manara}, {Natta}, {Sanchis}, {Tazzari}, {Stammler}, {Ricci}, \& {Testi}}]{pinilla2021}
{Pinilla}, P., {Kurtovic}, N.~T., {Benisty}, M., {et~al.} 2021, \aap, 649, A122, \dodoi{10.1051/0004-6361/202140371}

\bibitem[{{Pollack} {et~al.}(1996){Pollack}, {Hubickyj}, {Bodenheimer}, {Lissauer}, {Podolak}, \& {Greenzweig}}]{pollack1996}
{Pollack}, J.~B., {Hubickyj}, O., {Bodenheimer}, P., {et~al.} 1996, \icarus, 124, 62, \dodoi{10.1006/icar.1996.0190}

\bibitem[{{Rab} {et~al.}(2023){Rab}, {Weber}, {Picogna}, {Ercolano}, \& {Owen}}]{rab2023}
{Rab}, C., {Weber}, M.~L., {Picogna}, G., {Ercolano}, B., \& {Owen}, J.~E. 2023, \apjl, 955, L11, \dodoi{10.3847/2041-8213/acf574}

\bibitem[{{Ragusa} {et~al.}(2020){Ragusa}, {Alexander}, {Calcino}, {Hirsh}, \& {Price}}]{ragusa2020}
{Ragusa}, E., {Alexander}, R., {Calcino}, J., {Hirsh}, K., \& {Price}, D.~J. 2020, \mnras, 499, 3362, \dodoi{10.1093/mnras/staa2954}

\bibitem[{{Ribas} {et~al.}(2023){Ribas}, {Reiners}, {Zechmeister}, {Caballero}, {Morales}, {Sabotta}, {Baroch}, {Amado}, {Quirrenbach}, {Abril}, {Aceituno}, {Anglada-Escud{\'e}}, {Azzaro}, {Barrado}, {B{\'e}jar}, {Ben{\'\i}tez de Haro}, {Bergond}, {Bluhm}, {Calvo Ortega}, {Cardona Guill{\'e}n}, {Chaturvedi}, {Cifuentes}, {Colom{\'e}}, {Cont}, {Cort{\'e}s-Contreras}, {Czesla}, {D{\'\i}ez-Alonso}, {Dreizler}, {Duque-Arribas}, {Espinoza}, {Fern{\'a}ndez}, {Fuhrmeister}, {Galad{\'\i}-Enr{\'\i}quez}, {Garc{\'\i}a-L{\'o}pez}, {Gonz{\'a}lez-{\'A}lvarez}, {Gonz{\'a}lez Hern{\'a}ndez}, {Guenther}, {de Guindos}, {Hatzes}, {Henning}, {Herrero}, {Hintz}, {Huelmo}, {Jeffers}, {Johnson}, {de Juan}, {Kaminski}, {Kemmer}, {Khaimova}, {Khalafinejad}, {Kossakowski}, {K{\"u}rster}, {Labarga}, {Lafarga}, {Lalitha}, {Lamp{\'o}n}, {Lillo-Box}, {Lodieu}, {L{\'o}pez Gonz{\'a}lez}, {L{\'o}pez-Puertas}, {Luque}, {Mag{\'a}n}, {Mancini}, {Marfil}, {Mart{\'\i}n}, {Mart{\'\i}n-Ruiz}, {Molaverdikhani}, {Montes}, {Nagel}, {Nortmann},
  {Nowak}, {Pall{\'e}}, {Passegger}, {Pavlov}, {Pedraz}, {Perdelwitz}, {Perger}, {Ram{\'o}n-Ballesta}, {Reffert}, {Revilla}, {Rodr{\'\i}guez}, {Rodr{\'\i}guez-L{\'o}pez}, {Sadegi}, {S{\'a}nchez Carrasco}, {S{\'a}nchez-L{\'o}pez}, {Sanz-Forcada}, {Sch{\"a}fer}, {Schlecker}, {Schmitt}, {Sch{\"o}fer}, {Schweitzer}, {Seifert}, {Shan}, {Skrzypinski}, {Solano}, {Stahl}, {Stangret}, {Stock}, {St{\"u}rmer}, {Tabernero}, {Tal-Or}, {Trifonov}, {Vanaverbeke}, {Yan}, \& {Zapatero Osorio}}]{ribas2023}
{Ribas}, I., {Reiners}, A., {Zechmeister}, M., {et~al.} 2023, \aap, 670, A139, \dodoi{10.1051/0004-6361/202244879}

\bibitem[{{Rosotti} {et~al.}(2019){Rosotti}, {Booth}, {Tazzari}, {Clarke}, {Lodato}, \& {Testi}}]{rosotti2019}
{Rosotti}, G.~P., {Booth}, R.~A., {Tazzari}, M., {et~al.} 2019, \mnras, 486, L63, \dodoi{10.1093/mnrasl/slz064}

\bibitem[{{Sanchis} {et~al.}(2020){Sanchis}, {Testi}, {Natta}, {Manara}, {Ercolano}, {Preibisch}, {Henning}, {Facchini}, {Miotello}, {de Gregorio-Monsalvo}, {Lopez}, {Mu{\v{z}}i{\'c}}, {Pascucci}, {Santamar{\'\i}a-Miranda}, {Scholz}, {Tazzari}, {van Terwisga}, \& {Williams}}]{sanchis2020}
{Sanchis}, E., {Testi}, L., {Natta}, A., {et~al.} 2020, \aap, 633, A114, \dodoi{10.1051/0004-6361/201936913}

\bibitem[{{Schlecker} {et~al.}(2022){Schlecker}, {Burn}, {Sabotta}, {Seifert}, {Henning}, {Emsenhuber}, {Mordasini}, {Reffert}, {Shan}, \& {Klahr}}]{schlecker2022}
{Schlecker}, M., {Burn}, R., {Sabotta}, S., {et~al.} 2022, \aap, 664, A180, \dodoi{10.1051/0004-6361/202142543}

\bibitem[{{Skrutskie} {et~al.}(2006){Skrutskie}, {Cutri}, {Stiening}, {Weinberg}, {Schneider}, {Carpenter}, {Beichman}, {Capps}, {Chester}, {Elias}, {Huchra}, {Liebert}, {Lonsdale}, {Monet}, {Price}, {Seitzer}, {Jarrett}, {Kirkpatrick}, {Gizis}, {Howard}, {Evans}, {Fowler}, {Fullmer}, {Hurt}, {Light}, {Kopan}, {Marsh}, {McCallon}, {Tam}, {Van Dyk}, \& {Wheelock}}]{skrutskie2006}
{Skrutskie}, M.~F., {Cutri}, R.~M., {Stiening}, R., {et~al.} 2006, \aj, 131, 1163, \dodoi{10.1086/498708}

\bibitem[{{Somers} {et~al.}(2020){Somers}, {Cao}, \& {Pinsonneault}}]{somers20}
{Somers}, G., {Cao}, L., \& {Pinsonneault}, M.~H. 2020, \apj, 891, 29, \dodoi{10.3847/1538-4357/ab722e}

\bibitem[{{Stef{\'a}nsson} {et~al.}(2023){Stef{\'a}nsson}, {Mahadevan}, {Miguel}, {Robertson}, {Delamer}, {Kanodia}, {Ca{\~n}as}, {Winn}, {Ninan}, {Terrien}, {Holcomb}, {Ford}, {Zawadzki}, {Bowler}, {Bender}, {Cochran}, {Diddams}, {Endl}, {Fredrick}, {Halverson}, {Hearty}, {Hill}, {Lin}, {Metcalf}, {Monson}, {Ramsey}, {Roy}, {Schwab}, {Wright}, \& {Zeimann}}]{stefansson2023}
{Stef{\'a}nsson}, G., {Mahadevan}, S., {Miguel}, Y., {et~al.} 2023, Science, 382, 1031, \dodoi{10.1126/science.abo0233}

\bibitem[{{Suzuki} {et~al.}(2016){Suzuki}, {Ogihara}, {Morbidelli}, {Crida}, \& {Guillot}}]{suzuki2016}
{Suzuki}, T.~K., {Ogihara}, M., {Morbidelli}, A., {Crida}, A., \& {Guillot}, T. 2016, \aap, 596, A74, \dodoi{10.1051/0004-6361/201628955}

\bibitem[{{Tazzari} {et~al.}(2018){Tazzari}, {Beaujean}, \& {Testi}}]{tazzari2018}
{Tazzari}, M., {Beaujean}, F., \& {Testi}, L. 2018, \mnras, 476, 4527, \dodoi{10.1093/mnras/sty409}

\bibitem[{{Tazzari} {et~al.}(2021){Tazzari}, {Clarke}, {Testi}, {Williams}, {Facchini}, {Manara}, {Natta}, \& {Rosotti}}]{tazzari2021}
{Tazzari}, M., {Clarke}, C.~J., {Testi}, L., {et~al.} 2021, \mnras, 506, 2804, \dodoi{10.1093/mnras/stab1808}

\bibitem[{{Tobin} {et~al.}(2020){Tobin}, {Sheehan}, {Megeath}, {D{\'\i}az-Rodr{\'\i}guez}, {Offner}, {Murillo}, {van 't Hoff}, {van Dishoeck}, {Osorio}, {Anglada}, {Furlan}, {Stutz}, {Reynolds}, {Karnath}, {Fischer}, {Persson}, {Looney}, {Li}, {Stephens}, {Chandler}, {Cox}, {Dunham}, {Tychoniec}, {Kama}, {Kratter}, {Kounkel}, {Mazur}, {Maud}, {Patel}, {Perez}, {Sadavoy}, {Segura-Cox}, {Sharma}, {Stephenson}, {Watson}, \& {Wyrowski}}]{tobin2020}
{Tobin}, J.~J., {Sheehan}, P.~D., {Megeath}, S.~T., {et~al.} 2020, \apj, 890, 130, \dodoi{10.3847/1538-4357/ab6f64}

\bibitem[{{Trapman} {et~al.}(2019){Trapman}, {Facchini}, {Hogerheijde}, {van Dishoeck}, \& {Bruderer}}]{trapman2019}
{Trapman}, L., {Facchini}, S., {Hogerheijde}, M.~R., {van Dishoeck}, E.~F., \& {Bruderer}, S. 2019, \aap, 629, A79, \dodoi{10.1051/0004-6361/201834723}

\bibitem[{{Tripathi} {et~al.}(2017){Tripathi}, {Andrews}, {Birnstiel}, \& {Wilner}}]{tripathi2017}
{Tripathi}, A., {Andrews}, S.~M., {Birnstiel}, T., \& {Wilner}, D.~J. 2017, \apj, 845, 44, \dodoi{10.3847/1538-4357/aa7c62}

\bibitem[{{Tychoniec} {et~al.}(2020){Tychoniec}, {Manara}, {Rosotti}, {van Dishoeck}, {Cridland}, {Hsieh}, {Murillo}, {Segura-Cox}, {van Terwisga}, \& {Tobin}}]{tychoniec2020}
{Tychoniec}, {\L}., {Manara}, C.~F., {Rosotti}, G.~P., {et~al.} 2020, \aap, 640, A19, \dodoi{10.1051/0004-6361/202037851}

\bibitem[{{Valenti} {et~al.}(1993){Valenti}, {Basri}, \& {Johns}}]{valenti93}
{Valenti}, J.~A., {Basri}, G., \& {Johns}, C.~M. 1993, \aj, 106, 2024, \dodoi{10.1086/116783}

\bibitem[{{van der Marel} \& {Mulders}(2021)}]{vdMarel2021}
{van der Marel}, N., \& {Mulders}, G.~D. 2021, \aj, 162, 28, \dodoi{10.3847/1538-3881/ac0255}

\bibitem[{{van der Marel} {et~al.}(2016){van der Marel}, {Verhaar}, {van Terwisga}, {Mer{\'\i}n}, {Herczeg}, {Ligterink}, \& {van Dishoeck}}]{vdMarel2016}
{van der Marel}, N., {Verhaar}, B.~W., {van Terwisga}, S., {et~al.} 2016, \aap, 592, A126, \dodoi{10.1051/0004-6361/201628075}

\bibitem[{{van der Marel} {et~al.}(2013){van der Marel}, {van Dishoeck}, {Bruderer}, {Birnstiel}, {Pinilla}, {Dullemond}, {van Kempen}, {Schmalzl}, {Brown}, {Herczeg}, {Mathews}, \& {Geers}}]{vdMarel2013}
{van der Marel}, N., {van Dishoeck}, E.~F., {Bruderer}, S., {et~al.} 2013, Science, 340, 1199, \dodoi{10.1126/science.1236770}

\bibitem[{{van der Marel} {et~al.}(2018){van der Marel}, {Williams}, {Ansdell}, {Manara}, {Miotello}, {Tazzari}, {Testi}, {Hogerheijde}, {Bruderer}, {van Terwisga}, \& {van Dishoeck}}]{vdMarel2018}
{van der Marel}, N., {Williams}, J.~P., {Ansdell}, M., {et~al.} 2018, \apj, 854, 177, \dodoi{10.3847/1538-4357/aaaa6b}

\bibitem[{{van der Marel} {et~al.}(2022){van der Marel}, {Williams}, {Picogna}, {van Terwisga}, {Facchini}, {Manara}, {Zormpas}, {Ansdell}, \& {.}}]{vdMarel2022}
{van der Marel}, N., {Williams}, J.~P., {Picogna}, G., {et~al.} 2022, arXiv e-prints, arXiv:2204.08225, \dodoi{10.48550/arXiv.2204.08225}

\bibitem[{{van der Plas} {et~al.}(2016){van der Plas}, {M{\'e}nard}, {Ward-Duong}, {Bulger}, {Harvey}, {Pinte}, {Patience}, {Hales}, \& {Casassus}}]{vdPlas2016}
{van der Plas}, G., {M{\'e}nard}, F., {Ward-Duong}, K., {et~al.} 2016, \apj, 819, 102, \dodoi{10.3847/0004-637X/819/2/102}

\bibitem[{{Villenave} {et~al.}(2019){Villenave}, {Benisty}, {Dent}, {M{\'e}nard}, {Garufi}, {Ginski}, {Pinilla}, {Pinte}, {Williams}, {de Boer}, {Morino}, {Fukagawa}, {Dominik}, {Flock}, {Henning}, {Juh{\'a}sz}, {Keppler}, {Muro-Arena}, {Olofsson}, {P{\'e}rez}, {van der Plas}, {Zurlo}, {Carle}, {Feautrier}, {Pavlov}, {Pragt}, {Ramos}, {Sauvage}, {Stadler}, \& {Weber}}]{villenave2019}
{Villenave}, M., {Benisty}, M., {Dent}, W.~R.~F., {et~al.} 2019, \aap, 624, A7, \dodoi{10.1051/0004-6361/201834800}

\bibitem[{{Wafflard-Fernandez} \& {Lesur}(2023)}]{wafflard-fernandez2023}
{Wafflard-Fernandez}, G., \& {Lesur}, G. 2023, arXiv e-prints, arXiv:2305.11784, \dodoi{10.48550/arXiv.2305.11784}

\bibitem[{{Wang} {et~al.}(2019){Wang}, {Bai}, \& {Goodman}}]{wang2019mhd}
{Wang}, L., {Bai}, X.-N., \& {Goodman}, J. 2019, \apj, 874, 90, \dodoi{10.3847/1538-4357/ab06fd}

\bibitem[{{Wang} \& {Chen}(2019)}]{wang19}
{Wang}, S., \& {Chen}, X. 2019, \apj, 877, 116, \dodoi{10.3847/1538-4357/ab1c61}

\bibitem[{{Weber} {et~al.}(2023){Weber}, {P{\'e}rez}, {Zurlo}, {Miley}, {Hales}, {Cieza}, {Principe}, {C{\'a}rcamo}, {Garufi}, {K{\'o}sp{\'a}l}, {Takami}, {Kastner}, {Zhu}, \& {Williams}}]{weber2023}
{Weber}, P., {P{\'e}rez}, S., {Zurlo}, A., {et~al.} 2023, \apjl, 952, L17, \dodoi{10.3847/2041-8213/ace186}

\bibitem[{{Wright} {et~al.}(2010){Wright}, {Eisenhardt}, {Mainzer}, {Ressler}, {Cutri}, {Jarrett}, {Kirkpatrick}, {Padgett}, {McMillan}, {Skrutskie}, {Stanford}, {Cohen}, {Walker}, {Mather}, {Leisawitz}, {Gautier}, {McLean}, {Benford}, {Lonsdale}, {Blain}, {Mendez}, {Irace}, {Duval}, {Liu}, {Royer}, {Heinrichsen}, {Howard}, {Shannon}, {Kendall}, {Walsh}, {Larsen}, {Cardon}, {Schick}, {Schwalm}, {Abid}, {Fabinsky}, {Naes}, \& {Tsai}}]{wright2010}
{Wright}, E.~L., {Eisenhardt}, P. R.~M., {Mainzer}, A.~K., {et~al.} 2010, \aj, 140, 1868, \dodoi{10.1088/0004-6256/140/6/1868}

\bibitem[{{Yamaguchi} {et~al.}(2021){Yamaguchi}, {Tsukagoshi}, {Muto}, {Nomura}, {Nakazato}, {Ikeda}, {Tamura}, \& {Kawabe}}]{yamaguchi2021}
{Yamaguchi}, M., {Tsukagoshi}, T., {Muto}, T., {et~al.} 2021, \apj, 923, 121, \dodoi{10.3847/1538-4357/ac2bfd}

\bibitem[{{Zhang} {et~al.}(2015){Zhang}, {Blake}, \& {Bergin}}]{zhangke2015}
{Zhang}, K., {Blake}, G.~A., \& {Bergin}, E.~A. 2015, \apjl, 806, L7, \dodoi{10.1088/2041-8205/806/1/L7}

\bibitem[{{Zhang} {et~al.}(2023){Zhang}, {Kalscheur}, {Long}, {Zhang}, {Long}, {Bergin}, {Zhu}, \& {Trapman}}]{zhang2023}
{Zhang}, S., {Kalscheur}, M., {Long}, F., {et~al.} 2023, arXiv e-prints, arXiv:2305.03862, \dodoi{10.48550/arXiv.2305.03862}

\bibitem[{{Zhu} \& {Dong}(2021)}]{zhudong2021}
{Zhu}, W., \& {Dong}, S. 2021, \araa, 59, 291, \dodoi{10.1146/annurev-astro-112420-020055}

\bibitem[{{Zormpas} {et~al.}(2022){Zormpas}, {Birnstiel}, {Rosotti}, \& {Andrews}}]{zormpas2022}
{Zormpas}, A., {Birnstiel}, T., {Rosotti}, G.~P., \& {Andrews}, S.~M. 2022, \aap, 661, A66, \dodoi{10.1051/0004-6361/202142046}

\end{thebibliography}


\end{CJK*}
\end{document}